\documentclass[twocolumn]{aastex631}

\shorttitle{AASTeX v6.31}
\shortauthors{Huang et al.}
\graphicspath{{./}{figures/}}
\usepackage{times}
\usepackage{mathptmx}

\usepackage{graphicx}
\usepackage{natbib}
\hypersetup
{
	colorlinks=true,
	linkcolor=blue,
	filecolor=blue,
	urlcolor=blue,
	citecolor=blue,
}

\begin{document}

\title{GTC/OSIRIS Deep Spectroscopy of Galactic Compact Planetary Nebulae:  PN\,G048.5+04.2 and PN\,G068.7+14.8
}

\author[0009-0004-3407-0848]{Haomiao Huang}
\affiliation{CAS Key Laboratory of Optical Astronomy, National Astronomical Observatories, Chinese Academy of Sciences, Beijing 100101, P.~R.\ China}
\affiliation{School of Astronomy and Space Sciences, University of Chinese Academy of Sciences, Beijing 100049, P.~R.\ China}

\author[0000-0003-1286-2743]{Xuan Fang}
\affiliation{CAS Key Laboratory of Optical Astronomy, National Astronomical Observatories, Chinese Academy of Sciences, Beijing 100101, P.~R.\ China}
\affiliation{School of Astronomy and Space Sciences, University of Chinese Academy of Sciences, Beijing 100049, P.~R.\ China}
\affiliation{Xinjiang Astronomical Observatory, Chinese Academy of Sciences, 150 Science 1-Street, Urumqi, Xinjiang, 830011, P.~R.\ China}
\affiliation{Laboratory for Space Research, Faculty of Science, The University of Hong Kong, Pokfulam Road, Hong Kong, P.~R.\ China}

\author[0000-0003-4047-0309]{Letizia Stanghellini}
\affiliation{NSF's NOIRLab, 950~N.\ Cherry Ave., Tucson, AZ 85719, USA}

\author{Ting-Hui Lee}
\affiliation{Dept. of Physics and Astronomy, Western Kentucky University, 1906 College Heights Blvd \#11077, Bowling Green, KY 42101, USA}

\author[0000-0002-7759-106X]{Mart\'{i}n A. Guerrero}
\affiliation{Instituto de Astrof\'{i}sica de Andaluc\'{i}a (IAA-CSIC), Glorieta de la Astronom\'{i}a s/n, E-18008, Granada, Spain}

\author[0000-0003-2090-5416]{Xiaohu Li}
\affiliation{Xinjiang Astronomical Observatory, Chinese Academy of Sciences, 150 Science 1-Street, Urumqi, Xinjiang, 830011, P.~R.\ China}
\affiliation{Xinjiang Key Laboratory of Radio Astrophysics, 150 Science 1-Street, Urumqi, Xinjiang 830011, P.~R.\ China}

\author[0000-0003-4058-5202]{Richard Shaw}
\affiliation{Space Telescope Science Institute, 3700 San Martin Drive, Baltimore, MD 21218, USA}

\author{Jifeng Liu}
\affiliation{CAS Key Laboratory of Optical Astronomy, National Astronomical Observatories, Chinese Academy of Sciences, Beijing 100101, P.~R.\ China}
\affiliation{School of Astronomy and Space Sciences, University of Chinese Academy of Sciences, Beijing 100049, P.~R.\ China}
\affiliation{Institute for Frontiers in Astronomy and Astrophysics, Beijing Normal University, Beijing 102206, P.~R.\ China}
\affiliation{New Cornerstone Science Laboratory, National Astronomical Observatories, Chinese Academy of Sciences, Beijing 100012, P.~R.\ China}

\correspondingauthor{Xuan Fang}
\email{fangx@nao.cas.cn}

\begin{abstract}
We report deep long-slit spectroscopy of two Galactic compact planetary nebulae (PNe), PN\,G048.5+04.2 and PN\,G068.7+14.8, obtained using the 10.4m Gran Telescopio Canarias (GTC).  These spectra cover a broad wavelength range of $\sim$3630--10370\,{\AA}, enabling detection of numerous emission lines critical for nebular analysis, including [O\,{\sc ii}] $\lambda$3727 and [O\,{\sc iii}] $\lambda$4363 in the blue and [S\,{\sc iii}] $\lambda\lambda$6312,\,9069 in the red.  Also detected in the spectrum of PN\,G068.7+14.8 are broad C\,{\sc iii} and C\,{\sc iv} lines probably due to stellar emission, indicating that the central star of this PN is [WC]-type.  The GTC optical-NIR spectra were analyzed in details in conjunction with the archival \emph{Spitzer}/IRS mid-IR spectra, and elemental abundances of the two PNe determined.  Photoionization models were established for the two PNe using {\sc cloudy}, based on the ratios of line fluxes measured from the GTC spectra.  Our best photoionization modeling, in combination with state-of-the-art post-AGB evolutionary model tracks, shows that both PNe evolved from low-mass progenitors ($<$2\,$M_{\odot}$) with relatively young ages ($<$3\,Gyr), although their central cores are probably in different evolutionary stages -- the central star of PN\,G068.7+14.8 is still in the process of heating up, while that of PN\,G048.5+04.2 has already entered the white dwarf cooling track.  A comparison with AGB model predictions also constrains the progenitors of both PNe to be of low masses.  Moreover, the two PNe are in line with the radial oxygen gradients exhibited by the Galactic PN populations, using the most up-to-date distances. 
\end{abstract}

\keywords{Galaxy: abundances -- planetary nebulae: individual (PN\,G048.5+04.2, PN\,G068.7+14.8) -- stars: evolution}

\section{Introduction} \label{sec:intro}	

During the late-stage evolution of the low- to intermediate-mass ($\sim$1--8\,$M_{\odot}$) stars, they ascend the red giant branch (RGB) and subsequently the asymptotic giant branch (AGB) phases, and the newly formed elements are transported to the stellar surface through convection and other processes known as dredge-ups \citep{2000oepn.book.....K}.  These elements are then ejected into space by mass losses in forms of stellar winds.  The ejected material rich in heavy elements will gradually mix with the surrounding interstellar medium (ISM), altering chemical composition of the local environments and ultimately contributing to chemical enrichment of the host galaxy.  Therefore, studying the chemical composition of the circumstellar material helps to constrain the AGB nucleosynthesis, peer into the mass-loss processes experienced in the late-stage stellar evolution, and understand the chemical evolution of galaxies. 

Following the ending of AGB, stellar evolution enters the phase of planetary nebulae (PNe), which were first formed through the interacting stellar winds \citep[ISW; e.g.][]{1978ApJ...219L.125K,2000oepn.book.....K} and then photoionized by the central stars (usually a hot white dwarf with effective temperature $\gtrsim$5--20$\times$10$^{4}$\,K).  There are multiple tracers -- mainly stars, star clusters, and various ISM -- used to investigate the structure and evolution of a galaxy.  PNe are the only ISM that exist in almost every part of a galaxy, from the bulge to the disk and further to the halo regions; in comparison, the other ISM (H\,{\sc ii} regions, supernova remnants, molecular clouds, etc.) all reside in the galactic disk. 

Since narrow, bright emission lines dominate the UV-optical spectrum of a PN, nebular abundances of various ions can be calculated with high precision based on the measurements of line fluxes, ensuring reliable determination of chemical composition as well as derivation of the stellar population of this PN.  Moreover, Galactic structure and evolution have been investigated using PNe as tracers \citep[e.g.][]{2006ApJ...651..898S,2010ApJ...714.1096S}.  Previous chemical studies of PNe found that Galactic disk exhibits negative radial abundance gradients in oxygen and neon \citep[e.g.][]{2006MNRAS.372...45P,2006ApJ...651..898S,2010ApJ...714.1096S,SH_2018,Bucciarelli_2023}.  Comparing the abundance gradients derived from the PNe of different populations (e.g.\ of different main-sequence ages and/or metallicity) sheds light on the evolution of the Milky Way.  PNe thus are pivotal for advancing our knowledge of stellar evolution and galactic chemical enrichment. 

High-quality spectroscopy of Galactic PNe hitherto mainly focuses on nearby ($\lesssim$3--5\,kpc) and bright sources, due to limitation of the telescope apertures (small- and medium-apertures in the majority of observations) that were used.  On the other hand, deep optical spectroscopy of distant, \emph{compact} (and faint) PNe is extremely scarce.  This makes compact PNe seriously understudied, leading to systematic bias in the spectroscopic analysis of Galactic PNe.  Faint PNe at large distances thus need to be observed to form a more complete and statistically unbiased sample.  To accomplish this objective, it is necessary to use large (8--10\,m aperture) telescopes to obtain uniformly high-quality spectra of compact faint PNe to form an unbiased sample. 

We carried out optical spectroscopy of $\sim$20 angularly compact PNe, all northern objects, in the Galactic disk using the 10.4\,m Gran Telescopio Canarias (GTC).  Analysis of the GTC optical-NIR spectra of the whole sample will be reported separately (Fang et al.\ in preparation).  This GTC spectroscopic survey aims to expand the sample of the extant high-quality spectroscopy of compact PNe, and its immediate objectives are: (1) to compare with the AGB model predictions for the yields of low- and intermediate-mass stars, (2) to optimize the calculation of ionization correction factors (ICFs) of heavy elements in PNe in conjunction with the available archival IR and UV spectra, and (3) to derive the Galactic radial abundance gradients of heavy elements, using the \emph{Gaia} distances \citep[e.g.][]{Kimeswenger_2018,2020ApJ...889...21S,2021A&A...656A..51G,2021AJ....161..147B} or the distances derived with statistical methods \citep[e.g.][]{Frew_2016,2016ApJ...830...33S,Bucciarelli_2023}. 

In this paper, we report detailed spectral analyses and photoionization modeling of two compact PNe, PN\,G048.5$+$04.2 (a.k.a.\ K\,4-16) and PN\,G068.7$+$14.8 (a.k.a.\ Sp\,4-1), selected from the northern Galactic sample targeted by our GTC spectroscopy.  
The two PNe exhibit different nature of dust as indicated by the \emph{Spitzer} mid-IR spectra \citep[][]{2012ApJ...753..172S}, and were targeted by the \emph{Hubble Space Telescope} (\emph{HST}) STIS UV spectroscopy \citep[][]{2022ApJ...929..148S}.  Of the two objects, only PN\,G068.7$+$14.8 was spectroscopically observed with 2--2.5\,m optical telescopes \citep{2005MNRAS.362..424W,2021MNRAS.504..816B}; however, the spectral quality was limited given its faintness.  No optical spectroscopy of PN\,G048.5$+$04.2 has ever been reported; therefore, this paper is the first to report deep optical spectroscopy of the PN. 

This paper is arranged as follows:  In Section\,\ref{sec:data}, we describe the GTC long-slit spectroscopy and data reduction, as well as the archival data.  In Section\,\ref{sec:analysis}, we present analysis of GTC optical-NIR spectra, including line-flux measurements, plasma diagnostics, and ionic and elemental abundance determinations; the \emph{Spitzer}/IRS mid-IR archive spectra are also analyzed.  In Section\,\ref{sec:models}, we report photoionization models using the {\sc cloudy} code, based on the ratios of emission-line fluxes as measured from the GTC spectra, to derive the central star parameters of the two PNe.  We then discuss the stellar population of the two PNe, and compare the observed abundance ratios with AGB model predictions in Section\,\ref{sec:discussion}, where we also compare our targets with the Galactic abundance gradients of oxygen and neon represented by the Type\,II PNe.  We summarise our analyses and draw conclusions in Section\,\ref{sec:conclusion}.

\section{The Observational Data} 
\label{sec:data}

\subsection{HST Optical Images} 
\label{subsec:hst}

A sample of 51 compact (angular diameters $\lesssim$4\arcsec) PNe, including PN\,G048.5$+$04.2 and PN\,G068.7$+$14.8, was imaged with the \emph{HST} Wide Field Camera~3 (WFC3) in the F502N (centred at [O\,{\sc iii}] $\lambda$5007), F200LP, F350LP and F814W filters in a snapshop program \citep[GO~11657; PI: L.\ Stanghellini][]{2016ApJ...830...33S}, forming a morphological catalog of compact Galactic PNe.  In the \emph{HST}/WFC3 F502N narrowband image, PN\,G048.5$+$04.2 has an elliptical morphology with $\sim$3\arcsec$\times$4\arcsec\ in size, and PN\,G068.7$+$14.8 is round with $\sim$1\arcsec\ in diameter.  In the \emph{HST} images, both PNe show inner nebular structures (see Figure\,\ref{fig1}), although still with limited spatial resolution due to compactness.

\subsection{GTC Long-slit Spectroscopy and Data Reduction} 
\label{subsec:gtc}

Deep spectra of the two PNe were obtained in 2016 June--July using the Optical System for Imaging and low- to intermediate-Resolution Integrated Spectroscopy (OSIRIS) spectrograph, in the Long-Slit Spectroscopy mode, on the 10.4m GTC at Observatorio del Roque de los Muchachos (ORM, La Palma, Spain).  Observations were made under GTC program \#GTC66-16A (PI: X.\ Fang).  The OSIRIS grisms R1000B (blue) and R1000R (red) were used, covering wavelength ranges $\sim$3630–-7850\,{\AA} and $\sim$5080–-10,370{\AA} (optical to near-IR), respectively; slit width was 1\farcs0.  This instrument setup provides a spectral resolution $R\sim$1000.  The detector of OSIRIS consists of two CCDs\footnote{Note that in 2023, a new blue-sensitive monolithic 4k$\times$4k CCD was installed on the detector for OSIRIS, to replace the original two 2048$\times$4096 CCDs. \url{https://www.gtc.iac.es/instruments/osiris+/osiris+.php}}, each with 2048$\times$4096 pixels.  The pixel size is 15\,$\mu$m, corresponding to an angular size of 0\farcs127.  Spectroscopy was carried out in dark moons with excellent seeings $\sim$0\farcs6--0\farcs8.

\begin{figure*}
\begin{center}
\includegraphics[width=16.75cm,angle=0]{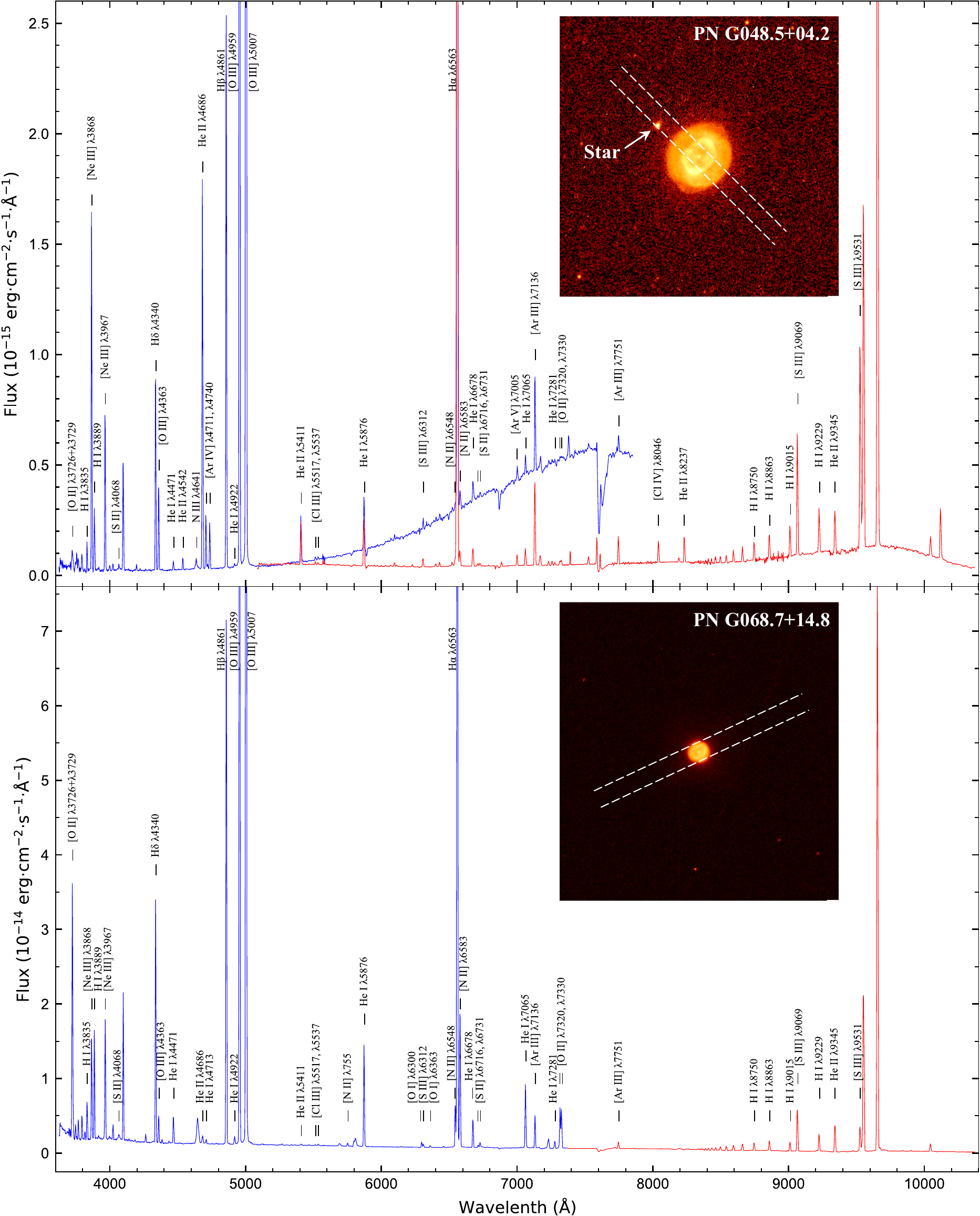}
\caption{Fully calibrated GTC/OSIRIS long-slit spectra of PN\,G048.5$+$04.2 (\emph{top}) and PN\,G068.7$+$14.8 (\emph{bottom}) obtained using the grisms R1000B (blue) and R1000R (red).  Abscissa ranges of both panels are set to accommodate the heights of H$\beta$.  Prominent emission lines in the spectra are identified and labeled.  Extinction has not been corrected for.  In the R1000B spectrum, the weak features between [O\,{\sc ii}] $\lambda\lambda$7320,7330 and [Ar\,{\sc iii}] $\lambda$7751 are the second-order contaminations; in the R1000R spectrum, the strong emission features redward of [S\,{\sc iii}] $\lambda$9531 are also affected by the second-order contamination.  The R1000B spectrum of PN\,G048.5$+$04.2 is contaminated by a field star, resulting in elevated continuum in the red region.  
Overlaid in each panel is the \emph{HST}/WFC3 F502N image (FoV$\sim$15\arcsec$\times$15\arcsec), along with the GTC long slit (1\arcsec\ width) marked as white-dashed lines.  The field star that contaminates the R1000B spectrum of PN\,G048.5$+$04.2 is also marked.} 
\label{fig1}
\end{center}
\end{figure*}

\begin{figure}[t]
\begin{center}
\includegraphics[width=8.2cm,angle=0]{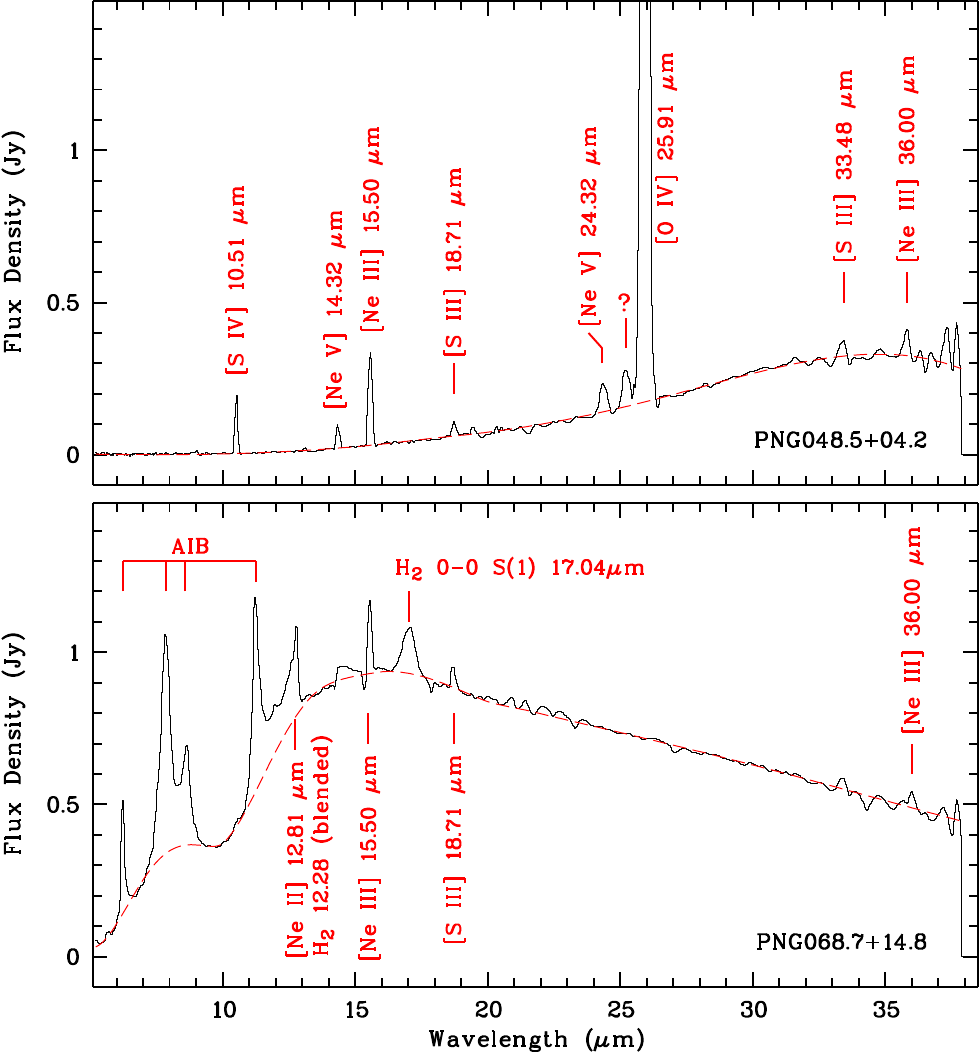}
\caption{\emph{Spitzer}/IRS spectrum of PN\,G048.5$+$04.2 (top) and PN\,G068.7$+$14.8 (bottom), with all emission features labeled and identified in red; an unidentified feature at $\sim$25.2\,$\mu$m in PN\,G048.5$+$04.2 is marked as ``?''.  The AIB bands at $\sim$6.2, 7.7, 8.6 and 11.2\,$\mu$m are also marked in the spectrum of PN\,G068.7$+$14.8 \citep{2012ApJ...753..172S}.  In each panel, the red-dashed curve is the empirically fitted continuum.} 
\label{fig2}
\end{center}
\end{figure}

During the spectroscopic observations, the long slit was placed along the parallactic angles to minimize light loss due to atmospheric diffraction.  Two long exposures and two short exposures were made on both PNe; here the long exposures are to ensure reliable flux measurements of weak emission lines in the spectra, and short exposures are to avoid saturation of strong nebular emission lines such as [O\,{\sc iii}] $\lambda\lambda$4959,5007 and H$\alpha$.  For observations of PN\,G048.5$+$04.2, two 600\,s long exposures were made, plus two 60\,s short exposures.  For PN\,G068.7$+$14.8, which is brighter among the two PNe, two 10\,s short exposures and two 30\,s long exposures were assigned.  For flux calibration of the targets, spectroscopy of two spectrophotometric standard stars, Feige\,66 (PN\,G048.5$+$04.2) and Ross\,640 (for PN\,G068.7$+$14.8), were obtained with a wider slit width of 2\farcs52 (to avoid light loss of a bright star). 

Data reduction of the GTC long-slit spectra followed the standard procedure using {\sc iraf}\footnote{{\sc iraf}, the Image Reduction and Analysis Facility, is distributed by the National Optical Astronomy Observatory, which is operated by the Association of Universities for Research in Astronomy under cooperative agreement with the National Science Foundation.} v2.16 \citep[see detailed description in][]{2015ApJ...815...69F}.  The raw data of PNe was first bias subtracted and flat-field corrected.  Wavelength calibration, as well as correction for geometry distortion (along the slit direction) in the two-dimensional (2D) spectrum, was then carried out using HgAr (for the R1000B spectrum) and HgAr$+$Xe (for the R1000R spectrum) arc lines.  We then subtracted the sky background from each single wavelength-calibrated (and geometry-rectified) 2D spectrum of a target, by fitting the background emission along the slit direction using polynomial functions \citep{2015ApJ...815...69F}.  The background subtracted 2D spectra of each target PN were then combined using the {\sc imcombine} package in {\sc iraf}, to remove cosmic rays. 
The combined, ``cleaned'' 2D spectral image of each target PN was then flux calibrated using the spectrum of a spectrophotometric standard star, whose GTC long-slit spectra were reduced following the same procedure aforementioned.  1D spectra were then extracted from the fully calibrated 2D frame for spectral analysis (Section\,\ref{sec:analysis}).  Figure\,\ref{fig1} shows the final extracted GTC spectra of the two PNe, along with emission line identifications. 

According to the emission features in the spectra (e.g., the large difference in the He\,{\sc ii} $\lambda$4686 line strength), we speculate that the two PNe exhibit totally different excitation natures: PN\,G048.5$+$04.2 is of high-excitation and PN\,G068.7$+$14.8 is a low-excitation PN.  This will be confirmed through detailed spectral analysis in Section\,\ref{sec:analysis}.

\subsection{The Archival Spitzer Spectra} 
\label{subsec:spitzer} 

The mid-IR spectra of 157 Galactic compact PNe, most of which are young and in the early post-AGB phase, were observed using the Infrared Spectrograph \citep[IRS,][]{2004ApJS..154...18H} on the \emph{Spitzer Space Telescope} \citep[GO prop.~ID: 50261; PI: L.\ Stanghellini;][]{2012ApJ...753..172S}, forming a sizeable mid-IR data set.  Both of the two compact PNe targeted by our GTC spectroscopy are in the \emph{Spitzer} sample.  We retrieved the archival \emph{Spitzer} spectra obtained in the Short-Low (SL, $\sim$5--14\,$\mu$m) and Long-Low (LL, $\sim$14--40\,$\mu$m) modules, with slit sizes of $\sim$3\farcs7$\times$57\arcsec\ and $\sim$11\arcsec$\times$168\arcsec, respectively.  Exposures of $\sim$12.6\,s and $\sim$44\,s were made for PN\,G048.5$+$04.2 and PN\,G068.7$+$14.8, respectively. 

In the \emph{Spitzer}/IRS spectra of the two PNe, several collisionally excited lines (CELs, usually also called forbidden lines) emitted by the ions of O, Ne and S are detected, as shown in Figure\,\ref{fig2}.  These mid-IR lines have never been analyzed yet, and are used to calculate ionic abundances in this work (see Section\,\ref{subsec:IRS}).  \citet{2012ApJ...753..172S} studied the dust properties of the two PNe using the \emph{Spitzer}/IRS spectra, revealing carbon-rich dust features in PN\,G068.7$+$14.8, with four AIBs band evident (Figure\,\ref{fig2}). PN\,G048.5$+$04.2 seems to exhibit aliphatic emission features.

\section{Spectral Analysis} \label{sec:analysis}

\subsection{Emission Line Measurements} \label{subsec:emission-line}

\begin{figure}[t]
\begin{center}
\includegraphics[width=8.0cm,angle=0]{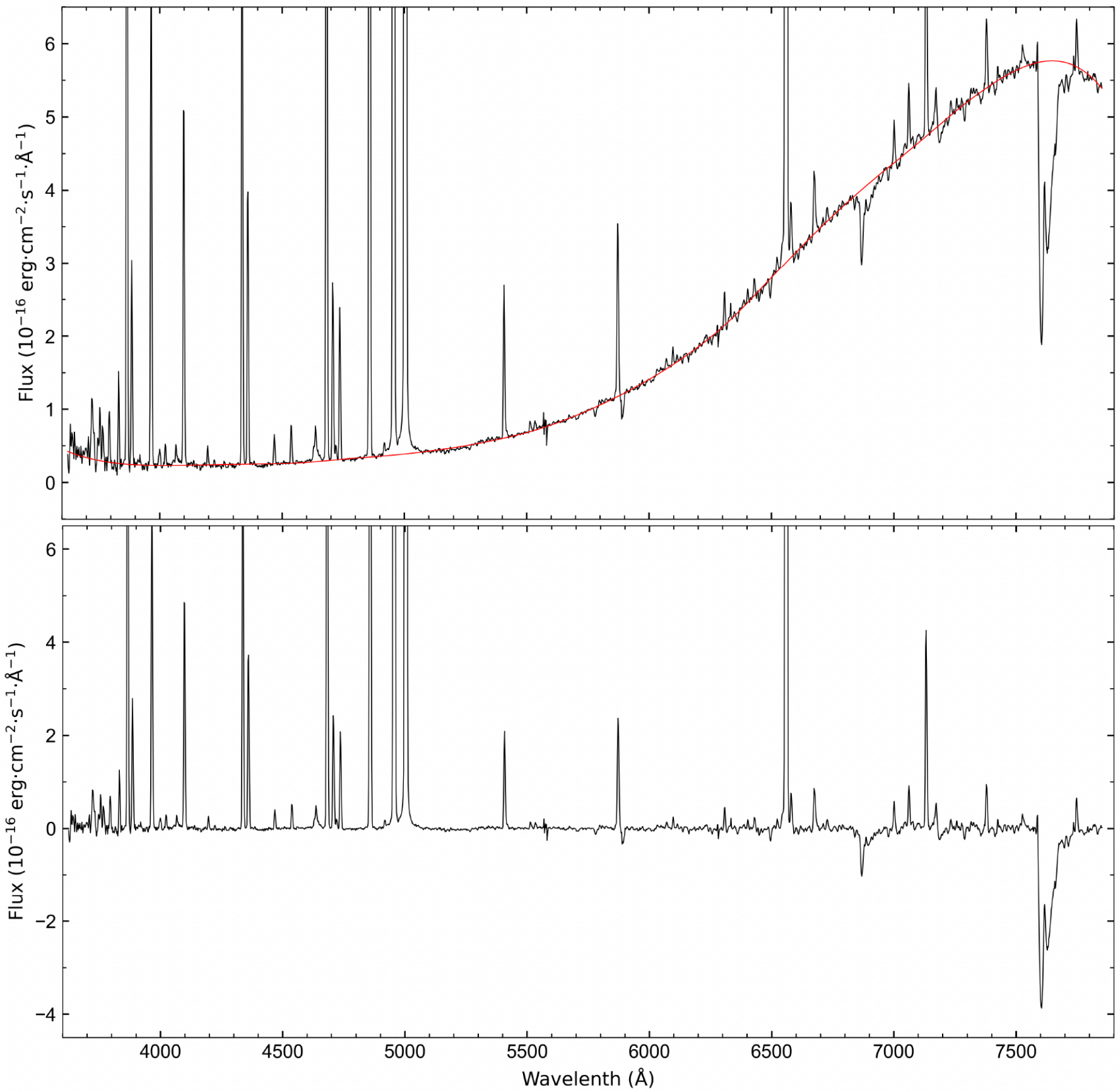}
\caption{\emph{Top}: GTC/OSIRIS R1000B spectrum of PN\,G048.5$+$04.2 overlaid with the segmented quartic polynomial fits of the continuum (red; see the description in text).  \emph{Bottom}: Continuum subtracted spectrum of this PN.} 
\label{fig3}
\end{center}
\end{figure}

Analysis of a PN spectrum, basically plasma diagnostics and abundance determination, mainly utilizes nebular emission lines.  Before measurements of the wavelengths and fluxes of the nebular lines detected in the calibrated GTC long-slit spectra, we need to remove the possible influence from the continuum emission.  This step is critical for PN\,G048.5$+$04.2, whose R1000B (and probably also R1000R) spectrum is contaminated by emission from a nearby field star (see Figure\,\ref{fig1}-top).  Such contamination results in a complex shape of the continuum in the GTC spectrum of PN\,G048.5$+$04.2, which is different from the usual smooth curve with mild slopes (Figure\,\ref{fig1}-bottom).  For both PNe, we divided the GTC spectrum into many wavelength regions (i.e.\ sections), and perform continuum fitting using a quartic polynomial function in each section; the adjacent spectral regions overlap in wavelength coverage so as to ensure smooth connection of the fitting curves (Figure\,\ref{fig3}-top).  Finally, the polynomial-fitted continuum was subtracted from the spectrum, resulting pure emission lines (Figure\,\ref{fig3}-bottom) for follow-up measurements.

We measured emission-line fluxes by integrating over the whole line profiles in the continuum-subtrated spectrum.  Fluxes of all emission lines were normalized to $F$(H$\beta$)=100.  The normalized fluxes of emission lines of both PNe, as well as their total H$\beta$ fluxes (in units of erg\,cm$^{-2}$\,s$^{-1}$) measured from the spectra, are presented in Table\,\ref{tab:lines}.  Fluxes of the strongest emission lines were measured from the short-exposure spectra, while fluxes of the weak lines were measured from the long-exposure spectra.  The flux differences of the emission lines in the overlapping region of R1000B and R1000R are mostly $\lesssim$5--10\%.  For PN\,G048.5$+$04.2, measurements of the emission lines in the overlapping region of the two OSIRIS grisms were made using the R1000R spectrum, to avoid influence from the nearby star.  For PN\,G068.7$+$14.8, emission lines redward of [O\,{\sc ii}] $\lambda\lambda$7320,7330 were measured using the R1000R spectrum, given that the second-order contamination affects its R1000B spectrum in the red region. 

Gaussian profile was used to fit each emission line to obtain its observed wavelength, which was then corrected (for systemic velocity of the PN) and compared with the laboratory wavelength so as to know the emitting ion and the upper and lower levels of this transition.  This task is called emission line identification, where the National Institute of Standards and Technology (NIST) Atomic Spectral Database\footnote{\url{https://www.nist.gov/pml/atomic-spectra-database}} was used.  The transitional informations of emission lines are summarized in Table\,\ref{tab:lines}. 

We derived the logarithmic extinction parameter $c$(H$\beta$) of the two PNe by comparing the measured flux ratios of H\,{\sc i} Balmer lines, including $I$(H$\alpha$)/$I$(H$\beta$) and $I$(H$\gamma$)/$I$(H$\beta$), with the Case~B theoretical calculations of \citet[][at an electron temperature of 10,000~K and a density of 10$^4$\,cm$^{-3}$]{1995MNRAS.272...41S}.  We then made extinction correction of line fluxes using the equation, 
\begin{equation} 
I ({\rm{\lambda}})=10^{c({\rm{H}} \beta)[1+f(\lambda)]} F(\lambda),
\end{equation} 
where $f$($\lambda$) is the standard Galactic extinction curve \citep{1989ApJ...345..245C}, with the total-to-selective extinction ratio $R_{V}$=3.1.  The extinction-corrected line intensities, normalized to $I$(H$\beta$)=100, are compiled in Table\,\ref{tab:lines}, where $c$(H$\beta$) values of the two PNe are also presented.  The uncertainties in line intensity were estimated based on the errors in line flux measurements. 

Although well detected in the GTC spectra of both PNe (Figure\,\ref{fig1}), the H\,{\sc i} Paschen lines were not used for reddening correction because they are generally fainter than the Balmer series and thus are prone to be affected by the uncertainty brought by measurements.  More specifically, the low-order ($n\rightarrow$3, with $n<$10) Paschen lines are affected by the second-order contamination beyond 9200\,{\AA} \citep{2018ApJ...853...50F}, and measurements of the high-order ($n\rightarrow$3, with $n>10$) Paschen lines are probably affected by the continuum, which was empirically fitted (Figure\,\ref{fig3}).  Moreover, the spectral continuum of PN\,G048.5$+$04.2 is contaminated by the emission from a nearby field star (Figure\,\ref{fig1}-top), which may bring extra uncertainty in continuum fits as well as line measurements.

\subsection{Plasma Diagnostics} 
\label{subsec:diagnostics}

\subsubsection{Plasma Diagnostics with CELs} 
\label{subsec:diagnostics:cels}

\begin{figure*}[ht!]
\begin{center}
\includegraphics[width=13cm, angle=0]{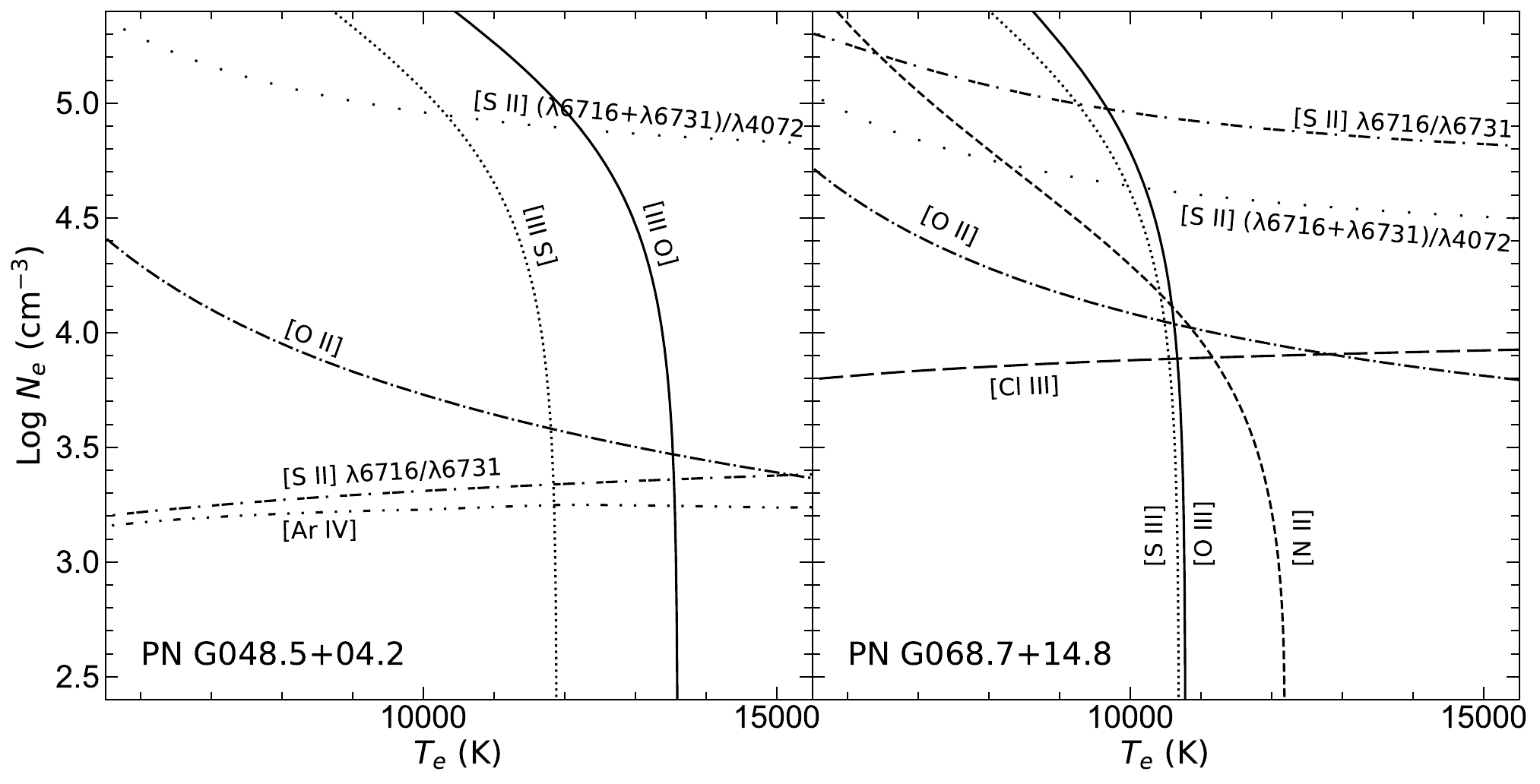} 
\caption{Plasma-diagnostic diagrams of PN\,G048.5+04.2 (left) and PN\,G068.7+14.8 (right) based on the CEL ratios.  Different line types represent the temperature or density diagnostics using different line ratios (see Table\,\ref{tab:temden}).} 
\label{fig4}
\end{center}
\end{figure*}

Plasma diagnostics of the two PNe were carried out using the extinction-corrected intensities of emission lines (most of which are CELs) from Table\,\ref{tab:lines}.  The {\sc pyneb} \citep{2015A&A...573A..42L} package was utilized to derive the nebular electron temperatures ($T_{\rm e}$'s) and densities ($N_{\rm e}$'s) using the intensity ratios of CELs.  References for the atomic data utilized in plasma diagnostics as well as ionic abundance determinations (see Section\,\ref{subsec:ionic}) are listed in Table\,\ref{tab:atomic_data}.  The diagnostic results are summarized in Table\,\ref{tab:temden} along with uncertainties, which were based on the measurement errors of line fluxes.  Plasma-diagnostic diagrams of the two PNe based on the CELs are shown in Figure\,\ref{fig4}, where the curves of different forbidden-line ratios intersect at different points, indicating ionization and thermal structures in the nebulae.  For PN\,G048.5$+$04.2, $N_{\rm e}$ was derived using the [S\,{\sc ii}] $I(\lambda 6716)$/$I(\lambda 6731)$ ratio, while $T_{\rm e}$ was determined using the [O\,{\sc iii}] $I$($\lambda$4959+$\lambda$5007)/$I$($\lambda$4363) nebular-to-auroral line ratio.  To ensure consistency of the results, we derive the [S\,{\sc ii}] density using the [O\,{\sc iii}] temperature as input, and derive the [O\,{\sc iii}] temperature using the [S\,{\sc ii}] density; we iterate this procedure until the final $T_{\rm e}$ and $N_{\rm e}$ values converge.  

Due to measurement errors, compact nature and contamination by the C\,{\sc iii} $\lambda$6730 emission from the central star (see Section\,3.5), the [S\,{\sc ii}] $\lambda\lambda$6716,6731 doublet cannot be used to derive the electron density of PN\,G068.7+14.8.  The [Ar\,{\sc iv}] $\lambda\lambda$4711,4740 doublet were not detected in the spectrum of this PN.  As an alternative, the [O\,{\sc ii}] $\lambda$3727/($\lambda$7320+$\lambda$7330) nebular-to-auroral line ratio was used to diagnose the electron density of PN\,G068.7+14.8, although this line ratio is more often used as a temperature diagnostic.  A similar iterative method was used to derive the converged [O\,{\sc iii}] temperature and [O\,{\sc ii}] density.  The [S\,{\sc ii}] ($\lambda$6716+$\lambda$6731)/$\lambda$4072 (= $\lambda$4068+$\lambda$4076) nebular-to-transauroral line ratio can also be used as a density diagnostic \citep[e.g.][]{1971BOTT....6...29P}; however, according to Figure\,\ref{fig4}, this [S\,{\sc ii}] nebular-to-transauroral line ratio yields very high electron densities in the two PNe (Table\,\ref{tab:temden}), indicating possible density inhomogeneity.  The [O\,{\sc ii}] $\lambda$3727/($\lambda$7320+$\lambda$7330) and [S\,{\sc ii}] ($\lambda$6716+$\lambda$6731)/$\lambda$4072 ratios have high sensitivity on electron density, making them more suitable to probe the high-density clumps in PNe and H\,{\sc ii} regions \citep[][]{Mendez_Delgado_2023}. 

The density-diagnostic [Ar\,{\sc iv}] $\lambda\lambda$4711,4740 doublet are only detected in the GTC spectrum of PN\,G048.5$+$04.2.  [Ar\,{\sc iv}] $\lambda$4711 is blended with He\,{\sc i} $\lambda$4713 and probably also [Ne\,{\sc iv}] $\lambda\lambda$4714,4716.  The flux contribution of He\,{\sc i} $\lambda$4713 was estimated using the measured He\,{\sc i} $\lambda$4471 line flux and the theoretical He\,{\sc i} $\lambda$4713/$\lambda$4471 emissivity ratio calculated by \citet[][in Case~B assumption]{2012MNRAS.425L..28P}.  The flux contribution of [Ne\,{\sc iv}] lines were estimated using the transition probabilities ($A$).  Finally, the [Ar\,{\sc iv}] $\lambda$4711 line intensity was corrected by removing the estimated fluxes of blended He\,{\sc i} and [Ne\,{\sc iv}] lines.  The [Cl\,{\sc iii}] $\lambda\lambda$5517,5537 doublet are detected in both PNe, but too faint to contribute reliable density diagnostics.  Although the [Cl\,{\sc iii}] lines of PN\,G048.5+04.2 are slightly stronger than in PN\,G068.7$+$14.8 (Table\,\ref{tab:lines}), the [Cl\,{\sc iii}] $\lambda$5517/$\lambda$5537 ratio of the former PN is beyond the diagnostic range of {\sc pyneb}, and thus cannot yield any density value.  This may be caused by the large uncertainties in line fluxes.  The electron density of PN\,G068.7$+$14.8 yielded by the [Cl\,{\sc iii}] line ratio, although a reasonable value, is of large uncertainty (Table\,\ref{tab:temden}).

\begin{table}
\begin{center}
\tablenum{2}
\caption{References for Atomic Data}
\label{tab:atomic_data}
\begin{tabular}{lll}
\hline\hline
Ion & \multicolumn{2}{c}{ORLs} \\
 & Effective recombination coeff. & Comments \\
\hline
H~{\sc i} & \citet{1995MNRAS.272...41S} & Case~B \\
He~{\sc i} & \citet{2012MNRAS.425L..28P, 2013MNRAS.433L..89P} & Case~B \\
 & \citet{2022MNRAS.513.1198D} & Case~B \\
He~{\sc ii} & \citet{1995MNRAS.272...41S} & Case~B \\
C~{\sc ii} & \citet{\detokenize{2000A&AS..142...85D}} & Case~B \\
\hline
Ion & \multicolumn{2}{c}{CELs} \\
 & Transition probabilities & Collision strengths \\
\hline
$[$N~{\sc ii}$]$ & \citet{2004ADNDT..87....1F} & \citet{2011ApJS..195...12T} \\
$[$O~{\sc ii}$]$ & \citet{1982MNRAS.198..111Z} & \citet{2009MNRAS.397..903K} \\
$[$O~{\sc iii}$]$ & \citet{2000MNRAS.312..813S} & \citet{2014MNRAS.441.3028S} \\
 & \citet{2004ADNDT..87....1F} &  \\
$[$Ne~{\sc iii}$]$ & \citet{\detokenize{1997A&AS..123..159G}} & \citet{2000JPhB...33..597M} \\
$[$Ne~{\sc iv}$]$ & \citet{1984JPhB...17..681G} & \citet{1981MNRAS.195P..63G} \\
$[$S~{\sc ii}$]$ & \citet{\detokenize{2019A&A...623A.155R}} & \citet{2010ApJS..188...32T} \\
$[$S~{\sc iii}$]$ & \citet{2006ADNDT..92..607F} & \citet{1999ApJ...526..544T} \\
$[$Cl~{\sc iii}$]$ & \citet{\detokenize{2019A&A...623A.155R}} & \citet{1989\detokenize{A&A...208..337B}} \\
$[$Cl~{\sc iv}$]$ & \citet{1986JPCRD..15..321K} & \citet{\detokenize{1995A&AS..111..347G}} \\
$[$Ar~{\sc iii}$]$ & \citet{\detokenize{2009A&A...500.1253M}} & \citet{\detokenize{2009A&A...500.1253M}} \\
$[$Ar~{\sc iv}$]$ & \citet{\detokenize{2019A&A...623A.155R}} & \citet{1997ADNDT..66...65R} \\
$[$Ar~{\sc v}$]$ & \citet{1986JPCRD..15..321K} & \citet{\detokenize{1995A&AS..111..347G}} \\
\hline
\end{tabular}
\end{center}
\end{table}

The [O\,{\sc ii}] $\lambda$3727/($\lambda$7320+$\lambda$7330) line ratio of PN\,G048.5$+$04.2 yields an abnormally high electron temperature, but an electron density consistent with that derived from the [S\,{\sc ii}] doublet (Table\,\ref{tab:temden}).  This [O\,{\sc ii}] line ratio works well in both temperature and density diagnostics for PN\,G068.7$+$14.8.  The [S\,{\sc iii}] ($\lambda$9069+$\lambda$9531)/$\lambda$6312 and [S\,{\sc ii}] ($\lambda$6716+$\lambda$6731)/$\lambda$4072 line ratios were also used to derive electron temperatures of the two PNe.  However, [S\,{\sc ii}] $\lambda$4072 suffers from line-blending issue, resulting in poorly constrained electron temperatures of the two PNe (see Table\,\ref{tab:temden}).  In both PNe, the measured line flux of [S\,{\sc iii}] $\lambda$9531 suffers from the second-order contamination and thus cannot be directly used.  We derived the intrinsic [S\,{\sc iii}] $\lambda$9531 line intensity by adopting the theoretical ratio $I$($\lambda$9531)/$I$($\lambda$9069) = 2.48 \citep{1982MNRAS.199.1025M} given that both lines decay from the same upper level ($^{1}$D$_{2}$) of S$^{2+}$.  The [N\,{\sc ii}] $\lambda$5755 auroral line was only detected in the GTC spectrum of PN\,G068.7$+$14.8, and was also used to derive the electron temperature of this PN.

In the classical stratified ionization structure of a PN, the [O\,{\sc ii}]-emitting region is expected to have a lower electron temperature than does the [O\,{\sc iii}]-emitting region.  However, the ionizing photons with energies higher than the hydrogen ionization threshold (13.6\,eV) have much lower probability of being absorbed and thus can travel further into the outer nebular regions where O$^{+}$ exists, resulting in an [O\,{\sc ii}] temperature higher than that of [O\,{\sc iii}]; this is the ``hardening effect'' in PNe \citep{OF2006}.  In the case of PN\,G048.5$+$04.2, the derived [O\,{\sc ii}] temperature is higher than the [O\,{\sc iii}] temperature; although this is consistent with the ``hardening effect'', we cannot rule out three other possibilities: (1) [O\,{\sc ii}] $\lambda$3727 is located at the blue end of the GTC/OSIRIS R1000B spectrum, where the detector efficiency is low, causing relatively large measurement uncertainty in line flux; (2) extra uncertainty might be introduced by reddening correction as well as the quality of flux calibration of the spectra given the wide wavelength separation between the nebular and auroral lines of O$^{+}$, despite that our data reduction was very careful (Section\,\ref{subsec:gtc}); and (3) in high-density environments \citep[$N_{\rm e}>$2$\times$10$^{4}$\,cm$^{-3}$,][]{OF2006,2017IAUS..323...43M}, the upper levels $^{2}$D$^{\rm o}_{5/2,3/2}$ of the [O\,{\sc ii}] $\lambda$3727 transitions can be collisionally de-excited by free electrons, thus suppressing the radiation of this line, resulting an underestimated [O\,{\sc ii}] nebular-to-auroral line ratio and producing a high electron temperature. 

It has been discussed that in PNe, the observed fluxes of the temperature-sensitive auroral lines, such as [O\,{\sc ii}] $\lambda\lambda$7320,7330 and [N\,{\sc ii}] $\lambda$5755, which decay from the metastable levels belonging to the 2p$^{n}$ electron configuration and were traditionally treated as being excited by electron collisions, may also be contributed by recombination processes \citep[e.g.][]{1986ApJ...309..334R,2000MNRAS.312..585L,Mendez_Delgado_2023}.  Using the Eq\,2 in \citet{2000MNRAS.312..585L}, we estimated that the recombination intensity of the [O\,{\sc ii}] $\lambda$7325 line in PN\,G048.5$+$04.2 is 0.0016 (relative to H$\beta$) when adopting the H\,{\sc i} Paschen jump (PJ) temperature of 9660\,K (Table\,\ref{tab:temden}) and O$^{2+}$/H$^{+}$ = 1.75$\times$10$^{-4}$ (Table\,\ref{tab:ionic}), as derived from the CELs; this contribution is negligible compared to the observed total intensity of 0.89.  In PN\,G068.7$+$14.8, the recombination contribution to the total intensity of this [O\,{\sc ii}] auroral line is $<$0.02\%.  No [N\,{\sc iii}] mid-IR lines are detected in the \emph{Spitzer}/IRS spectra of the two PNe (Figure\,\ref{fig2}), and thus we cannot estimate the recombination contribution to the total intensity of [N\,{\sc ii}] $\lambda$5755.  Moreover, no heavy-element ORLs are detected in the GTC deep spectra (even the strongest O\,{\sc ii} M1 $\lambda$4650 lines), probably due to faintness of the two PNe as well as low spectral resolution; such non-detection may also indicate that the recombination contribution to the emission of auroral forbidden lines -- at least to [O\,{\sc ii}] $\lambda\lambda$7320,7330 -- is indeed negligible. 

Besides the [O\,{\sc ii}] $\lambda\lambda$7320,7330 and [N\,{\sc ii}] $\lambda$5755 auroral lines, recombination may also play a role in exciting the [S\,{\sc ii}] $\lambda\lambda$4068,4076 transauroral lines ($\lambda$4072 in our GTC low-resolution spectra).  However, the fluxes of the two [S\,{\sc ii}] lines might be uncertain, as both of them are blended with strong recombination lines from the O\,{\sc ii} M10 (3p\,$^{4}$D$^{\rm o}$ -- 3d\,$^{4}$F) multiplet: in particular, [S\,{\sc ii}] $\lambda$4076 is blended with the much stronger O\,{\sc ii} M10 $\lambda$4075.86 (3p\,$^{4}$D$^{\rm o}_{7/2}$ -- 3d\,$^{4}$F$_{9/2}$) line \citep{2000MNRAS.312..585L}.  Moreover, the [S\,{\sc ii}] $\lambda$4068 line might also be blended with the C\,{\sc iii} M16 $\lambda$4069 (4f\,$^{3}$F$^{\rm o}$ -- 5g\,$^{3}$G) emission from the [WC]-type central star of PN\,G068.7$+$14.8 (see Section\,\ref{subsec:central-stars}). 

Although it is reasonable to assume that better correlation between temperature and density diagnostics can be obtained when ions coexist in the same ionization zone, it has been demonstrated by multiple observational and theoretical studies \citep[e.g.][]{1989ApJS...69..897R,Mendez_Delgado_2023} that the dominant factor in density diagnostics is the sensitivity range of a line ratio.  For a nebula with density inhomogeneity, density diagnostics are inherently biased toward the range to which a line ratio is most sensitive.  Our density diagnostics for the two Galactic PNe as described above are also subject to this factor.

\subsubsection{Plasma Diagnostics with ORLs} 
\label{subsec:diagnostics:orls}

\begin{figure}[ht!]
\begin{center}
\includegraphics[width=8.4cm,angle=0]{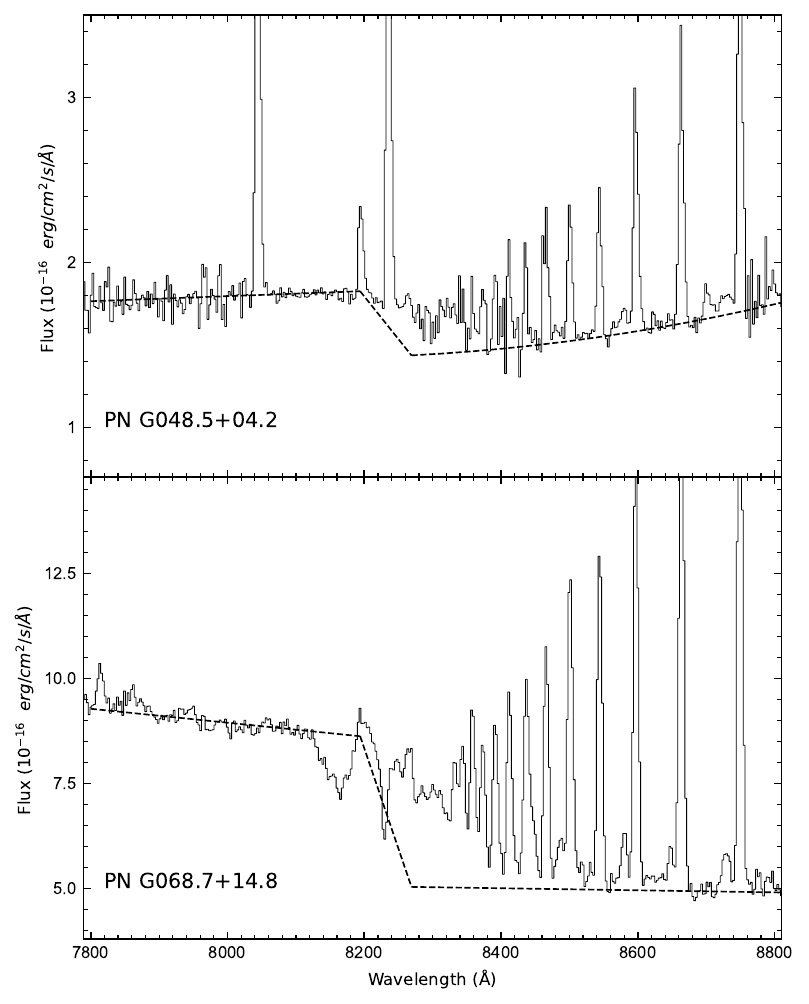}
\caption{GTC/OSIRIS R1000R spectra of PN\,G048.5+04.2 (top) and PN\,G068.7+14.8 (bottom) in the wavelength region of 7800--9000\,{\AA}, showing the Paschen discontinuity at $\sim$8204\,{\AA} and the high-order Paschen decrements.  The dashed curves overplotted are empirical fits to the continuum on both sides of the Paschen jump; note that for PN\,G048.5+04.2, the continuum redward of the Paschen jump of is obviously elevated due to contamination from a field star (as shown in Figure\,\ref{fig1}-top).  Both spectra have been corrected for extinction and radial velocities, as determined through measurements of H\,{\sc i} Balmer and Paschen lines.} 
\label{fig5}
\end{center}
\end{figure}

The optical recombination lines (ORLs) of H\,{\sc i} and He\,{\sc i}, as well as the Paschen jump/discontinuity (at 8204\,{\AA}) of the H\,{\sc i} recombination continuum, are well detected in the GTC/OSIRIS R1000R spectra (Figure\,\ref{fig5}), and can be used to diagnose the electron temperature and density \citep{Zhang_2004,2005MNRAS.358..457Z}.  The blue cutoff the GTC/OSIRIS R1000B spectra is at $\sim$3630\,{\AA}, right to the blue side of the Balmer jump (at 3646\,{\AA}); reliable measurement of the height of Balmer jump is thus impossible due to the low signal-to-noise (S/N) in this region.  Therefore, we derive the electron temperature using the Paschen jump (Figure\,\ref{fig5}), adopting the fitting formula Eq\,7 in \citet{2011MNRAS.415..181F}, 
where the $\rm{He^+/ H^+}$ and $\rm{He^{2+}/ H^{+}}$ abundance ratios were adopted from Table\,\ref{tab:ionic}.  Although many high-order Paschen lines are well detected, the noise level near the Paschen jump is non-negligible (Figure\,\ref{fig5}).  We measured the continuum levels $I_{\rm c}$(8194\,{\AA}) and $I_{\rm c}$(8269\,{\AA}) using the method of \citet{2000MNRAS.312..585L} through extrapolations from the shorter and longer wavelengths, respectively (see the continuum fits in Figure \ref{fig5}).  The electron temperature yielded by the Paschen jump is 9660\,K in PN\,G048.5+04.2 and 9390\,K in PN\,G068.7+14.8 (Table\,\ref{tab:temden}), both lower than those yielded by the CEL ratios.

\begin{figure}[ht!]
\begin{center}
\includegraphics[width=7.85cm,angle=0]{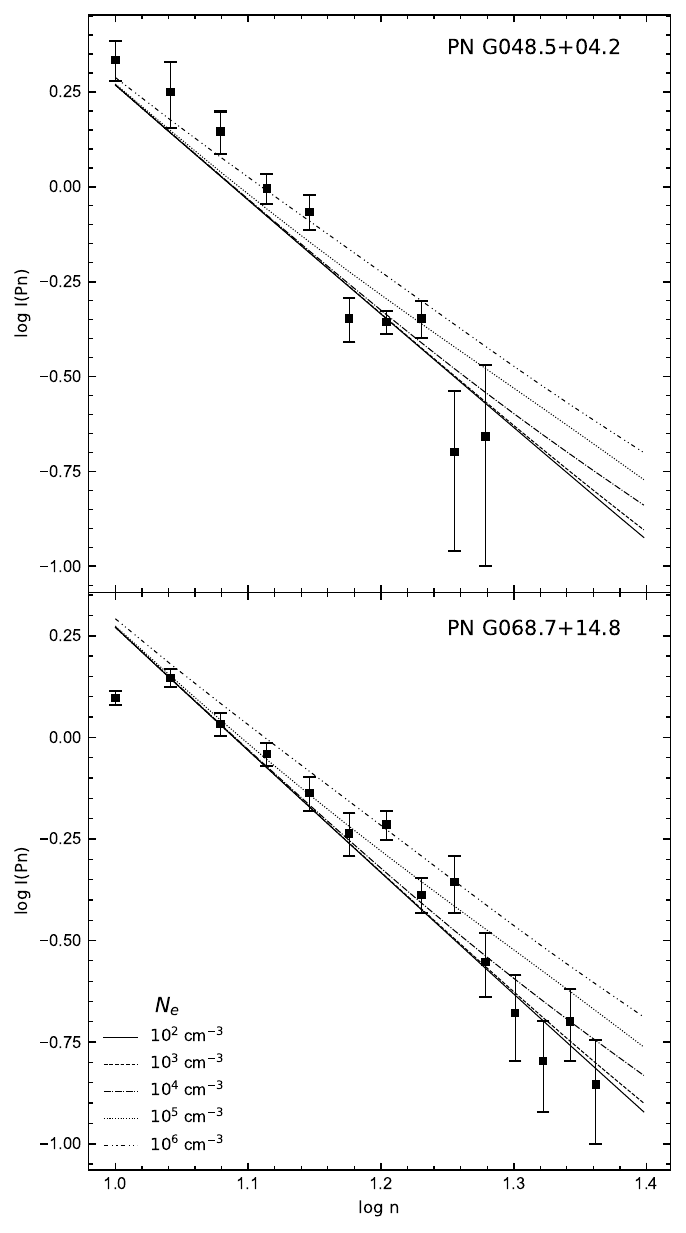}
\caption{Observed intensities (normalized to $I(\rm{H}_{\beta})=100$) of high-order Paschen lines P$n$ ($n\geq10$) of H\,{\sc i}, as measured from the GTC spectrum of PN\,G048.5+04.2 (top) and PN\,G068.7+14.8 (bottom), as a function of principal quantum number $n$.  The various curves show the predicted Paschen decrements for electron densities from 10$^{2}$ to 10$^{6}$\,cm$^{-3}$, calculated at Paschen jump temperatures of 9660\,K (for PN\,G048.5+04.2, top) and 9390\,K (for PN\,G068.7+14.8, bottom).  Measurement uncertainties of the Paschen lines in PN\,G048.5+04.2 are relatively larger.  Measurement of the P10 line in PN\,G068.7+14.8 is underestimated, probably due to over-subtraction of nebular continuum.  Intensities of P16 and P18 might be overestimated due to unknown blending.} 
\label{fig6}
\end{center}
\end{figure}

\begin{table}
\begin{center}
\tablenum{3}
\caption{Plasma Diagnostics}
\label{tab:temden}
\begin{tabular}{lll}
\hline\hline
Diagnostic Ratio & PN\,G048.5$+$04.2 & PN\,G068.7$+$14.8 \\
\hline
 & \multicolumn{2}{c}{$T_{\rm e}$ (K)} \\
$[$O\,{\sc iii}$]$ $(\lambda4959+\lambda5007)/\lambda4363$ & 13,500$\pm$100 & 10,600$\pm$200 \\
$[$N\,{\sc ii}$]$ $(\lambda6548+\lambda6583)/\lambda5755$ & $\cdots$ & 10,900$\pm$1000 \\
$[$O\,{\sc ii}$]$ $\lambda3727/(\lambda7320+\lambda7330)$ & 15,300$\pm$4000 & 10,600$\pm$300 \\
$[$S\,{\sc iii}$]$ $(\lambda9069+\lambda9531)/\lambda6312$ & 11,800$\pm$800 & 10,500$\pm$300 \\
$[$S\,{\sc ii}$]$ $(\lambda6716+\lambda6731)/\lambda4072$\tablenotemark{\rm{\scriptsize a}} & $>$20,000 & $>$20,000 \\
He\,{\sc i} $\lambda7281/\lambda6678$ & 7800$\pm$2000 & 11,200$\pm$800  \\
He\,{\sc i} $\lambda7281/\lambda5876$ & 6600$\pm$2200 & 11,000$\pm$1200 \\
H I PJ/P11 & 9660$\pm$5000 & 9390$\pm$3300\\
\cline{2-3}
 & \multicolumn{2}{c}{$N_{\rm e}$ (cm$^{-3}$)} \\
$[$S\,{\sc ii}$]$ $\lambda6716/\lambda6731$ & 2300$\pm1300$ & $>$20,000 \\
$[$S\,{\sc ii}$]$ $(\lambda6716+\lambda6731)/\lambda4072$\tablenotemark{\rm{\scriptsize a}} & $\sim$80,000\tablenotemark{\rm{\scriptsize b}} & $\sim$40,000\tablenotemark{\rm{\scriptsize b}}\\
    & $\cdots$ & $\sim$160,000\tablenotemark{\rm{\scriptsize c}}\\
$[$Ar\,{\sc iv}$]$ $\lambda4711/\lambda4740$ & 1750$\pm340$ & $\cdots$ \\
$[$Cl\,{\sc iii}$]$ $\lambda5517/\lambda5537$ & $\cdots$ & 7700$\pm$4300\tablenotemark{\rm{\scriptsize d}} \\
$[$O\,{\sc ii}$]$ $\lambda3727/(\lambda7320+\lambda7330)$ & 3000$\pm1900$ & 10,900$\pm500$ \\
Paschen decrement & $\cdots$ & $\sim 10^4$ \\
\hline
\end{tabular}
\begin{description}
NOTE. -- ``:” indicates that the uncertainty of the temperature or density is probably larger than 100\%; ``$\cdots$" means that the value cannot be derived. The symbol of intensity ``$I$" is omitted in the table. \\
\vspace{3mm}
\tablenotemark{\rm{\scriptsize a}} A blend of [S\,{\sc ii}] $\lambda\lambda$4068,4076; also blended with O\,{\sc ii} M10 3p\,$^{4}$D$^{\rm o} $ –- 3d\,$^{4}$F and possibly C\,{\sc iii} M16 4f\,$^{3}$F$^{\rm o}$ –- 5g\,$^{3}$G. \\
\tablenotemark{\rm{\scriptsize b}} At the intersection point with the [O\,{\sc iii}] $T_{\rm e}$-diagnostic curve in Figure\,\ref{fig4}.\\
\tablenotemark{\rm{\scriptsize c}} At the intersection point with the [N\,{\sc ii}] $T_{\rm e}$-diagnostic curve in Figure\,\ref{fig4}-right.\\
\tablenotemark{\rm{\scriptsize d}} Probably with much larger uncertainty due to faintness of the [Cl\,{\sc iii}] lines.
\end{description}
\end{center}
\end{table}

Relative fluxes of the high-order Paschen (and also Balmer) lines are insensitive to electron temperature but have dependence on electron density, making them another density diagnostic \citep{Zhang_2004}.  However, it is impossible to make reliable measurements of the high-order Balmer lines at the blue end of the GTC/OSIRIS R1000B spectrum; only Paschen decrements in the R1000R spectrum are well observed (Figure\,\ref{fig1}).  We present theoretical intensities of the high-order Paschen lines, in units where $I$(H$\beta$) = 100, calculated for various density cases at an H\,{\sc i} Paschen jump temperature of 9660\,K (for PN\,G048,5+04.2) and 9390\,K (for PN\,G068.7+14.8), using the effective recombination coefficients of \citet[][in Case~B assumption]{1995MNRAS.272...41S}, in Figure\,\ref{fig6}, where the Paschen line intensities measured from our GTC spectra are overplotted.  The intensities of the high-order Paschen lines in PN\,G048.5+04.2 have relatively larger measurement errors as well as scattering, as visually discernible in Figure\,\ref{fig6}, and thus cannot well constrain the nebular density.  The electron density of PN\,G068.7+14.8, after excluding a few data points with large deviations, is estimated to be $\sim$10$^4$\,cm$^{-3}$ (Table\,\ref{tab:temden}). 

He\,{\sc i} recombination lines can also be used to determine the electron temperatures of PNe \citep{2005MNRAS.358..457Z}.  We determined the nebular electron temperatures using the He\,{\sc i} $\lambda7281/\lambda6678$ and He\,{\sc i} $\lambda7281/\lambda5876$ line ratios.  Compared to other ratios of the strong He\,{\sc i} lines, $\lambda7281/\lambda6678$ is more sensitive to electron temperature but less sensitive to density; moreover, both He\,{\sc i} lines are singlet transitions, and thus are unaffected by the optical depth effects of the metastable 1s2s\,$^{3}$S state \citep{Benjamin_1999,Benjamin_2002}.  The nebular temperature determined using the He\,{\sc i} $\lambda7281/\lambda6678$ ratio is expected to be more reliable.  The He\,{\sc i} line ratios involving $\lambda4471$ are not used in temperature diagnostics, because this line is probably blended with the recombination lines of O\,{\sc ii} (M86c 3d\,$^{2}$P$_{1/2}$ -- 4f\,D[1]$^{\rm o}_{3/2}$ $\lambda$4469.48, \citealt{OIIcoe}) and O\,{\sc iii} (M49c 4f\,D[2]$_{1}$ -- 5g\,F[3]$^{\rm o}_{2}$ $\lambda$4471.02, \citealt{1999A&AS..137..157K}), which may make non-negligible flux contributions.  The He\,{\sc i} temperatures of the two PNe were derived using the theoretical He\,{\sc i} emissivities calculated by \citet{2022MNRAS.513.1198D}, and are summarized in Table\,\ref{tab:temden}.  Uncertainties in the He\,{\sc i} temperatures were propagated from the measurement errors in line fluxes.

\begin{table}
\begin{center}
\tablenum{4}
\caption{Ionic Abundances Derived from the GTC Optical Spectra}
\label{tab:ionic}
\begin{tabular}{llcc}
\hline\hline
Ion & Line & \multicolumn{2}{c}{Abundance (X$^{i+}$/H$^+$)} \\
\cline{3-4}
 & (\AA) & PN\,G048.5+04.2 & PN\,G068.7+14.8\\
\hline
He$^+$ & 4471 & 0.036$\pm$0.004 & 0.096$\pm$0.003 \\
 & 5876 & 0.037$\pm$0.002 & 0.100$\pm$0.001 \\
 & 6678 & 0.035$\pm$0.002 & 0.097$\pm$0.002 \\
Adopted\tablenotemark{\rm{\scriptsize a}} & & 0.037$\pm$0.002 & 0.100$\pm$0.001 \\
He$^{2+}$ & 5411 & 0.064$\pm$0.002 & 2.15($\pm$0.64)$\times10^{-3}$ \\
 & 4686 & 0.062$\pm$0.001 & 1.87($\pm$0.13)$\times10^{-3}$ \\
Adopted\tablenotemark{\rm{\scriptsize b}} & & 0.062$\pm$0.001 & 1.87($\pm$0.13)$\times10^{-3}$ \\
C$^{2+}$ & 4267 & 1.62($\pm$:)$\times10^{-4}$ & 1.36($\pm$0.15)$\times10^{-3}$ \\
N$^+$ & 5755 & $\cdots$ & 3.67($\pm$0.61)$\times10^{-6}$ \\
 & 6548 & 5.96($\pm$0.45)$\times10^{-7}$ & 3.68($\pm$0.05)$\times10^{-6}$ \\
 & 6583 & 4.24($\pm$0.19)$\times10^{-7}$ & 3.76($\pm$0.02)$\times10^{-6}$ \\
Adopted\tablenotemark{\rm{\scriptsize c}} & & 4.24($\pm$0.19)$\times10^{-7}$ & 3.76($\pm$0.02)$\times10^{-6}$ \\
O$^+$ & 3727 & 4.20($\pm$1.09)$\times10^{-6}$ & 4.29($\pm$0.12)$\times10^{-5}$ \\
 & 7325 & 4.40($\pm$0.55)$\times10^{-6}$ & 4.30($\pm$0.08)$\times10^{-5}$ \\
Adopted\tablenotemark{\rm{\scriptsize d}} & & 4.22($\pm$1.00)$\times10^{-6}$ & 4.29($\pm$0.11)$\times10^{-5}$ \\
O$^{2+}$ & 4363 & 1.75($\pm$0.03)$\times10^{-4}$ & 1.77($\pm$0.08)$\times10^{-4}$ \\
 & 4959 & 1.69($\pm$0.01)$\times10^{-4}$ & 1.72($\pm$0.01)$\times10^{-4}$ \\
 & 5007 & 1.75($\pm$0.01)$\times10^{-4}$ & 1.74($\pm$0.01)$\times10^{-4}$ \\
Adopted\tablenotemark{\rm{\scriptsize e}} & & 1.75($\pm$0.01)$\times10^{-4}$ & 1.74($\pm$0.01)$\times10^{-4}$ \\
Ne$^{2+}$ & 3868 & 3.53($\pm$0.05)$\times10^{-5}$ & 1.68($\pm$0.03)$\times10^{-5}$ \\
 & 3967 & 3.61($\pm$0.10)$\times10^{-5}$ & 1.93($\pm$0.26)$\times10^{-5}$ \\
Adopted\tablenotemark{\rm{\scriptsize f}} & & 3.53($\pm$0.05)$\times10^{-5}$ & 1.68($\pm$0.08)$\times10^{-5}$ \\
Ne$^{3+}$ & 4725 & 5.51($\pm$0.75)$\times10^{-5}$ & $\cdots$ \\
S$^+$ & 6716 & 2.42($\pm$0.39)$\times10^{-8}$ & 6.10($\pm$0.94)$\times10^{-8}$ \\
 & 6731 & 2.35($\pm$0.33)$\times10^{-8}$ & 7.02($\pm$0.67)$\times10^{-8}$ \\
Adopted\tablenotemark{\rm{\scriptsize g}} & & 2.38($\pm$0.25)$\times10^{-8}$ & 6.10($\pm$0.55)$\times10^{-8}$ \\
S$^{2+}$ & 6312 & 9.72($\pm$1.02)$\times10^{-7}$ & 8.35($\pm$0.52)$\times10^{-7}$ \\
 & 9069 & 9.66($\pm$0.18)$\times10^{-7}$ & 8.27($\pm$0.14)$\times10^{-7}$ \\
 & 9531 & 6.65($\pm$0.13)$\times10^{-7}$ & 1.98($\pm$0.04)$\times10^{-7}$ \\
Adopted\tablenotemark{\rm{\scriptsize h}} & & 9.66($\pm$0.18)$\times10^{-7}$ & 8.27($\pm$0.14)$\times10^{-6}$ \\
Cl$^{2+}$ & 5517 & 2.41($\pm$0.42)$\times10^{-8}$ & 2.92($\pm$0.49)$\times10^{-8}$ \\
 & 5537 & 1.64($\pm$0.35)$\times10^{-8}$ & 2.92($\pm$0.41)$\times10^{-8}$ \\
Adopted\tablenotemark{\rm{\scriptsize i}} & & 2.10($\pm$0.29)$\times10^{-8}$ & 2.92($\pm$0.32)$\times10^{-8}$ \\
Cl$^{3+}$ & 7530 & 1.01($\pm$0.09)$\times10^{-7}$ & $\cdots$ \\
 & 8046 & 6.99($\pm$0.35)$\times10^{-8}$ & $\cdots$ \\
Adopted\tablenotemark{\rm{\scriptsize j}} & & 6.99($\pm$0.35)$\times10^{-8}$ & $\cdots$ \\
Ar$^{2+}$ & 7136 & 3.87($\pm$0.02)$\times10^{-7}$ & 3.05($\pm$0.08)$\times10^{-7}$ \\
 & 7751 & 4.03($\pm$0.83)$\times10^{-7}$ & 2.99($\pm$0.45)$\times10^{-7}$ \\
Adopted\tablenotemark{\rm{\scriptsize k}} & & 3.87($\pm$0.02)$\times10^{-7}$ & 3.05($\pm$0.08)$\times10^{-7}$ \\
Ar$^{3+}$ & 4711 & 1.04($\pm$0.02)$\times10^{-6}$ & $\cdots$ \\
 & 4740 & 1.04($\pm$0.03)$\times10^{-6}$ & $\cdots$ \\
Adopted\tablenotemark{\rm{\scriptsize l}} & & 1.04($\pm$0.02)$\times10^{-7}$ & $\cdots$ \\
Ar$^{4+}$ & 6435 & 1.32($\pm$0.15)$\times10^{-7}$ & $\cdots$ \\
 & 7006 & 1.08($\pm$0.07)$\times10^{-7}$ & $\cdots$ \\ 
\hline
\end{tabular}
\caption{(Continued)}
\end{center}
\end{table}

\addtocounter{table}{-1}
\begin{table}
\begin{center}
\tablenum{4}
\caption{(Continued)}
\label{ionic}
\begin{tabular}{llcc}
\hline\hline
Ion & Line & \multicolumn{2}{c}{Abundance (X$^{i+}$/H$^+$)} \\
\cline{3-4}
 & (\AA) & PN\,G048.5+04.2 & PN\,G068.7+14.8\\
\hline
Adopted\tablenotemark{\rm{\scriptsize l}} & & 1.15($\pm$0.08)$\times10^{-7}$ & $\cdots$ \\
\hline
\end{tabular}
\begin{description}
NOTE. -- ``:” indicates that the uncertainty is larger than 100$\%$, and ``$\cdots$" means that the corresponding line was not detected. The weights used for all weighted average results are the intensities of the corresponding lines. \\
\tablenotemark{\rm{\scriptsize a}} The He$^+$/H$^+$ ratio derived from the He\,{\sc i} $\lambda$5876 line is adopted. \\
\tablenotemark{\rm{\scriptsize b}} The He$^{2+}$/H$^+$ ratio derived from the He\,{\sc ii} $\lambda$4686 line is adopted. \\
\tablenotemark{\rm{\scriptsize c}} The N$^+$/H$^+$ ratio derived from the [N\,{\sc ii}] $\lambda$6583 line is adopted, because the [N\,{\sc ii}] $\lambda$6548 is contaminated by the wing of H$\alpha$.    \\
\tablenotemark{\rm{\scriptsize d}} A weighted average value of the O$^+$/H$^+$ ratio derived from the [O\,{\sc ii}] $\lambda3727$ nebular and $\lambda7325$ auroral lines is adopted. \\
\tablenotemark{\rm{\scriptsize e}} The O$^{2+}$/H$^+$ ratio derived from [O\,{\sc iii}] $\lambda$5007 is adopted. \\
\tablenotemark{\rm{\scriptsize f}} The Ne$^{2+}$/H$^+$ ratio derived from [Ne\,{\sc iii}] $\lambda$3868 is adopted. \\
\tablenotemark{\rm{\scriptsize g}} A weighted average value of the S$^+$/H$^+$ ratio derived from the [S\,{\sc ii}] $\lambda6716$ and $\lambda6731$ lines is adopted for PN\,G048.5+04.2. The ratio derived only from [S\,{\sc ii}] $\lambda6716$ is adopted for PN\,G068.7+14.8 because [S\,{\sc ii}] $\lambda6731$ is contaminated by the emission from central stars in this PN\\
\tablenotemark{\rm{\scriptsize h}} The S$^{2+}$/H$^+$ ratio derived from [S\,{\sc iii}] $\lambda9069$ is adopted. \\
\tablenotemark{\rm{\scriptsize i}} A weighted average value of the Cl$^{2+}$/H$^+$ ratio derived from the [Cl\,{\sc iii}] $\lambda5517$ and $\lambda5537$ lines is adopted. \\
\tablenotemark{\rm{\scriptsize j}} The Cl$^{3+}$/H$^+$ ratios derived from [Cl\,{\sc iv}] $\lambda8046$ is adopted. \\
\tablenotemark{\rm{\scriptsize k}} The Ar$^{2+}$/H$^+$ ratio derived from [Ar\,{\sc iii}] $\lambda$7136 is adopted. \\
\tablenotemark{\rm{\scriptsize l}} A weighted average value of the Ar$^{3+}$/H$^+$ ratio derived from the [Ar\,{\sc iv}] $\lambda4711$ and $\lambda4740$ lines is adopted. \\
\tablenotemark{\rm{\scriptsize m}} A weighted average value of the Ar$^{4+}$/H$^+$ ratio derived from the [Ar\,{\sc v}] $\lambda6435$ and $\lambda7006$ lines is adopted. \\
\end{description}
\end{center}
\end{table}

The He\,{\sc i} temperatures of PN\,G068.7+14.8 are higher than that of PN\,G048.5+04.2 whose excitation is much higher, as can be seen in the very strong He\,{\sc ii} $\lambda$4686 line (Figure\,\ref{fig1}); on the contrary, the CEL temperatures of PN\,G068.7+14.8 are systematically lower than those of PN\,G048.5+04.2 (Table\,\ref{tab:temden}).  This inconsistency in temperature difference, on the one hand, might be caused by measurement errors, given the faintness of the He\,{\sc i} $\lambda$7281 line.  On the other hand, it does not align with the statistical results of \citet[][Figure\,7 therein]{2005MNRAS.358..457Z}, who reported a weakly positive correlation between the He~{\sc i} temperature and excitation class (EC) in PNe.  Moreover, for PNe with a low-mass Wolf-Rayet (denoted as [WR]) central star, nebular emission lines can be contaminated by emission from stellar winds.  As discussed in Section\,\ref{subsec:central-stars}, PN\,G068.7+14.8 probably has a [WR]-type central star, and the nebular He\,{\sc i} line might be affected by the central-star emission, resulting in overestimated He\,{\sc i} temperature of the nebula. 

Our plasma diagnostics with He\,{\sc i} lines utilized the atomic recombination data that were calculated based on the Case\,B\footnote{Here Case\,B means the nebula is optically thick to all the singlet He\,{\sc i} transitions of $n$\,$^{1}$P$^{\rm o}\rightarrow$ 1\,$^{1}$S.} assumption (Table\,\ref{tab:atomic_data}).  However, recent analyses of a large sample of H\,{\sc ii} regions and PNe indicate that the singlet He\,{\sc i} transitions in ionized nebulae may actually deviate from Case\,B, affecting the He\,{\sc i} singlet transitions originating from the $n$\,$^{1}$P$^{\rm o}_{1}$ levels and resulting in unrealistic temperatures diagnosed with these He\,{\sc i} lines \citep{2025ApJ...986...74M}.  Although worth careful investigation, this issue is beyond the scope of this paper given the low resolution ($R\sim$1000) of our GTC spectra, and can only be further studied through follow-up high-dispersion spectroscopy.

\begin{table*}
\begin{center}
\tablenum{5}
\caption{Elemental Abundances Derived Using ICFs\,\tablenotemark{\rm{\scriptsize a}}}
\label{tab:elemental}
\begin{tabular}{llrlr|lrlr|lr}
\hline\hline
Elem. & \multicolumn{10}{c}{X/H} \\
\cline{2-11}
 & \multicolumn{4}{c}{PN\,G048.5+04.2} & \multicolumn{4}{c}{PN\,G068.7+14.8} & \multicolumn{2}{c}{~~~~~~~~~~Solar\,\tablenotemark{\rm{\scriptsize c}}} \\
\hline
He & 0.099$\pm$0.003 & 11.00 & 0.099$\pm$0.003 & 11.00 & 0.100$\pm$0.002 & 11.00 & 0.102$\pm$0.002 & 11.00 & 0.85 & 10.93 \\
C  & 3.20($\pm$:)$\times10^{-4}$ & 8.50 & 3.99($\pm$:)$\times10^{-4}$ & 8.60 & 1.72($\pm$0.20)$\times10^{-3}$ & 9.23 & 1.61($\pm$0.18)$\times10^{-3}$ & 9.21 & 2.69$\times10^{-4}$ & 8.43 \\
N  & 3.47($\pm$0.86)$\times10^{-5}$ & 7.54 & 2.74($\pm$0.93)$\times10^{-5}$ & 7.44 & 2.32($\pm$0.07)$\times10^{-5}$ & 7.37 & 2.39($\pm$0.06)$\times10^{-5}$ & 7.38 & 6.76$\times10^{-5}$ & 7.83 \\
O  & 3.45($\pm$0.16)$\times10^{-4}$ & 8.54 & 3.63($\pm$0.12)$\times10^{-4}$ & 8.56 & 2.20($\pm$0.05)$\times10^{-4}$ & 8.34 & 2.19($\pm$0.03)$\times10^{-4}$ & 8.34 & 4.89$\times10^{-4}$ & 8.69 \\
Ne & 6.97($\pm$0.12)$\times10^{-5}$ & 7.84 & 7.35($\pm$0.28)$\times10^{-5}$ & 7.87 & 2.12($\pm$0.11)$\times10^{-5}$ & 7.33 & 2.02($\pm$0.21)$\times10^{-5}$ & 7.31 & 8.51$\times10^{-5}$ & 7.93 \\
Ne\,\tablenotemark{\rm{\scriptsize b}} & 9.04($\pm$0.76)$\times10^{-5}$ & 7.96 & $\cdots$ & $\cdots$ & $\cdots$ & $\cdots$ & $\cdots$ & $\cdots$ & & \\
S  & 2.99($\pm$0.42)$\times10^{-6}$ & 6.48 & 4.46($\pm$0.39)$\times10^{-6}$ & 6.65 & 1.14($\pm$0.05)$\times10^{-6}$ & 6.06 & 1.06($\pm$0.02)$\times10^{-6}$ & 6.03 & 1.32$\times10^{-5}$ & 7.12 \\
Cl & 6.50($\pm$0.90)$\times10^{-8}$ & 4.82 & 9.90($\pm$1.52)$\times10^{-8}$ & 5.00 & 4.03($\pm$0.42)$\times10^{-8}$ & 4.60 & 4.27($\pm$0.47)$\times10^{-8}$ & 4.63 & 3.16$\times10^{-7}$ & 5.50 \\
Cl\,\tablenotemark{\rm{\scriptsize b}} & 9.09($\pm$0.48)$\times10^{-8}$ & 4.96 & $\cdots$ & $\cdots$ & $\cdots$ & $\cdots$ & $\cdots$ & $\cdots$ & & \\
Ar & 1.56($\pm$0.04)$\times10^{-6}$ & 6.19 & 1.31($\pm$0.05)$\times10^{-6}$ & 6.12 & 5.70($\pm$1.37)$\times10^{-7}$ & 5.76 & 3.89($\pm$0.11)$\times10^{-7}$ & 5.59 & 2.51$\times10^{-6}$ & 6.40 \\
\hline
\end{tabular}
\begin{description}
NOTE. -- ``:” indicates the uncertainty is $>$100\%; for each element, abundance values in both linear (left) and logarithmic, 12\,+\,log(X/H), are presented. \\
\tablenotemark{\rm{\scriptsize a}} For each PN, abundance values (in both linear and logarithmic) on the left were derived using the ICFs given by \citet[][see Appdendix\,A therein]{1994MNRAS.271..257K}, and the abundance values on the right were derived using the ICFs from \citet{2014MNRAS.440..536D}. \\
\tablenotemark{\rm{\scriptsize b}} Derived via direct summation of ionic abundances (see text in Section\,\ref{subsec:element}). \\
\tablenotemark{\rm{\scriptsize c}} Solar abundances from \citet{2009ARAA..47..481A}. 
\end{description}
\end{center}
\end{table*}

\subsection{Ionic Abundances} \label{subsec:ionic}

Based on the plasma diagnostics (Section\,\ref{subsec:diagnostics}), we calculated the ionic abundances of the two PNe using the extinction-corrected intensities of emission lines from Table\,\ref{tab:lines}.  The {\sc pyneb} package was used in CELs abundance calculations.  Different diagnostic line ratios yield different electron temperatures/densities (Table\,\ref{tab:temden}), indicating the presence of ionization structures in a PN (Figure\,\ref{fig4}).  In the determination of ionic abundances, different electron temperatures and densities were assumed according to the ionization stages of the species.  The ionic abundances of the two PNe are presented in Table\,\ref{tab:ionic} along with uncertainties, which were estimated using the measurement errors in line fluxes and including propagation of uncertainties in plasma diagnostics. 

The [O\,{\sc iii}] temperature was used to derive the abundances of the doubly ionized species (O$^{2+}$, Ne$^{2+}$, Ar$^{2+}$ and Cl$^{2+}$).  In the spectrum of PN\,G048.5+04.2, we detected highly ionized species including Ne$^{3+}$, Cl$^{3+}$, Ar$^{3+}$ and Ar$^{4+}$; $T_{\rm e}$([O\,{\sc iii}]) was also assumed when calculating their ionic abundances.  In the calculations of N$^+$/H$^+$ and O$^+$/H$^+$, we adopted the [N\,{\sc ii}] temperature.  The [N\,{\sc ii}] $\lambda5755$ auroral line is not detected in the spectrum of PN\,G048.5+04.2 (probably due to high excitation of the nebula); therefore, we utilized the method proposed by \citet{2015ApJ...803...23D} based on the work of \citet{2001ApJ...562..804K}, to estimate the electron temperature of the low-ionization region: for a nebula with He\,{\sc ii} $\lambda$4686 detected, $T_{\rm e}$([N\,{\sc ii}]) = 10,300\,K derived by \citet{1986ApJ...308..322K} was adopted.  In neither of the two PNe can the [S\,{\sc ii}] temperature be obtained, we adopted the [N\,{\sc ii}] temperature to calculate the S$^+$/H$^+$ ratio.  For the calculation of S$^{2+}$/H$^{+}$, we used the [S\,{\sc iii}] temperature. 

Electron density has influence on the ionic abundance determinations.  For PN\,G048.5+04.2, we used the [S\,{\sc ii}] density to calculate the abundances of low-ionization species, and adopted the [Ar\,{\sc iv}] density for the high-ionization species.  [Ar\,{\sc iv}] lines are not detected in the spectra of PN\,G068.7+14.8, and the [S\,{\sc ii}] density of this PN is unusable (Table\,\ref{tab:temden}).  The [Cl\,{\sc iii}] and [O\,{\sc ii}] densities were adopted for PN\,G068.7+14.8 in the calculations of the high- and low-ionization species, respectively. 

The He$^{+}$/H$^{+}$, He$^{2+}$/H$^{+}$ and C$^{2+}$/H$^{+}$ abundance ratios were derived using the ORL intensities.  The theoretical emissivities of the He\,{\sc i} lines calculated by \citet{2022MNRAS.513.1198D} were used to derive the He$^{+}$/H$^{+}$ ratios.  The effective recombination coefficients of the H\,{\sc i} and He\,{\sc ii} lines calculated by \citet[][in Case~B assumption]{1995MNRAS.272...41S} were used to calculate the He$^{2+}$/H$^{+}$ ratio.  For carbon, only C\,{\sc ii} $\lambda$4267 is detected in the GTC optical spectra; the effective recombination coefficients calculated by \citet[][at $T_{\rm e}$=10,000\,K and $N_{\rm e}$=10$^{4}$\,cm$^{-3}$, in Case~B assumption]{2000AAS..142...85D} were used to derive the C$^{2+}$/H$^{+}$ ratio.  The ionic abundances of the two PNe are summarized in Table\,\ref{tab:ionic}.

\begin{table*}
\begin{center}
\tablenum{6}
\caption{Ionization Correction Factors\tablenotemark{\rm{\scriptsize a}} \label{tab:ICF}}
\begin{tabular}{lccccccccc}
\hline\hline
PN & \multicolumn{7}{c}{ICFs} \\
\cline{2-10}
 & C & N & O & Ne\tablenotemark{\rm{\scriptsize b}} & Ne\tablenotemark{\rm{\scriptsize c}} & S & Cl\tablenotemark{\rm{\scriptsize b}} & Cl\tablenotemark{\rm{\scriptsize c}} & Ar \\
\hline
PN\,G048.5+04.2 & 1.974 & 81.85 & 1.927 & 1.974 & 1.000 & 3.023 & 3.096 & 1.000 & 1.012 \\
PN\,G068.7+14.8 & 1.263 & 5.124 & 1.014 & 1.263 & $\cdots$ & 1.278 & 1.378 & $\cdots$ & 1.870 \\
\hline
\end{tabular}
\begin{description}
\tablenotemark{\rm{\scriptsize a}} ICFs were calculated using the methodology of \citet{1994MNRAS.271..257K}. \\ 
\tablenotemark{\rm{\scriptsize b}} The corresponding Ne and Cl abundances were calculated using ICFs. \\
\tablenotemark{\rm{\scriptsize c}} The corresponding Ne and Cl abundances were derived by direct summing of ionic abundances. \\
\end{description}
\end{center}
\end{table*}

For the same ion, different lines yield different abundance values.  For PN\,G048.5+04.2, we found that the O$^{+}$/H$^{+}$ ratios yielded by the [O\,{\sc ii}] $\lambda$3727 (=$\lambda$3726+$\lambda$3729) line and the [O\,{\sc ii}] $\lambda$7325 (=$\lambda$7320+$\lambda$7330) line are significantly different when the same electron temperature and density were adopted.  One possible reason for this discrepancy is that emission of [O\,{\sc ii}] $\lambda$3727 is suppressed due to collisional de-excitation when the local electron density exceeds the critical density\footnote{At $T_{\rm e}$=10,000\,K, the critical densities of [O\,{\sc ii}] $\lambda\lambda$3726,\,3729 are 1.5$\times$10$^{4}$ and 3.4$\times$10$^{3}$\,cm$^{-3}$, respectively \citep{OF2006}.} of the [O\,{\sc ii}] nebular line, 
resulting in overestimated [O\,{\sc ii}] temperature.  In order to maintain consistency in abundance determination, we adopted the [S\,{\sc ii}] density when using the [O\,{\sc ii}] $\lambda$3727 line intensity to calculate the abundance of O$^{+}$, and assumed a density of 10$^{4}$ cm$^{-3}$ when using [O\,{\sc ii}] $\lambda$7325, so that the difference between the two abundance values is smaller.  For PN\,G068.7+14.8, because of its high-density nature, the two [O\,{\sc ii}] lines yield close values of O$^{+}$/H$^{+}$; the [O\,{\sc ii}] density was used.  The final adopted O$^{+}$/H$^{+}$ ratios of the two PNe are a weighted average, with the weights proportional to the [O\,{\sc ii}] line intensities. 

If the abundances yielded by different emission lines of a certain ion are different, which occurs to several ionic species in the two PNe, we adopted the abundance value derived using the strongest lines or a weighted-average value (with the weight proportional to line intensity).  We adopted the He$^{+}$/H$^{+}$ and He$^{2+}$/H$^{+}$ abundance ratios derived using the He\,{\sc i} $\lambda$5876 and He\,{\sc ii} $\lambda$4686 lines, respectively.  The N$^+$/H$^+$ ratio was derived using [N\,{\sc ii}] $\lambda$6583, which is the strongest line of N$^{+}$ in the optical and less affected by the much stronger H$\alpha$.  For Ne$^{2+}$/H$^{+}$, S$^{2+}$/H$^{+}$, Ar$^{3+}$/H$^{+}$ and Ar$^{4+}$/H$^{+}$, the flux weighted-average value calculated using the lines of the corresponding ion as presented in Table\,\ref{tab:ionic} was adopted.  The adopted S$^{2+}$/H$^{+}$ ratio was derived using [S\,{\sc iii}] $\lambda$9069 (the $\lambda$9531 line was affected by the second-order contamination of the R1000R grism; Figure\,\ref{fig1}).  We adopted the weighted average value of Cl$^{2+}$ from the abundances of the [Cl\,{\sc iii}] $\lambda$5517,5537 doublet.  The [Cl\,{\sc iv}] $\lambda$8046 line was used to derive the Cl$^{3+}$/H$^{+}$ ratio because [Cl\,{\sc iv}] $\lambda$7530 is close to the atmospheric absorption A-Band.  [Ar\,{\sc iii}] $\lambda$7136 was used to derive Ar$^{2+}$/H$^{+}$, while the [Ar\,{\sc iii}] $\lambda$7751 line is close to the red end of R1000B grism and is affected by the second-order contamination.

\subsection{Elemental Abundances} \label{subsec:element}

A direct calculation of elemental abundances is summing over the ionic abundances of all ionization stages detected in a nebula.  The helium abundance He/H = He$^{+}$/H$^{+}$ + He$^{2+}$/H$^{+}$.  For a heavy element, which has more than two ionization states and only one or two are detected in the optical spectrum, we utilize the classical, model-based ionization correction factors (ICFs) of \citet{1994MNRAS.271..257K} to estimate the elemental abundances.  The updated ICFs of \citet{2014MNRAS.440..536D}, who used a much larger grid of photoionization models that spans over a wide range of parameters, are also used in elemental abundance determination.  These are summarized in Table\,\ref{tab:elemental}, where solar values are presented for purpose of comparison.  Below we mainly describe the methodologies of \citet{1994MNRAS.271..257K}, and the ICFs hence derived are presented in Table\,\ref{tab:ICF}.

\begin{figure}[htp!]
\begin{center}
\includegraphics[width=8.65cm, angle=0]{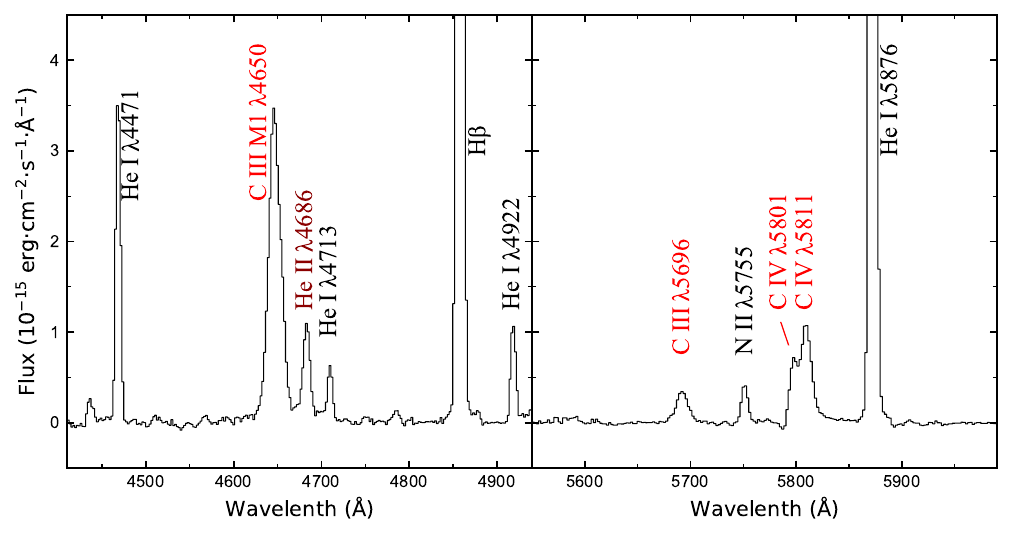}
\caption{GTC/OSIRIS spectrum of PN\,G068.7+14.8; probable emission features from its [WR]-type central star are labeled in red.} 
\label{fig7}
\end{center}
\end{figure}

For both PNe, the oxygen abundance was calculated via O/H = ICF(O)$\times$(O$^{+}$/H$^{+}$\,+\,O$^{2+}$/H$^{+}$), where ICF(O) was derived based on the ionic abundances of helium \citep[][Eq\,A9 therein]{1994MNRAS.271..257K}.  For carbon and nitrogen, only C$^{2+}$/H$^+$ and N$^+$/H$^+$ ratios were derived in the two PNe, therefore we adopted ICF(C) = O/O$^{2+}$ and ICF(N)=O/O$^{+}$ (\citealt{1969BOTT....5....3P}; \citealt{1994MNRAS.271..257K}, Eqs\,A1 and A13 therein).  In PN\,G068.7+14.8, only Ne$^{2+}$ was detected for neon, and we adopted ICF(Ne) = O/O$^{2+}$ (\citealt{1969BOTT....5....3P}; \citealt{1994MNRAS.271..257K}, Eq\,A28 therein).  Both the Ne$^{2+}$ and Ne$^{3+}$ ions were observed in PN\,G048.5+04.2, but for such case no ICF(Ne) formula is given in \citet{1994MNRAS.271..257K} or any other literatures; we directly sum the two ionic species, i.e., Ne/H = Ne$^{2+}$/H$^{+}$\,+\,Ne$^{3+}$/H$^{+}$, which is a reasonable approximation given that Ne$^{2+}$ and Ne$^{3+}$ are probably the two dominant ionization stages of neon in this high-excitation PN.  We also use the Ne$^{2+}$/H$^+$ ratio and ICF(Ne) of \citet[][Eq\,A28 therein]{1994MNRAS.271..257K} to derive the Ne/H ratio of PN\,G048.5+04.2 for purpose of comparison with the summed value. 

For sulfur, only S$^{+}$ and S$^{2+}$ were observed in the two PNe, and we used S/H = ICF(S)$\times$(S$^{+}$/H$^{+}$\,+\,S$^{2+}$/H$^{+}$), where ICF(S) was given by \citet[][Eq\,A36 therein]{1994MNRAS.271..257K}.   In the GTC spectrum of PN\,G068.7+14.8, only Cl$^{2+}$ was detected for chlorine; considering the similar ionization potentials of S and Cl, we estimated the Cl/H ratio based on the assumption that Cl/Cl$^{2+}\approx$S/S$^{2+}$ as suggested by \citet{2007MNRAS.381..669W}.  The Cl/H abundance ratio of PN\,G048.5+04.2 was derived using the same method, but we also derived a lower limit using the sum of two ionic abundances, i.e., Cl/H = Cl$^{2+}$/H$^{+}$\,+\,Cl$^{3+}$/H$^{+}$, given that the [Cl\,{\sc iv}] lines were also observed in this PN (Table\,\ref{tab:lines}).  Nebular lines of Ar$^{2+}$, Ar$^{3+}$ and Ar$^{4+}$ were detected in PN\,G048.5+04.2, and we adopted ICF(Ar) of \citet[][Eq\,A30 therein]{1994MNRAS.271..257K} to derive total Ar/H ratio of this PN.  In PN\,G068.7+14.8, only the lines of Ar$^{2+}$ were detected, we therefore used the approximation ICF(Ar)=1.87 as suggested in \citet[][Eq\,A32 therein]{1994MNRAS.271..257K} to determine the total Ar/H ratio. 

For PN\,G048.5+04.2, the Ne/H and Cl/H ratios obtained by direct summation of ionic abundances are larger than those obtained using ICFs; this could be due to inadequate ICFs of the two elements, although possible line-blending issue in the [Cl\,{\sc iv}] and [Ne\,{\sc iv}] lines that leads to overestimated ionic abundances cannot be totally excluded.  It is thus difficult to decide which method produces more reliable abundances.  Considering that the elemental abundances obtained via direct summation is closer to the solar values (Table\,\ref{tab:elemental}), we adopted these abundances in the following discussion.

\begin{table}
\begin{center}
\tablenum{7}
\caption{[WR] Central Star Emission of PNG068.7$+$14.8 
\label{tab:[WR] lines}}
\begin{tabular}{lcc}
\hline\hline
lines & FWHM & {$I$($\lambda$)\tablenotemark{\rm{\scriptsize a}}}\\
\hline
C\,{\sc iii} $\lambda4650$ & $\sim$16.7\,{\AA} & 362 \\
C\,{\sc iii} $\lambda5696$ & $\sim$13.1\,{\AA} & 25\\
C\,{\sc iv}  $\lambda\lambda$5801,5011 & $\sim$20.5\,{\AA} & 100\\
\hline
\end{tabular}
\begin{description}
NOTE. -- Intensities were obtained by integration over the emission-line profile, extinction corrected. \\
\tablenotemark{\rm{\scriptsize a}} Intensities normalized such that $I$(C\,{\sc iv} $\lambda5805$)=100
\end{description}
\end{center}
\end{table}

The uncertainties in elemental abundances presented in Table\,\ref{tab:elemental} were determined using error propagation from the uncertainties in ionic abundances.  Systematic uncertainties in ICFs were not considered.  The oxygen abundance is the best determined among all heavy elements, given that ICF(O) is always close to 1.  The ICFs of C, N, Ne and S (and consequently Ar) are empirically based on the ionic and elemental abundances of oxygen, and more or less have issues in accuracy.  For nitrogen, only N$^{+}$ is detected in the optical spectrum of a PN, where N$^{2+}$ is the dominant ionization stage (as is the case of high-excitation PN\,G048.5+04.2; see Table\,\ref{tab:ICF}); thus the total N/H abundance derived using the ICF method might be reliable.  The ICF-derived Cl/H abundance of PN\,G068.7+14.8 could be of significant uncertainty, because only the faint [Cl\,{\sc iii}] lines are observed in its spectrum and the assumption of Cl/Cl$^{2+}\approx$S/S$^{2+}$ could be problematic.

\begin{table*}
\begin{center}
\tablenum{8}
\caption{Fluxes and Intensities of the Emission Lines Detected in the \emph{Spitzer}/IRS Spectra}
\label{tab:spitzer_lines}
\begin{tabular}{lllrrrr}
\hline\hline
Ion & $\lambda$ & Transition & \multicolumn{2}{c}{\underline{~~~~~~~~~~~~~~~~~~~~~~~PN\,G048.5+04.2~~~~~~~~~~~~~~~~~~~~~~~}} & \multicolumn{2}{c}{\underline{~~~~~~~~~~~~~~~~~~~~~~~PN\,G068.7+14.8~~~~~~~~~~~~~~~~~~~~~~~}} \\
  & ($\mu$m) & (Lower--Upper) & $F$($\lambda$) [10$^{-13}$ erg\,s$^{-1}$\,cm$^{-2}$] & $I$($\lambda$) & $F$($\lambda$) [10$^{-13}$ erg\,s$^{-1}$\,cm$^{-2}$] & $I$($\lambda$) \\
\hline
$[$S\,{\sc iv}$]$   & 10.51 & \textrm{$3p\ ^2P^{\rm{o}}_{1/2}-3p\ ^2P^{\rm{o}}_{3/2}$}     & 6.62$\pm$0.13 & 172.43$\pm$3.28 & $\cdots$    & $\cdots$ \\
$[$S\,{\sc ii}$]$   & 12.81\tablenotemark{\rm{\scriptsize a}} & \textrm{$3p^5\ ^2P^{\rm{o}}_{3/2}-3p^5\ ^2P^{\rm{o}}_{1/2}$} & $\cdots$      & $\cdots$        & 11.78$\pm$: & 39.16$\pm$: \\
$[$Ne\,{\sc v}$]$   & 14.32 & \textrm{$2p^2\ ^3P_1-2p^2\ ^3P_2$}                           & 1.62$\pm$0.15 & 42.11$\pm$3.83 & $\cdots$ & $\cdots$ \\
$[$Ne\,{\sc iii}$]$ & 15.55 & \textrm{$2p^4\ ^3P_2-2p^4\ ^3P_1$} & 7.32$\pm$0.33 & 190.68$\pm$8.55 & 5.88$\pm$0.61 & 19.53$\pm$2.02 \\
$[$S\,{\sc iii}$]$  & 18.71 & \textrm{$3p^2\ ^3P_1-3p^2\ ^3P_2$} & 0.75$\pm$0.19 & 19.42$\pm$5.08 & 1.27$\pm$0.38 & 4.23$\pm$1.28 \\
$[$Ne\,{\sc v}$]$   & 24.32 & \textrm{$2p^2\ ^3P_0-2p^2\ ^3P_1$} & 1.47$\pm$0.19 & 38.34$\pm$4.92 & $\cdots$ & $\cdots$ \\
$[$S\,{\sc i}$]$?   & 25.25 & \textrm{$3p^4\ ^3P_2-3p^4\ ^3P_1$} & 2.31$\pm$0.41 & 60.20$\pm$10.77 & $\cdots$ & $\cdots$ \\
$[$O\,{\sc iv}$]$   & 25.91 & \textrm{$2p\ ^2P^{\rm{o}}_{1/2}-2p\ ^2P^{\rm{o}}_{3/2}$} & 75.04$\pm$1.34 & 1954.27$\pm$34.93 & $\cdots$ & $\cdots$ \\
$[$S\,{\sc iii}$]$  & 33.48 & \textrm{$3p^2\ ^3P_0-3p^2\ ^3P_1$} & 0.35$\pm$0.20 & 9.21$\pm$5.33 & 0.27$\pm$0.12 & 0.90$\pm$0.40 \\
$[$Ne\,{\sc iii}$]$ & 36.00 & \textrm{$2p^4\ ^3P_1-2p^4\ ^3P_0$} & 0.51$\pm$: & 13.17$\pm$: & 0.34$\pm$0.12 & 1.14$\pm$0.40 \\
\hline
\end{tabular}
\begin{description}
NOTE. -- Line intensities $I$($\lambda$) are normalized such that $I$(H$\beta$)=100, using the total H$\beta$ line fluxes of the two PNe given by \citet{1992secg.book.....A}; extinction uncorrected for given that reddening of infrared emission is negligible compared to optical emission.  The meanings of ``:'' and ``$\cdots$'' are the same as in Table\,\ref{tab:lines}; ``?'' means identification of the line is uncertain. \\
\smallskip
\tablenotemark{\rm{\scriptsize a}} Probably blended with PAH emission at 12.7\,$\mu$m; also affected by AIB band at 11.2\,$\mu$m. 
\end{description}
\end{center}
\end{table*}

\subsection{Emission from [WR] Central Star of PN\,G068.7+14.8} 
\label{subsec:central-stars}

The broad emission features of C\,{\sc iii} $\lambda$4650,\,5696 and C\,{\sc iv} $\lambda\lambda$5801.35,5811.97\footnote{Rest wavelengths adopted from Atomic Line List v3.00b5 \citep{2018vanhoof}; \url{https://linelist.pa.uky.edu/newpage/}} are detected in the GTC spectrum of PN\,G068.7+14.8 (Figure\,\ref{fig7}), which probably come from the [WR]-type central star of this PN.  Measurements of these carbon lines are summarized in Table\,\ref{tab:[WR] lines}.  The intensity ratio $I($C\,{\sc iii} $\lambda$4650)/$I($C\,{\sc iv} $\lambda$5805), where $\lambda$5805 = $\lambda$5801.35+$\lambda$5811.97, signifies a [WC7-8]-type central star according to the criteria of \citet{2003AA...403..659A}, while the $I$(C\,{\sc iii} $\lambda$5696)/$I$(C\,{\sc iv} $\lambda$5805) ratio indicates a stellar type between [WC5-6] and [WC7-8].  However, the C\,{\sc iii} $\lambda$4650 line flux is probably overestimated, as it is blended with the nebular O\,{\sc ii} M1 $\lambda\lambda$4649,4651 (3s\,$^{4}$P -- 3p\,$^{4}$D$^{\rm o}$) recombination lines due to low resolution ($R\sim$1000) of the GTC spectra.  The central star of PN\,G068.7+14.8 thus might lie between types-[WC5-6] and [WC7-8], but with doubts given that the FWHMs of the C\,{\sc iii} lines are narrower than the criteria of \citet{2003AA...403..659A}. 

The central star of PN\,G068.7+14.8 was classified by \citet{1993A&AS..102..595T} and \citet{2011A&A...526A...6W} as a weak emission-line star because its carbon lines are weaker and narrower than those of most other [WR]-type central stars of Galactic PNe. 
Our GTC spectroscopy also reveals these emission features, with no clear detection of other central-star emission lines except C\,{\sc iii} $\lambda$5696 (Figure\,\ref{fig7}); this is similar to the nature of weak emission-line stars ([WELSs]).  \citet{2021MNRAS.504..816B} discovered other [WELSs] features in their optical spectra of PN\,G068.7+14.8, such as emission at 5470\,{\AA} (a blend of the C\,{\sc iv} and O\,{\sc iv} lines) and C\,{\sc ii} $\lambda$5892.  We detected a weak emission feature at 5470\,{\AA} as well, but no C\,{\sc ii} $\lambda$5892. 

Emission from the [WR]-type central star of PN\,G068.7+14.8 may affect spectral-line analysis.  The nebular [S\,{\sc ii}] $\lambda$6731 line of PN\,G068.7+14.8 is blended with C\,{\sc iii} $\lambda$6730, leading to an underestimated [S\,{\sc ii}] $\lambda$6716/$\lambda$6731 ratio and consequently an extremely high electron density (Table\,\ref{tab:temden}).  The He\,{\sc ii} $\lambda$4686 line may also be contaminated by stellar emission.  Therefore, the He$^{2+}$/H$^{+}$ ratio of PN\,G068.7+14.8 might be overestimated, although does not have a significant impact on the total He/H (as well as the ionization correction of some heavy elements) given that the H\,{\sc ii} emission of this PN is very weak.  Emission from the [WR]-type central star also influences our photoionizaiton modeling of PN\,G068.7+14.8 (see Section\,\ref{sec:models}). 

\begin{table}
\begin{center}
\tablenum{9}
\caption{Ionic Abundances Derived from the \emph{Spitzer}/IRS spectra}
\label{tab:spitzer_ionic}
\begin{tabular}{llcc}
\hline\hline
Ion & Line & \multicolumn{2}{c}{Abundance (X$^{i+}$/H$^+$)} \\
\cline{3-4}
 & ($\mu$m) & PN\,G048.5+04.2 & PN\,G068.7+14.8\\
\hline
O$^{3+}$ & 25.91 & 3.93($\pm$0.08)$\times10^{-4}$ & $\cdots$ \\
Ne$^{2+}$ & 15.55 & 1.08($\pm$0.05)$\times10^{-4}$ & 1.32($\pm$0.14)$\times10^{-5}$ \\
 & 36.00 & 8.70($\pm$:)$\times10^{-5}$ & 9.90($\pm$3.47)$\times10^{-6}$ \\
Ne$^{4+}$ & 14.32 & 2.73($\pm$0.25)$\times10^{-6}$ & $\cdots$ \\
 & 24.32 & 2.89($\pm$0.37)$\times10^{-6}$ & $\cdots$ \\
S$^{2+}$ & 18.71 & 1.65($\pm$0.44)$\times10^{-6}$ & 5.91($\pm$1.79)$\times10^{-7}$ \\
 & 33.48 & 1.64($\pm$0.95)$\times10^{-6}$ & 5.93($\pm$2.64)$\times10^{-7}$ \\
S$^{3+}$ & 10.51 & 4.32($\pm$0.09)$\times10^{-6}$ & $\cdots$ \\
\hline
\end{tabular}
\end{center}
\end{table}

\subsection{Analysis of the \emph{Spitzer}/IRS Spectra} 
\label{subsec:IRS}

Nebular lines were detected in the \emph{Spitzer}/IRS mid-IR spectra, including [Ne\,{\sc iii}] 15.55,\,36.0\,$\mu$m and [S\,{\sc iii}] 18.71,\,33.48\,$\mu$m in both PNe, [Ne\,{\sc ii}] 12.81\,$\mu$m in PN\,G068.7+14.8, and [S\,{\sc iv}] 10.51\,$\mu$m, [O\,{\sc iv}] 25.91\,$\mu$m and [Ne\,{\sc v}] 14.32,\,24.32\,$\mu$m in PN\,G048.5+04.2 (see Figure\,\ref{fig2}). 
In the \emph{Spitzer} spectrum of PN\,G048.5+04.2, we identified an emission feature near the strong [O\,{\sc iv}] 25.91\,$\mu$m line as [S\,{\sc i}] 25.25\,$\mu$m, but with doubts.  Line fluxes measured using Gaussian-profile fits are presented in Table\,\ref{tab:spitzer_lines}, where uncertainties in line fluxes were estimated based on the errors of Gaussian-profile fits combined with measurement errors on each data point.  The angular sizes of both PNe are larger than the slit width (1\arcsec) of our GTC/OSIRIS spectroscopy, but smaller than (or comparable to) the apertures of the \emph{Spitzer}/IRS spectroscopy (see Section\,\ref{subsec:spitzer}); thus the H$\beta$ fluxes from Table\,\ref{tab:lines} cannot be used in conjunction with the mid-IR line fluxes in Table\,\ref{tab:spitzer_lines} to derive the ionic abundances.  Therefore we adopted the total H$\beta$ fluxes of the two PNe from \citet{1992secg.book.....A}; extinction was corrected for using the $c$(H$\beta$) parameters in Table\,\ref{tab:lines}.  

Ionic abundances of O$^{3+}$, Ne$^{2+}$, Ne$^{4+}$, S$^{2+}$ and S$^{3+}$ were calculated using the mid-IR CELs in Table\,\ref{tab:spitzer_lines}.  Since most of the mid-IR lines are emitted by the doubly and triply ionized species, the [O\,{\sc iii}] temperature was assumed in abundance calculations except for sulphur, whose ionic abundance was derived using the [S\,{\sc iii}] temperature.  In PN\,G048.5+04.2, the [S\,{\sc ii}] density was used to derive Ne$^{2+}$/H$^{+}$ and S$^{2+}$/H$^{+}$, while the [Ar\,{\sc iv}] density was used for the calculations of O$^{3+}$/H$^{+}$, Ne$^{4+}$/H$^{+}$ and S$^{3+}$/H$^{+}$.  The [Cl\,{\sc iii}] density was used for PN\,G068.7+14.8.  Although clearly detected in the \emph{Spitzer} spectrum of PN\,G068.7+14.8, [Ne\,{\sc ii}] 12.81\,$\mu$m was not used to determine the Ne$^{+}$/H$^{+}$ ratio, as this line is probably blended with the polycyclic aromatic hydrocarbon (PAH) emission at 12.7\,$\mu$m \citep[e.g.][]{1973ApJ...183...87G,2020NatAs...4..339L}, and may also be affected by the AIB emission band at 11.2\,$\mu$m (Figure\,\ref{fig2}-bottom).  Ionic abundances derived from the mid-IR CELs are summarized in Table\,\ref{tab:spitzer_ionic}. 

For PN\,G048.5+04.2, the Ne$^{2+}$/H$^{+}$ and S$^{2+}$/H$^{+}$ ratios derived from the mid-IR CELs are systematically higher than the corresponding values derived from the optical CELs, while for PN\,G068.7+14.8 this is opposite.  PN\,G068.7+14.8 is of relatively low excitation (compared with PN\,G048.5+04.2), and the [Ne\,{\sc iii}] and [S\,{\sc iii}] line mission probably comes from the nebular centre.  Therefore, the mid-IR CEL ionic abundances of PN\,G068.7+14.8, as derived using the total H$\beta$ flux, are an average over the entire nebula and thus might be underestimated, i.e., lower than the optical CEL abundances, which are an average over the aperture covered by the GTC long slit.  For the high-excitation PN\,G048.5+04.2, as signified by its strong He\,{\sc ii} $\lambda$4686 line, emission of the singly ionized species is relatively weak, and thus the distribution of the [Ne\,{\sc iii}] and [S\,{\sc iii}] mission is expected to be more homogeneous across the nebula.

\begin{table}
\begin{center}
\tablenum{10}
\caption{Elemental Abundance Comparison for PN\,G048.5+04.2: Optical v.s. mid-IR}
\label{tab:elemental_IR}
\begin{tabular}{llclc}
\hline\hline
Elem. & \multicolumn{4}{c}{X/H} \\
\cline{2-5}
 & \multicolumn{2}{c}{Optical\,\tablenotemark{\rm{\scriptsize a}}} & \multicolumn{2}{c}{Optical + mid-IR}\\
\hline
O  & 3.45($\pm$0.16)$\times10^{-4}$ & 8.54 & 5.72($\pm$0.09)$\times10^{-4}$ & 8.76\tablenotemark{\rm{\scriptsize b}} \\
Ne & 6.97($\pm$0.12)$\times10^{-5}$ & 7.84\tablenotemark{\rm{\scriptsize c}} & 5.73($\pm$0.11)$\times10^{-5}$ & 7.76\tablenotemark{\rm{\scriptsize c}} \\
Ne & 9.04($\pm$0.76)$\times10^{-5}$ & 7.96\tablenotemark{\rm{\scriptsize b}} & 9.32($\pm$0.77)$\times10^{-5}$ & 7.97\tablenotemark{\rm{\scriptsize b}} \\
S  & 2.99($\pm$0.42)$\times10^{-6}$ & 6.48 & 5.31($\pm$0.10)$\times10^{-6}$ & 6.73\tablenotemark{\rm{\scriptsize b}} \\
\hline
\end{tabular}
\tablecomments{Optical: GTC spectroscopy; mid-IR: \emph{Spitzer}/IRS data.}
\begin{description}
\tablenotemark{\rm{\scriptsize a}} Abundances from Table\,\ref{tab:elemental} (from GTC spectroscopy). \\
\tablenotemark{\rm{\scriptsize b}} Direct summation of ionic abundances. \\
\tablenotemark{\rm{\scriptsize c}} Using the ICFs given by \citet{1994MNRAS.271..257K}.
\end{description}
\end{center}
\end{table}

However, the discrepancy in the optical/mid-IR CEL abundances may not be entirely due to the aperture difference between the GTC and \emph{Spitzer} observations.  The emissivities of the optical CELs are highly temperature-sensitive, while those of the mid-IR CELs are less temperature-sensitive given their much lower excitation energies\footnote{The IR emission lines detected in PNe are mostly transitions between the fine-structure levels within the ground spectral term, and thus their excitation energies are very low, mostly $<$0.1\,eV.} \citep[e.g.][]{OF2006}.  In classical paradigm, if there is temperature fluctuation in a photoionized gaseous nebula, emission of the optical CELs is more biased to the high-temperature regions (but heavily suppressed in the low-temperature regions); 
therefore, the temperatures probed with optical CELs can be overestimated and consequently the ionic abundances underestimated \citep[e.g.][]{1967ApJ...150..825P}. 

We thus estimated the root-mean-square of temperature fluctuations of a nebula, $t^{2}$, using the electron temperatures derived with different CEL ratios.  According to \citet[][Eq\,15 therein]{1967ApJ...150..825P} and using the excitation energies of CELs from the NIST\footnote{\url{https://physics.nist.gov/PhysRefData/ASD/levels_form.html}}, we have equations,  
\begin{equation}
\label{te_oiii}
T_{\rm e}([{\rm O\ III}])= T_0[1+({91303\over T_0}-3){t^2\over 2}] 
\end{equation} 
and
\begin{equation}
\label{te_siii}
T_{\rm e}([{\rm S\ III}])= T_0[1+({55369\over T_0}-3){t^2\over 2}],
\end{equation} 
where $T_0$ is the average temperature weighted by the square of nebular density over the volume considered \citep[][defined by Eq\,9 therein]{1967ApJ...150..825P}.  Eq\,\ref{te_oiii} is slightly different from Eq\,18 in \citet{1967ApJ...150..825P} maybe due to updated atomic data, but the two equations give similar results. 
Using the CEL temperatures and the corresponding uncertainties, we derived $T_{0}$ = 10,700$^{+1500}_{-1800}$ and $t^{2}$ = 0.095$\pm$0.47 for PN\,G048.5+04.2.  Under this condition and using $T_{\rm e}$([O\,{\sc iii}]) = 13,500, we predicted a ratio of the Ne$^{2+}$/H$^{+}$ ionic abundances yielded by the [Ne\,{\sc iii}] $\lambda$3868 and [Ne\,{\sc iii}] 15.5\,$\mu$m lines to be 0.636$^{+0.302}_{-0.320}$, while the observed ratio of PN\,G048.5+04.2 is 0.327$\pm$0.017.  The predicted ratio of the S$^{2+}$/H$^{+}$ ionic abundances yielded by the [S\,{\sc iii}] $\lambda$9069 and [S\,{\sc iii}] 18.71\,$\mu$m lines is 0.797$^{+0.465}_{-0.625}$, which is close to the observed ratio 0.558$\pm$0.147 as calculated adopting $T_{\rm e}$([S\,{\sc iii}]) = 11,800.  Hence some of the discrepancy in the optical/mid-IR ionic (and maybe also elemental) abundances might be explained by temperature fluctuations.

\begin{figure}[ht!]
\begin{center}
\includegraphics[width=8.5cm,angle=0]{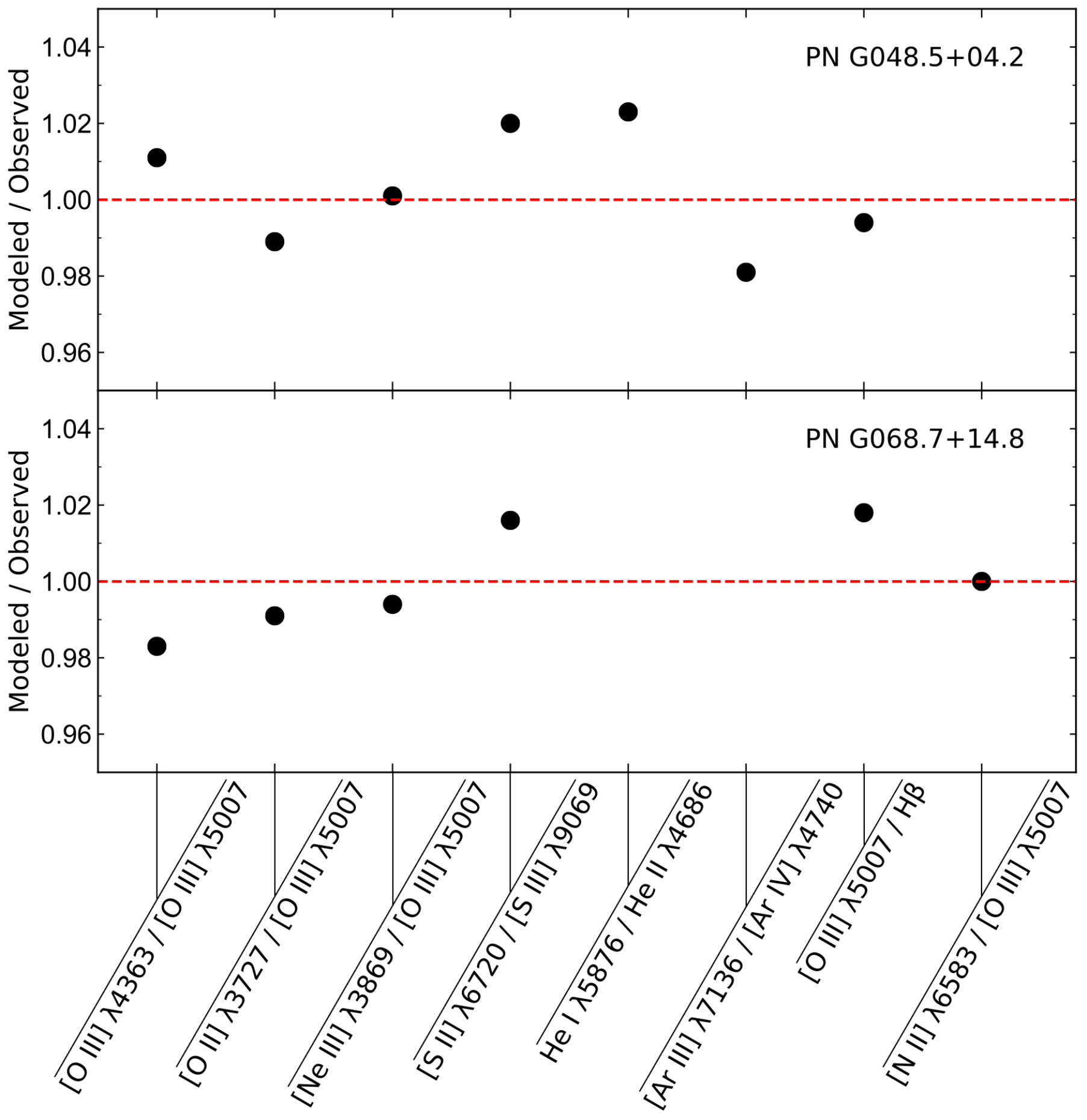} 
\caption{Comparison of the observed and the model-predicted line ratios of PN\,G048.5+04.2 (top) and PN\,G068.7+14.8 (bottom) as listed in Table\,\ref{tab:ratio}; [S\,{\sc ii}] $\lambda$6720 is a combination of the $\lambda\lambda$6716,\,6731 doublet . 
Differences are all $<$2.5\%.} 
\label{fig8}
\end{center}
\end{figure}

Our GTC optical-NIR and the \emph{Spitzer}/IRS mid-IR observations of the two PNe yield different ionic abundances in Ne$^{2+}$ and S$^{2+}$ (see Tables\,\ref{tab:ionic} and \ref{tab:spitzer_ionic}).  Spectroscopic analysis of a distant star-forming galaxy, Mrk\,71 ($D\sim$3.4\,Mpc), revealed that the O$^{2+}$/H$^{+}$ abundance ratio derived using the [O\,{\sc iii}] $\lambda\lambda$52,\,88\,$\mu$m far-IR fine-structure lines is consistent with that yielded by the [O\,{\sc iii}] optical nebular lines \citep{2023NatAs...7..771C}, although this conclusion has been proposed to depend on the effects of electron density \citep{2024NatAs...8..275M}.  More recent observations suggest that far-IR [O\,{\sc iii}] CELs tend to yield higher abundances than does the [O\,{\sc iii}] optical lines \citep[e.g.][]{2025arXiv250509186H}.  However, no definite conclusion can be drawn yet given that high-quality observations, in particular the far-IR spectroscopy (e.g., with \emph{JWST}), so far mainly focus on star-forming galaxies (H\,{\sc ii} regions) and that the sample is also limited.  It is worthwhile to carry out a similar comparison study of the far-IR/optical abundances in PNe, which is still lacking\footnote{Neither PN\,G048.5+04.2 nor PN\,G068.7+14.8 are found in the archives of \emph{SOFIA} (to detect [O\,{\sc iii}] 52\,$\mu$m using FIFI-LS) and \emph{Herschel} (to detect [O\,{\sc iii}] 88\,$\mu$m using PACS)}. 

Elemental abundances of O, Ne, and S in PN\,G048.5+04.2 were calculated using the \emph{Spitzer}/IRS mid-IR ionic abundances in combination with our GTC optical results, and are summarized in Table\,\ref{tab:elemental_IR}, where the optical-based results (from Table\,\ref{tab:elemental}) are presented for purpose of comparison.  The oxygen abundance derived by summing over the optical (O$^+$ O$^{2+}$) and mid-IR (O$^{3+}$) ionic abundances is higher than the ICF-based value by 0.22\,dex.  When using the optical Ne$^{2+}$ and mid-IR Ne$^{4+}$ ionic abundances as well as the ICF(Ne) given by \citet[][Eq\,A27 therein]{1994MNRAS.271..257K}, the neon abundance is lower than the optical-based value by 0.08\,dex; the value increases by direct summation.  For sulfur, the two abundances differ by 0.25\,dex.

\section{Photoionization Models}
\label{sec:models} 

In order to better understand the stellar population of the two PNe, we need to constrain the properties of their central stars, i.e.\ effective temperature ($T_{\rm eff}$) and luminosity ($L_{\ast}$).  These stellar parameters can be obtained indirectly through photoionization modeling based on the observed fluxes of nebular emission lines of a PN.  This is the only method of investigation for a compact PN whose central star can not be directly observed (such as PN\,G068.7+14.8).  The stellar parameters given by the best models of the two PNe, as well as their initial masses and main-sequence ages, are summarized in Table\,\ref{tab:Central stars}.  Here we describe our method of modeling.

\subsection{Line Ratios Used to Constrain Models} 
\label{sec:models:part1} 

To obtain $T_{\rm eff}$ and $L_{\ast}$ of the central stars of the two PNe, we construct photoionization models using the code {\sc cloudy} \citep{1998PASP..110..761F,2017RMxAA..53..385F}, based on the extinction-corrected fluxes of the nebular emission lines detected in the GTC optical-NIR spectra.  We estimated the ECs, according to the definition of \citet[][Eqs\,2.1a and 2.1b therein]{1990ApJ...357..140D}, to be 8.3 and 2.73 for PN\,G048.5+04.2 and PN\,G068.7+14.8, respectively, suggesting that PN\,G048.5+04.2 is a highly excited PN and PN\,G068.7+14.8 is of low excitation.  The $T_{\rm eff}$ and $L_{\ast}$ of the central stars of the two PNe were estimated using the methods of \citet[][Eqs\,3.1 and 3.2 therein]{1991ApJ...377..480D}, who provided empirical fits to the $\log{T_{\rm eff}}$-EC and $\log{(L_{\ast}/L_{\odot})}$-EC correlations based on the photoionization modeling of a sample of Magellanic PNe \citep{1991ApJ...367..115D}. 
Since the nebular H$\beta$ luminosity is required to estimate the luminosity of the PN central star, we adopted the H$\beta$ fluxes, the distances and the angular diameters of the two PNe from \citet{2016ApJ...830...33S}, in conjunction with the slit width ($1^{\prime\prime}$) of our GTC spectroscopy, to estimate the luminosities of the PN central stars.  The above estimations help to narrow down the ranges of $T_{\rm eff}$ and $L_{\ast}$ in photoionization modeling.

\begin{table}
\begin{center}
\tablenum{11}
\caption{Parameters of PN Central Stars yielded by Photoionization Modeling, and Properties of the Main-sequence Progenitors} 
\label{tab:Central stars} 
\begin{tabular}{lccccc}
\hline\hline
PN & log$T_{\rm{eff}}$ & log($L_{\ast}/L_{\odot}$) & $M_{\rm{fin}}$\tablenotemark{\rm{\scriptsize a}} & $M_{\rm{ini}}$\tablenotemark{\rm{\scriptsize b}} & $t_{\rm{ms}}$\tablenotemark{\rm{\scriptsize c}}\\
\hline
PN\,G048.5+04.2 & 5.162 & 2.536 & 0.606 & 1.90 & 1.35\\
PN\,G068.7+14.8 & 4.793 & 3.674 & 0.558 & 1.49 & 2.80 \\
\hline
\end{tabular}
\begin{description}
\tablenotemark{\rm{\scriptsize a}} Core mass (in units of $M_{\odot}$) was interpolated from the post-AGB model tracks of \citet{2016AA...588A..25M}. \\
\tablenotemark{\rm{\scriptsize b}} Initial mass (in units of $M_{\odot}$) was derived using the initial-final mass relation of \citet[][]{2008MNRAS.387.1693C}. \\
\tablenotemark{\rm{\scriptsize c}} Main-sequence age $t_{\rm{ms}}$ (Gyr) was derived based on the model grids computed by \citet{1992AAS...96..269S}. 
\end{description}
\end{center}
\end{table}

Photoionization models were constrained by comparing the model-predicted and the observed nebular line ratios, including $I$([O\,{\sc iii}] $\lambda4363$)/$I$([O\,{\sc iii}] $\lambda5007$), $I$([S\,{\sc ii}] $\lambda$6716+$\lambda$6731)/$I$([S\,{\sc iii}] $\lambda9069$), $I$([O\,{\sc ii}] $\lambda3727$)/$I$([O\,{\sc iii}] $\lambda5007$), $I$([Ne\,{\sc iii}] $\lambda3868$)/$I$([O\,{\sc iii}] $\lambda5007$) and $I$([O\,{\sc iii}] $\lambda5007$)/$I$(H$\beta$); these line ratios to some extent reflect the excitation degree, physical conditions, and relative elemental abundances of a PN, and help to constrain the properties of its central star.  For the high-excitation PN\,G048.5+04.2, two additional line ratios, $I$(He\,{\sc i} $\lambda5876$)/$I$(He\,{\sc ii} $\lambda4686$) and $I$([Ar\,{\sc iii}] $\lambda7136$)/$I$([Ar\,{\sc iv}] $\lambda4740$), were also used.  If the model-predicted line ratios are close to the observed values, it means that the photoionization models are likely to be close to the real nebular condition.

\begin{table*}
\begin{center}
\tablenum{12}
\caption{Comparison of the Observed and {\sc cloudy} Modeled Line Ratios \label{tab:ratio}}
\begin{tabular}{lccclccc}
\hline\hline
Line Ratios & \multicolumn{3}{c}{PN\,G048.5+04.2} &  & \multicolumn{3}{c}{PN\,G068.7+14.8} \\
\cline{2-4}
\cline{6-8}
 & Observed & Modeled & Mod./Obs. & & Observed & Modeled & Mod./Obs. \\
\hline
$I$([O\,{\sc iii}] $\lambda4363$)/$I$([O\,{\sc iii}] $\lambda5007$) & 1.54$\times10^{-2}$ & 1.56$\times10^{-2}$ & 1.011 & & 8.24$\times10^{-3}$ & 8.10$\times10^{-3}$ & 0.983\\
$I$([O\,{\sc ii}] $\lambda3727$)/$I$([O\,{\sc iii}] $\lambda5007$) & 7.50$\times10^{-3}$ & 7.42$\times10^{-3}$ & 0.989 & & 9.40$\times10^{-2}$ & 9.31$\times10^{-2}$ & 0.991\\
$I$([Ne\,{\sc iii}] $\lambda3868$)/$I$([O\,{\sc iii}] $\lambda5007$) & 7.64$\times10^{-2}$ & 7.64$\times10^{-2}$ & 1.001 & & 3.22$\times10^{-2}$ & 3.20$\times10^{-2}$ & 0.994\\
$I$([S\,{\sc ii}] $\lambda$6716+31)/$I$([S\,{\sc iii}] $\lambda9069$) & 8.39$\times10^{-2}$ & 8.57$\times10^{-2}$ & 1.020 & & 0.187 & 0.190 & 1.016\\
$I$(He\,{\sc i} $\lambda5876$)/$I$(He\,{\sc ii} $\lambda4686$) & 7.62$\times10^{-2}$ & 7.80$\times10^{-2}$ & 1.023 & & $\cdots$ & $\cdots$ & $\cdots$\\
$I$([Ar\,{\sc iii}] $\lambda7136$)/$I$([Ar\,{\sc iv}] $\lambda4740$) & 1.07 & 1.05 & 0.981 & & $\cdots$ & $\cdots$ & $\cdots$\\
$I$([O\,{\sc iii}] $\lambda5007$)/$I$(H$\beta$) & 12.63 & 12.55 & 0.994 & & 6.077 & 6.186 & 1.018\\
$I$([N\,{\sc ii}] $\lambda6583$)/$I$([O\,{\sc iii}] $\lambda5007$) & $\cdots$ & $\cdots$ & $\cdots$ & & 3.38$\times10^{-2}$ & 3.38$\times10^{-2}$ & 1.000 \\
\hline
\end{tabular}
\begin{description}
NOTE. -- Line ratios are retained to only 3 significant digits, and Modeled/Observed ratios are retained to three decimal places. 
\end{description}
\end{center}
\end{table*}

\begin{table*}
\begin{center}
\tablenum{13}
\caption{Comparison of the Observed and {\sc cloudy} Modeled Line Intensities, All Normalized to $I$(H$\beta$) = 100} 
\label{tab:lineintensity} 
\begin{tabular}{lrrrcrrr}
\hline\hline
 & \multicolumn{3}{c}{PN\,G048.5+04.2} &  & \multicolumn{3}{c}{PN\,G068.7+14.8} \\
\cline{2-4}
\cline{6-8}
Line ID & Observed & Modeled & Mod./Obs. & & Observed & Modeled & Mod./Obs. \\
\hline
$[$O\,{\sc ii}$]$ $\lambda3727$ & 9.47 & 9.31 & 0.983 & & 57.11 & 57.62 & 1.009 \\
$[$Ne\,{\sc iii}$]$ $\lambda3868$ & 96.57 & 95.97 & 0.994 & & 19.56 & 19.80 & 1.012 \\
H\,{\sc i} $\lambda4101$ & 27.54 & 26.23 & 0.952 & & 27.71 & 26.30 & 0.949 \\
H\,{\sc i} $\lambda4340$ & 44.62 & 47.45 & 1.063 & & 44.99 & 47.35 & 1.052 \\
$[$O\,{\sc iii}$]$ $\lambda4363$ & 19.48 & 19.54 & 1.003 & & 5.01 & 5.01 & 1.000 \\
He\,{\sc i} $\lambda4471$ & 1.84 &1.83 & 0.995 & & 4.98 & 5.24 & 1.052\\
He\,{\sc ii} $\lambda4686$ & 71.75 & 69.13 & 0.963 & & $\cdots$ & $\cdots$ & $\cdots$ \\
$[$Ar\,{\sc iv}$]$ $\lambda4740$ & 8.30 & 8.31 & 1.001 & & $\cdots$ & $\cdots$ & $\cdots$ \\
H\,{\sc i} $\lambda4861$ & 100.00 & 100.00 & 1.000 & & 100.00 & 100.00 & 1.000 \\
$[$O\,{\sc iii}$]$ $\lambda4959$ & 409.20 & 420.77 & 1.028 & & 201.35 & 207.35 & 1.030 \\
$[$O\,{\sc iii}$]$ $\lambda5007$ & 1263.2 & 1255.4 & 0.994 & & 607.69 & 618.64 & 1.018 \\
He\,{\sc ii} $\lambda5411$ & 5.87 & 5.92 & 1.009 & & $\cdots$ & $\cdots$ & $\cdots$ \\
$[$Cl\,{\sc iii}$]$ $\lambda5517$ & 0.41 & 0.27 & 0.659 & & 0.18 & 0.17 & 0.944 \\
$[$Cl\,{\sc iii}$]$ $\lambda5537$ & 0.28 & 0.45 & 1.607 & & 0.29 & 0.36 & 1.241 \\
He\,{\sc i} $\lambda5876$ & 5.47 & 5.39 & 0.985 & & 15.33 & 15.30 & 0.998\\
$[$S\,{\sc iii}$]$ $\lambda6312$ & 0.86 & 1.02 & 1.186 & & 0.48 & 0.43 & 0.896 \\
$[$N\,{\sc ii}$]$ $\lambda6548$ & 1.05 & 0.89 & 0.848 & & 6.84 & 7.10 & 1.038\\
H\,{\sc i} $\lambda6563$ & 300.74 & 275.80 & 0.917 & & 274.82 & 277.84 & 1.011\\
$[$N\,{\sc ii}$]$ $\lambda6583$ & 2.20 & 2.62 & 1.191 & & 20.54 & 20.92 & 1.019\\
He\,{\sc i} $\lambda6678$ & 1.38 & 1.34 & 0.971 & & 3.94 & 3.80 & 0.964 \\
$[$S\,{\sc ii}$]$ $\lambda6716$ & 0.32 & 0.28 & 0.875 & & 0.39 & 0.38 & 0.974 \\ 
$[$S\,{\sc ii}$]$ $\lambda6731$ & 0.43 & 0.49 & 1.140 & & 0.87 & 0.78 & 0.897 \\ 
He\,{\sc i} $\lambda7065$ & 2.02 & 2.44 & 1.208 & & 11.00 & 11.43 & 1.039\\
$[$Ar\,{\sc iii}$]$ $\lambda7136$ & 8.89 & 8.74 & 0.983 & & 4.23 & 4.09 & 0.967 \\
$[$O\,{\sc ii}$]$ $\lambda7325$ & 0.89 & 1.32 & 1.483 & & 10.51 & 10.97 & 1.044\\
$[$Ar\,{\sc iii}$]$ $\lambda7751$ & 2.23 & 2.07 & 0.928 & & 1.00 & 0.97 & 0.970 \\
$[$S\,{\sc iii}$]$ $\lambda9069$ & 8.94 & 8.98 & 1.004 & & 6.19 & 6.10 & 0.985 \\
\hline 
\end{tabular}
\begin{description}
NOTE. -- Only the emission lines not affected by the second-order contamination of the GTC/OSIRIS instrument are selected for comparison.  Nebular lines whose fluxes have been corrected for blending are also presented; ``$\cdots$'' means the emission line was undetected in the spectrum or contaminated by emission from the central star.  Values of the Modeled/Observed line ratios are retained to three decimal places.  The majority of the model-predicted line intensities are in good agreement with the observed values; only a small number of weak lines have the modeled intensities deviating significantly from the observations. 
\end{description}
\end{center}
\end{table*}

\begin{table}
\begin{center}
\tablenum{14}
\caption{Comparison of Elemental Abundances: Observations vs. Photoionization Modeling}
\label{tab:comparison1}
\begin{tabular}{lrrcrr}
\hline\hline
Elem.  & \multicolumn{2}{c}{PN\,G048.5+04.2} & & \multicolumn{2}{c}{PN\,G068.7+14.8} \\
\cline{2-3}
\cline{5-6}
 & Observed & Modeled & & Observed & Modeled \\
\hline
He & 11.00 & 11.03 & & 11.00 & 11.00 \\
C & 8.50 & 8.95 & & 9.23 & 9.11 \\
N & 7.54 & 7.72 & & 7.37 & 7.48 \\
O & 8.54 & 8.58 & & 8.34 & 8.35 \\
Ne & 7.85 & 8.00 & & 7.34 & 7.47 \\
S & 6.48 & 6.37 & & 6.06 & 6.02 \\
Cl & 4.82 & 5.00 & & 4.60 & 4.60 \\
Ar & 6.19 & 6.24 & & 5.76 & 5.55 \\
\hline
\end{tabular}
\begin{description}
NOTE. -- All abundances are in logarithm, $\log{\rm (X/H)}$+12. 
\end{description}
\end{center}
\end{table}

For the low-excitation PN\,G068.7+14.8, we did not use the $I$(He\,{\sc i} $\lambda$5876)/$I$(He\,{\sc ii} $\lambda$4686) and $I$([Ar\,{\sc iii}] $\lambda$7136)/$I$([Ar\,{\sc iv}] $\lambda$4740) ratios in model constraint, because the [Ar\,{\sc iv}] lines are not detected in the GTC spectrum and the He\,{\sc ii} line may be contaminated by emission from the [WR]-type central star of this PN (see discussion in Section\,\ref{subsec:central-stars}).  Due to the low resolution of the GTC spectra and weakness of the He\,{\sc ii} $\lambda$4686 line in PN\,G068.7+14.8 (Table\,\ref{tab:lines}), it is difficult for us to distinguish the stellar component from the nebular emission.  The He\,{\sc ii} $\lambda$4686 nebular emission is thus probably overestimated.  Moreover, He\,{\sc ii} emission may also be shock-excited in the inner regions of PNe \citep[e.g.][]{2022MNRAS.517.5166R}, which cannot be simulated by {\sc cloudy}.  Therefore, it is difficult obtain the model-predicted He\,{\sc i}/He\,{\sc ii} line ratio that agrees with GTC observations.  We thus added $I$([N\,{\sc ii}] $\lambda$6583)/$I$([O\,{\sc iii}] $\lambda$5007) in the list of predicted line ratios as a compensation for PN\,G068.7+14.8.

\begin{figure}[ht!]
\begin{center}
\includegraphics[width=8.5cm,angle=0]{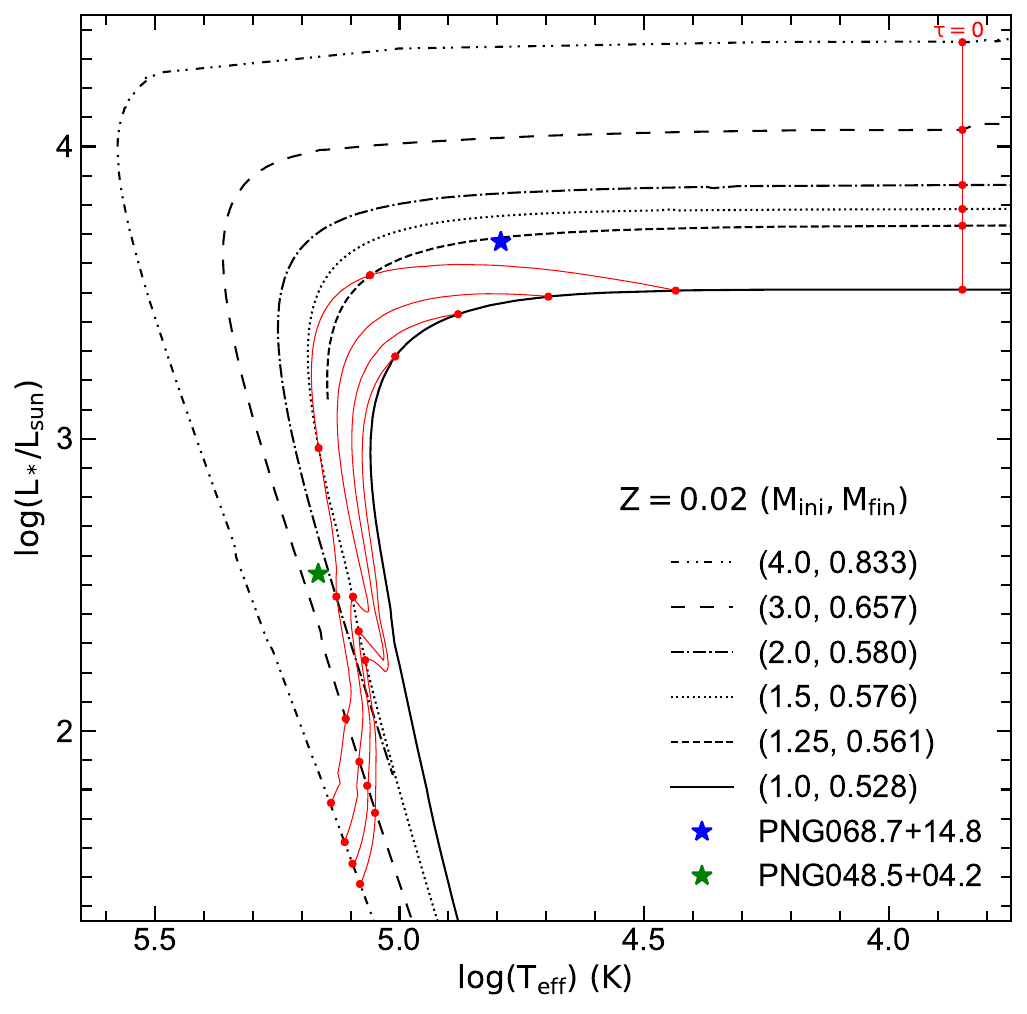}
\caption{Central star positions of the two PNe in the H-R diagram. Overplotted in the diagram are the model tracks of the H-burning post-AGB sequences calculated by \citep[][at $Z$ = 0.02]{2016AA...588A..25M}.  Different line types represent different different initial and final masses.  Red curves are the isochrones for the evolutionary ages ($\tau$ = 0, 5000, 10,000, 15,000, and 20,000\,yr) since the beginning point of the post-AGB defined at $\log{T_{\rm eff}}$ = 3.85.} 
\label{fig9}
\end{center}
\end{figure}

Our photoionization modeling of the two PNe is mainly based on the GTC optical-NIR spectra.  Although the archival \emph{Spitzer}/IR mid-IR and \emph{HST}/STIS UV spectra (GO prop.~ID: 15211; PI: L.\ Stanghellini) are available, they were not used to constrain the {\sc cloudy} models.  For PN\,G048.5+04.2, the data quality of the \emph{HST}/STIS UV spectrum is poor, with no lines detected (probably due to faintness of this PN); for PN\,G068.7+14.8, only two emission lines were detected in the STIS NUV spectrum, with limited S/N's \citep{2022ApJ...929..148S}.  The slit aperture of the \emph{Spitzer}/IRS spectroscopy are different from that of the GTC observations, and the number of mid-IR lines are small.

\begin{table*}
\begin{center}
\tablenum{15}
\caption{Abundance Comparison for PN\,G068.7+14.8}
\label{tab:comparison2}
\begin{tabular}{lrrcrrcrr}
\hline\hline
Elem.  & \multicolumn{2}{c}{this work} & & \multicolumn{2}{c}{\citet{2021MNRAS.504..816B}} & & \multicolumn{2}{c}{\citet{2005MNRAS.362..424W}}\\
\cline{2-3}
\cline{5-6}
\cline{8-9}
 & Observed & Modeled & & Observed & Modeled & & CEL & ORL\\
\hline
He & 11.00 & 11.00 & & 11.06 & 11.08 & & $\cdots$ & 10.97\\
C & 9.23 & 9.11 & & $\cdots$ & 9.13 & & $\cdots$ & 9.21\\
N & 7.37 & 7.48 & & 7.63 & 7.41 & & 7.49 & 9.55\\
O & 8.34 & 8.35 & & 8.35 & 8.22 & & 8.20 & 8.67\\
Ne & 7.34 & 7.47 & & 7.58 & 7.27 & & 7.02 & $\cdots$\\
S & 6.06 & 6.02 & & 6.05 & 6.12 & & 6.03 & $\cdots$\\
Cl & 4.60 & 4.60 & & 4.66 & 4.66 & & $\cdots$ & $\cdots$\\
Ar & 5.76 & 5.55 & & 5.61 & 5.56 & & 5.65 & $\cdots$\\
\hline
\end{tabular}
\begin{description}
NOTE. -- All abundances are in logarithm, $\log{\rm (X/H)}$+12. \\
\end{description}
\end{center}
\end{table*}

\subsection{The Parameters Considered in Modeling} 
\label{sec:models:part2}

The inputs to the {\sc cloudy} photoionization model for a PN mainly consist of three factors:  (1) the incident radiation field of the central ionizing source (i.e.\ a white dwarf) as defined by $T_{\rm eff}$ and $L_{\ast}$, (2) chemical composition of the nebula, including the gaseous and dust components, and (3) radial density profile of hydrogen, and geometry of the PN: 

(1) In order for a more realistic incident field, Rauch atmospheres \citep{2003AA...403..709R} were adopted.  The H-Ni atmospheric models of \citet{2003AA...403..709R} were used for PN\,G048.5+04.2.  For PN\,G068.7+14.8, realistic H-poor atmospheric models should be used, given that emission features from the [WR]-type central star were observed (Figure\,\ref{fig7}). 
The only probable central-star emission detected in the \emph{HST}/STIS UV spectrum of PN\,G068.7+14.8 is C\,{\sc iv} $\lambda\lambda$1548,1551 \citep{2022ApJ...929..148S}.  In the low-resolution ($R\sim$1000) GTC optical-NIR spectrum of PN\,G068.7+14.8, the emission lines from its [WR]-type central star are only from carbon (Figure\,\ref{fig7}); therefore it is impossible to constrain the chemical composition of stellar atmosphere, and consequently difficult to generate a theoretical SED for the [WR] central star using the NLTE Potsdam Wolf-Rayet Star Models \citep[PoWR,][]{2002AA...387..244G,2004AA...427..697H}.  We adopted the PG\,1159 grids of Rauch models \citep{2003AA...403..709R} as the incident central-star SEDs in the photoionization modeling for PN\,G068.7+14.8.  This approximation is acceptable, given that the central star of PN\,G068.7+14.8 is probably [WELS], some of which are considered as a transitional stage between the [WCE] and the PG\,1159 type. 

(2) The ICF-derived elemental abundances (from Table\,\ref{tab:elemental}, see also Section\,\ref{subsec:element}) were used as the initial input to the model runs; the other elements that were not observed in our GTC optical-NIR spectra were not included in the modeling.  According to the uncertainties in elemental abundances, especially those from the faint emission lines (e.g., C\,{\sc ii} $\lambda$4267 in PN\,G048.5+04.2), we adjusted the input abundance values so that the model-predicted intensity ratios of nebular emission lines agree with those measured from the GTC optical spectra, within a given deviation level (e.g.\ $<$2.5\%).  A comparison between the observed abundances and those given by our best-fit model is discussed in Section\,\ref{subsec:versus}.  The default parameters of dust in {\sc cloudy} were used for the two PNe because of a lack of dust information in our spectra. 

(3) For the nebular geometry and matter distribution of the models, we assumed that both PNe have spherical shell structures with inner boundaries/radii $\log{r_{\rm in}}$ (=16.4 and 16.8 for PN\,G048.5+04.2 and PN\,G068.7+14.8, respectively, in units of cm).  Due to the optically thin nature of PN\,G048.5+04.2 \citep{2016ApJ...830...33S}, we set an outer boundary $\log{r_{\rm out}}$ =17.5 (in unit of cm) as a truncation suggested by \citet{1991ApJ...377..480D} in the modeling of this PN.  Compared to PN\,G048.5+04.2, the angularly compact nebula PN\,G068.7+14.8 could be more optically thick \citep{2016ApJ...830...33S}.  We could not obtain optimal results for PN\,G068.7+14.8 by setting an outer truncation; thus instead of setting an outer boundary for this PN, we let the model running of this PN stops when the default lowest temperature allowed by {\sc cloudy} was reached.  The density distribution of hydrogen also matters; thus we adopted \textit{isobaric} models for PN\,G068.7+14.8, and a power-law radial density distribution (with a power-law index of $-$2.3) for PN\,G048.5+04.2.  The number density of hydrogen ($n_{\rm H}$) at the inner radii of PN\,G048.5+04.2 and PN\,G068.7+14.8 was set to be $\sim$20,000 and $\sim$10,000\,cm$^{-3}$, respectively; these values make the average densities close to observations. 

In addition to the above three main factors, observational/aperture effects were also considered in our photoionization modeling.  The angular diameter of PN\,G068.7+14.8 is slightly larger than the width (1\arcsec) of the GTC long slit, and thus the loss in line flux is expected to be negligible.  The angular diameter ($\sim$4\arcsec) of PN\,G048.5+04.2 is much larger than the slit width, and this was considered the models of this PN.  The filling factor also affects the model results, and was treated as a free parameter to optimize our models.

\subsection{Model Grids and the Results} 
\label{sec:models:part3} 

Grids of photoionization models were built for the two PNe based on their central star parameters ($T_{\rm eff}$ and $L_{\ast}$).  The initial ranges of temperature and luminosity for the central star of PN\,G048.5+04.2 were set to be $\log{T_{\rm eff}}$ = 4.7--5.3 [K] with a step of 0.05, and $\log{(L_{\ast}/L_{\odot})}$ = 2.0--4.2 with a step of 0.2; for the central star of PN\,G068.7+14.8, these two parameter ranges were 4.65--5.0 and 3.3--4.3, respectively, with the same steps.  Photoionization models of the two PNe were then run on these grids; on each grid, the model-predicted nebular line ratios were compared with the observed line ratios.  Once ``good'' grids were reached via general comparison, much finer grids, with much narrower ranges and smaller steps, were then set in the vicinity of these ``good'' grids.  Through the fine-grid model running and more detailed comparison, we achieved the best models that have the best agreement with the observations.  The predicted line ratios from the best models and the observed line ratios all differ by $<$2.5\% (see Table\,\ref{tab:ratio} and Figure \ref{fig8}). 

To further verify the agreement between the photoionizaiton models and the GTC spectroscopic observations, we compared the observed line intensities, as measured from the GTC optical-NIR spectra and normalized to $I$(H$\beta$) = 100, with the counterparts from our best models; these lines are presented in Table\,\ref{tab:lineintensity}.  The nebular emission lines selected for comparison have not blending issues and unaffected by the second-order contamination.  The modeled/observed intensity difference of some weak lines (e.g., [Cl\,{\sc iii}]) are significant, maybe due to large measurement errors (and of course, uncertainties in photoionization models).  The model-predicted and the observed values of other lines agree well, with differences mostly $<$10\%. 

The $T_{\rm eff}$ and $L_{\ast}$ of the centrals stars yielded by our best photoionization models were adopted for the two PNe (Table\,\ref{tab:Central stars}).  The central star positions of the two PNe in the Hertzsprung-Russell (H-R) diagram are shown in Figure\,\ref{fig9}, where model tracks of the H-burning post-AGB evolutionary sequences calculated by \citet[][at $Z$ = 0.02]{2016AA...588A..25M} are presented for purpose of comparison.  Using linear interpolations between the post-AGB model tracks, we derived the masses of the central stars (i.e.\ the final core mass $M_{\rm fin}$):  0.606\,$M_{\odot}$ for PN\,G048.5+04.2, and 0.558\,$M_{\odot}$ for PN\,G068.7+14.8.  
The initial masses ($M_{\rm ini}$) of their progenitors were then estimated using the semi-empirical initial–final mass relation (IFMR) given by \citet[][Eq\,1 therein]{2008MNRAS.387.1693C}.  Finally, the lifetimes of the main-sequence progenitors ($t_{\rm ms}$) of the two PNe were obtained using the initial masses, based on the model grids of \citet{1992AAS...96..269S}.  These are summarized in Table\,\ref{tab:Central stars}.

\begin{figure*}[ht!]
\begin{center}
\includegraphics[width=15cm,angle=0]{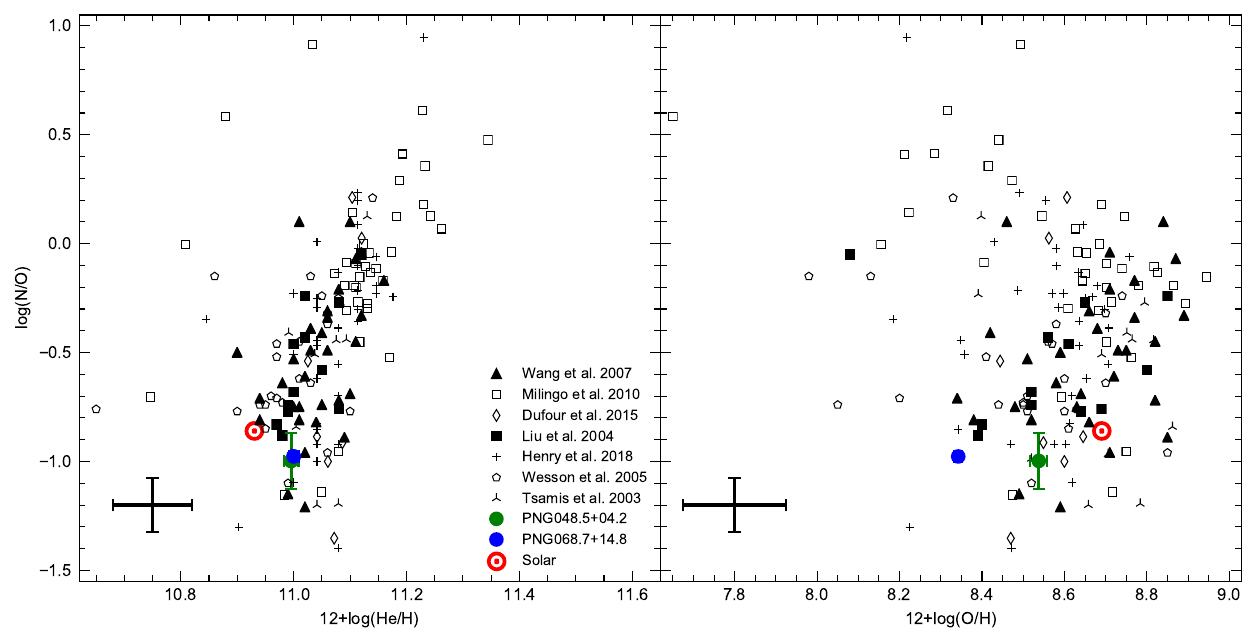}
\caption{N/O vs. O/H (left) and He/H (right), both displayed in logarithm.  PN\,G048.5+04.2 (large green dot) and PN\,G068.7+14.8 (large blue dot) are compared with the samples from literature as represented by different symbols (see the legend).  Solar abundances are adopted from \citep{2009ARAA..47..481A}.  Typical error bars of the literature samples are indicated in the bottom-left corner.} 
\label{fig10}
\end{center}
\end{figure*}

Photoionization modeling of PN\,G068.7+14.8 using {\sc cloudy} was also reported by \citet{2021MNRAS.504..816B}, whose model setups differ significantly from ours, mainly in the spectrum of the central star and density distribution/profile of the nebula.  The {\sc tlusty} atmosphere \citep{2003ApJS..146..417L} was used in their best photoionization model, while we adopted the PG\,1159 grid of \citet{2003AA...403..709R} to simulate the incident radiation field.  Although some of the [WELSs] are hydrogen-rich \citep{2003IAUS..209..235F}, the central star of PN\,G068.7+14.8 is more likely to be H-poor because the Balmer lines in the GTC spectrum are plainly narrower than the C\,{\sc iii} and C\,{\sc iv} lines (Figure \ref{fig7}).  The SED of the PG\,1159 atmosphere thus may reasonably represent the central star radiation of PN\,G068.7+14.8. 

\citet{2021MNRAS.504..816B} utilized the density profile created using the 3D modeling code {\sc shape} \citep{2011ITVCG..17..454S} based on the \emph{HST} [O\,{\sc iii}] narrowband image.  However, PN\,G068.7+14.8 is so compact ($\theta<$2\arcsec, see Figure\,\ref{fig1}) that it is difficult to discern any nebular structures in details.  
We assumed an isobaric nebular model for PN\,G068.7+14.8 and set the Str$\rm{\ddot{o}}$mgren sphere as the outer boundary. 
Although our model settings and assumptions for PN\,G068.7+14.8 are quite different from those of \citet{2021MNRAS.504..816B}, the $T_{\rm eff}$ and $L_{\ast}$ of the central star we obtained for this PN are not too far from the model results of \citet{2021MNRAS.504..816B}:  $\Delta{T_{\rm eff}}$ = 0.12\,dex and $\Delta{L_{\ast}}$ = 0.07\,dex.  

\section{Discussion} 
\label{sec:discussion}

\subsection{Abundance Comparison: Observations vs. Models} 
\label{subsec:versus}

During photoionization modeling, the input elemental abundances elements were adjusted so that the model-predicted line ratios agree with observations (Section\,\ref{sec:models:part2}).  A comparison between the observed abundances and our best-model predictions is summarized in Table\,\ref{tab:comparison1}, where for most elements, the differences between the modeled and the observed abundances are $<$0.1\,dex.  The modeled/observed abundance difference is obvious for those elements that have been only one single ionization stage observed in the GTC spectroscopy, such as C in PN\,G048.5+04.2, Ne and Ar in PN\,G068.7+14.8, and N in both PNe.  This could be due to large uncertainty introduced by the ICFs. 

\citet{2005MNRAS.362..424W} and \citet{2021MNRAS.504..816B} also conducted spectroscopic observations of PN\,G068.7+14.7 and derived abundances for this PN.  A comparison of the elemental abundances obtained in this work and those reported in \citet{2021MNRAS.504..816B} and \citet{2005MNRAS.362..424W} is presented in Table\,\ref{tab:comparison2}.  Our abundance values generally are similar to those of \citet{2021MNRAS.504..816B}.  The Ne/H ratios we obtained for PN\,G068.7+14.7, from both observations and photoionization modeling, are much higher than that of \citet{2005MNRAS.362..424W}.  The abundances of the other elements determined using CELs by \citet{2005MNRAS.362..424W} do not differ much from our results, considering different observations.  The abundances of He and C given by \citet{2005MNRAS.362..424W}, as derived from ORLs, are close to this work.

\begin{figure*}[ht!]
\begin{center}
\includegraphics[width=15cm,angle=0]{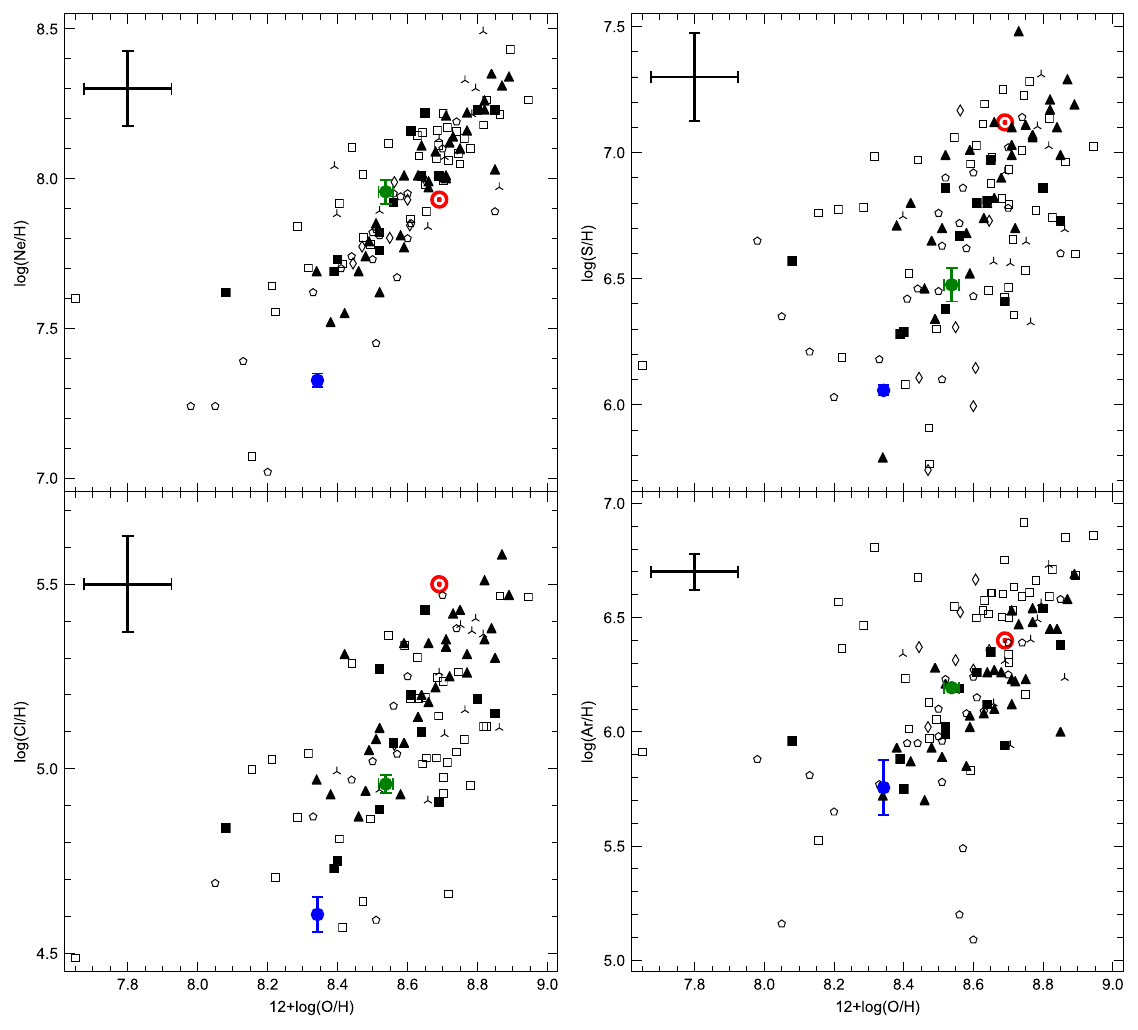}
\caption{Ne/H vs. O/H (top-left), S/H vs. O/H (top-right), Cl/H vs. O/H (bottom-left), and Ar/H vs. O/H (bottom-right), all displayed in logarithm.  Symbols are the same as in Figure\,\ref{fig10}.  Typical error bars of the literature samples are indicated in the top-left corner.} 
\label{fig11}
\end{center}
\end{figure*}

\subsection{Evolutionary Stages and Possible Stellar Population} 
\label{subsec:population}

Our best {\sc cloudy} models show that the central star of PN\,G048.5+04.2 has a higher effective temperature but lower luminosity than does the central star of PN\,G068.7+14.8 (Table\,\ref{tab:Central stars}), although the hotter central core of PN\,G048.5+04.2 is discernible from the strong He\,{\sc ii} $\lambda$4686 emission line in the GTC spectrum.  In comparison with the theoretical H-burning post-AGB evolutionary tracks of \citet{2016AA...588A..25M}, the central stars of the two PNe are in different evolution stages -- the central star of PN\,G048.5+04.2 has entered the white dwarf cooling track, while the central star of PN\,G068.7+14.8 is probably still in the process of heating up (Figure\,\ref{fig9}).  The high $T_{\rm eff}$ ($\sim$14,5000) of PN\,G048.5+04.2's central star produces a very hard radiation field with a large portion of UV ionizing photons, which is responsible for high-ionization species and results in a higher average electron temperature derived from the CELs of this PN (Table\,\ref{tab:temden}). 

The progenitor of PN\,G068.7+14.8 is of low mass (1.49\,$M_{\odot}$, compared with that of PN\,G048.5+04.2); its central star thus evolves much slower, and seems to be still in the early stage of post-AGB evolution, which is consistent with the smaller angular diameter of the nebula \citep[][see also Figure\,\ref{fig1}-bottom]{2016ApJ...830...33S} and the higher electron density as diagnosed using the CELs (Table\,\ref{tab:temden}).  Although the central star of PN\,G048.5+04.2 has entered the cooling track (Figure\,\ref{fig9}), its surrounding nebula still can be well observed in optical emission.  This is because the central star of PN\,G048.5+04.2 evolves faster given its relatively higher initial mass (1.90\,$M_{\odot}$): it took a short time ($<$5000\,yr) to reach the cooling track.  At this stage, the surrounding PN has not dispersed into the space and its optical emission not faded. 

The main-sequence ages of PN\,G048.5+04.2 and PN\,G068.7+14.8 are 1.35\,Gyr and 2.80\,Gyr, respectively.  The two PNe thus evolved from the low-mass progenitors ($\lesssim$2\,$M_{\odot}$) of young population ($<$3\,Gyr).  These are the results we derived using the GTC deep spectroscopy and the post-AGB evolutionary model tracks based on the assumption of single-star evolution.  However, there are always uncertainties inherited in our photoionization modeling and also in the initial-final mass relation of stellar evolution, in particular for the low-mass range; moreover, the post-AGB evolution models used in our analysis are H-burning, which might not be suitable for the central star of PN\,G068.7+14.8, which is probably [WC]-type (i.e., H-poor). 

\begin{figure*}[ht!]
\begin{center}
\includegraphics[width=17.5cm,angle=0]{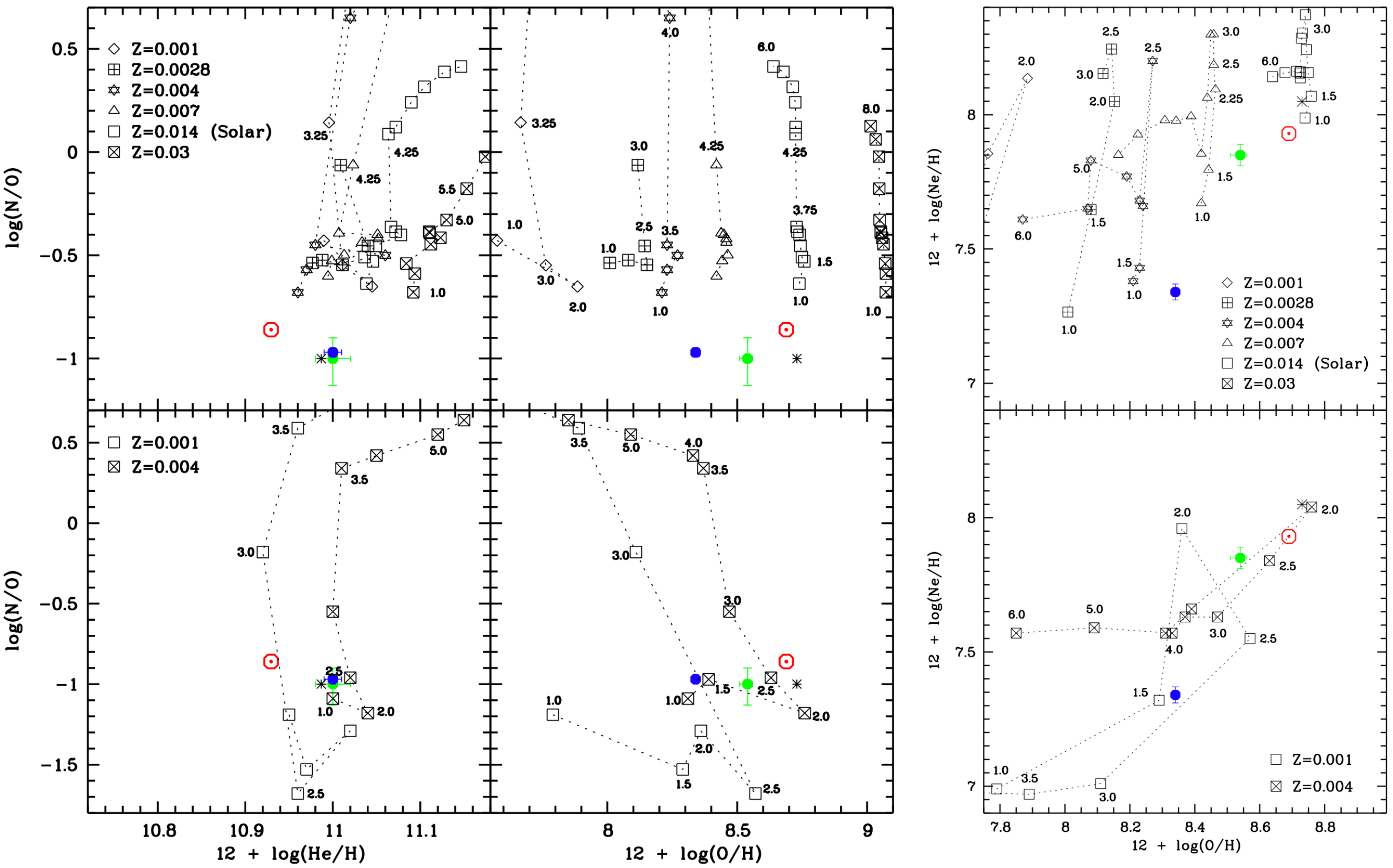}
\caption{\emph{Left}: N/O versus He/H (left) and O/H (right) of PN\,G048.5+04.2 (green dot) and PN\,G068.7+14.8 (blue dot), along with values of the Sun \citep[red $\odot$,][]{2009ARAA..47..481A} and the Orion Nebula \citep[black asterisk,][]{2004MNRAS.355..229E}.  Symbols and color-coding are the same as in Figure\,\ref{fig10}.  \emph{Right}: Same as \emph{Left} but for Ne/H versus O/H.  In the \emph{top} panels: AGB model predictions from \citet[][$Z$=0.004]{Karakas_2010}, \citet[][$Z$=0.001]{Fishlock_2014} and \citet[][$Z$=0.007, 0.014 and 0.03]{KL_2016} for the surface abundances at different metallicities are overplotted for purpose of comparison.  In the \emph{bottom} panels: AGB model predictions from \citet[][$Z$=0.001]{Ventura_2013} and \citet[][$Z$=0.004]{Ventura_2014} are overplotted.  Different symbols represent different metallicities; symbols of the same metallicity are linked by dotted lines to aid visualization.  Initial mass (in units of $M_{\sun}$) of the progenitor star is labeled for each model.  The vertical scales of the top and bottom panels are different to accommodate the AGB model grids.} 
\label{fig12}
\end{center}
\end{figure*}

Statistical investigation revealed that the population of PNe is correlated with nebular morphology as well as their spatial distribution in the Milky Way \citep[e.g.,][]{2002ApJ...576..285S,2006ApJ...651..898S}.  The Galactic latitudes and scale heights of the round and elliptical PNe are higher \citep{2000ASPC..199...17M} in avarage, and their progenitors are probably low-mass stars \citep{2002ApJ...576..285S}.  PNe with bipolar morphology have high nitrogen and helium abundances and likely evolve from high-mass progenitors ($>$3\,$M_{\odot}$); they are mostly located in the disk and tend to have low Galactic latitudes \citep{1983IAUS..103..233P,1995AA...293..871C,2000ASPC..199...17M,2002ApJ...576..285S,2006ApJ...651..898S}.  Although both PNe are in line with the nebular morphology--progenitor mass correlation (i.e., they are round/elliptical and both evolve from low-mass progenitors, as derived from photoionization modeling), the stark difference in the evolution stages of their central stars (see Figure\,\ref{fig9}) indicate that these two PNe are probably more complicated than their appearance.  In the \emph{HST}/WFC3 [O\,{\sc iii}] narrowband image, PN\,G048.5+04.2 is elliptical with a brighter inner shell (Figure\,\ref{fig1}-top); although PN\,G068.7+14.8 is very compact in angular size, it seems to be round, with an inner bright elliptical ring (Figure\,\ref{fig1}-bottom). 

We also attempted to constrain the population of the two PNe using their Galactic locations.  Because of the difficulty in distance measurements of Galactic PNe, there are significant discrepancy in the results of different methods, especially for the distant objects.  If the distances derived using the standard statistical techniques of \citet{2008ApJ...689..194S} were adopted, the Galactic heights of PN\,G048.5+04.2 and PN\,G068.7+14.8 are $\sim$1.1\,kpc and $\sim$5.5\,kpc, respectively \citep{2016ApJ...830...33S}.  If the new distances based on the revised distance scale of \citet[][$D$ = 7.906\,kpc for PN\,G048.5+04.2 and 14.251\,kpc for PN\,G068.7+14.8]{Bucciarelli_2023} were adopted, the Galactic heights of the two PNe are then $\sim$0.58\,kpc (for PN\,G048.5+04.2) and 3.64\,kpc (for PN\,G068.7+14.8).  Either method, the Galactic heights indicate that PN\,G048.5+04.2 is located in the thick disk, and PN\,G068.7+14.8 is probably in the halo.  Hence the stellar population, morphology, and spatial location of the two PNe seem to be consistent with the general correlations found in the Galactic PNe.

\subsection{Abundance Comparison with Other Galactic PNe} 
\label{subsec:comparison} 

Elemental abundances were used to constrain the population of PNe, in particular the N/O ratio, which is a indicator of the initial masses of PNe progenitors \citep[e.g.][]{Karakas_2010,KL_2016}.  Type\,I PNe evolved from the stars with higher initial masses that tend to yield higher N/O and He/H abundance ratios, whereas the Type\,II PNe generally have lower N/O ratios and their progenitor masses are lower \citep{1978IAUS...76..215P,1994MNRAS.271..257K}.  In Figures\,\ref{fig10} and \ref{fig11}, we compare the elemental abundance ratios of our two PNe with those of other samples of Galactic PNe from literature \citep{2003MNRAS.345..186T, 2004MNRAS.353.1251L, 2005MNRAS.362..424W, 2007MNRAS.381..669W, 2010ApJ...711..619M, 2015ApJ...803...23D, 2018MNRAS.473..241H}; the solar abundances of \citet{2009ARAA..47..481A} are also included.  In some of the literatures \citep[e.g.][]{2003MNRAS.345..186T,2004MNRAS.353.1251L,2005MNRAS.362..424W,2007MNRAS.381..669W}, elemental abundances were derived using both CELs and ORLs; only the CEL abundances were used in Figures\,\ref{fig10} and \ref{fig11} for comparison. 

The N/O ratios of PN\,G048.5+04.2 and PN\,G068.7+14.8 are slightly lower than the solar value, and obviously lower than the majority of the literature samples (Figure\,\ref{fig10}).  The two PNe have N/O ratios lower than the Galactic Type\,I PNe of \citet{2010ApJ...711..619M}.  The helium abundances of both PNe are lower than most of the Type\,I PNe.  Therefore, the chemical compositions of the two PNe are closer to Type\,II, suggesting that they probably evolved from low-mass stars.  Abundance correlations between $\alpha$ elements (Ne, S, Cl and Ar) and oxygen are shown in Figure\,\ref{fig11}.  Although the abundances of these $\alpha$ elements generally have positive correlations with oxygen, obvious scatter exists in S, Cl and Ar versus O.  The abundances of PN\,G048.5+04.2 are similar to those of most Galactic samples, but the abundances of PN\,G068.7+14.8 are lower than the majority.  Both PNe have lower abundance ratios than the Sun.

\begin{figure}[ht!]
\begin{center}
\includegraphics[width=8.5cm,angle=0]{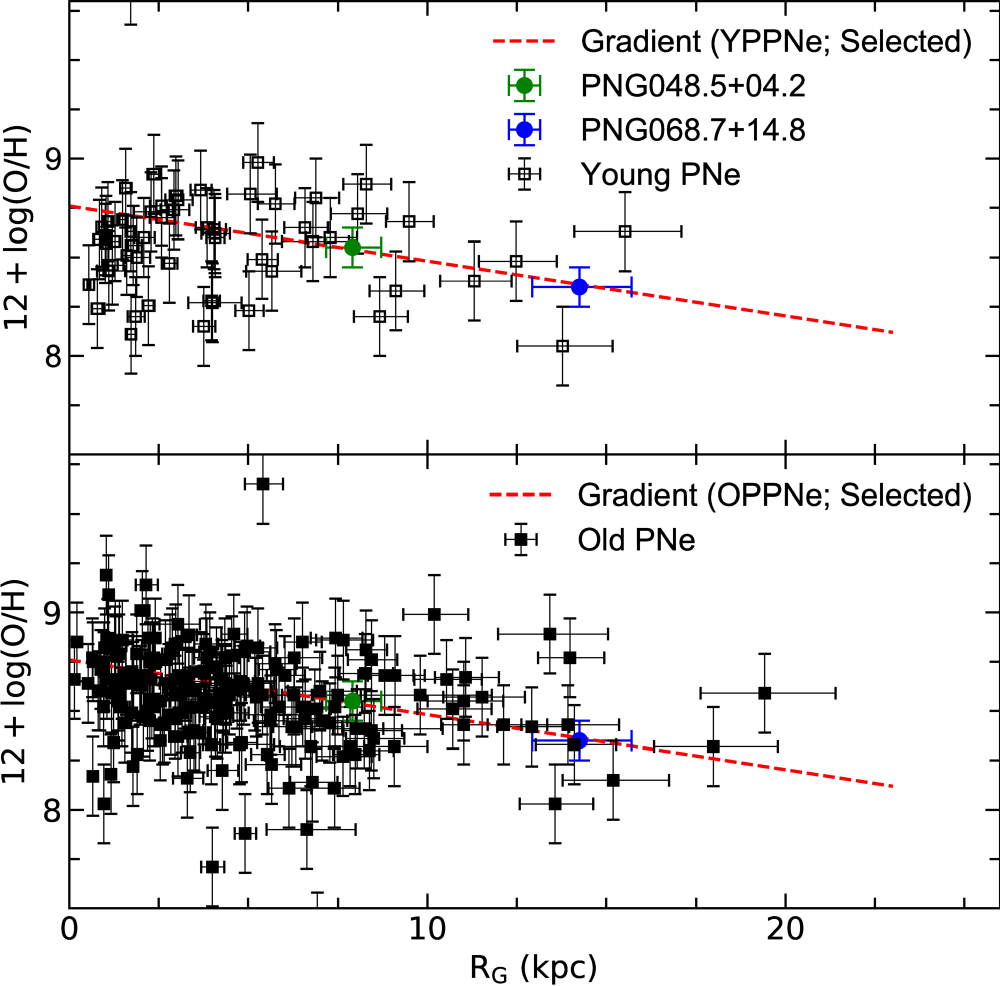} 
\caption{Galactic radial oxygen gradients of the PNe \citep[black symbols, samples adopted from][]{Bucciarelli_2023} with young progenitors (YPPNe, top panel) and old progenitors (OPPNe, bottom panel) as defined according to \citet{SH_2018}.  PN\,G048.5+04.2 (green dot) and PN\,G068.7+14.8 (blue dot) are overplotted, with 0.1\,dex error bar in oxygen abundance set for both; their $R_{\rm G}$ values are adopted from \citet{Bucciarelli_2023}.  In each panel, the red-dashed line is a linear fit to the PN sample of \citet{Bucciarelli_2023}.} 
\label{fig13}
\end{center}
\end{figure}

\subsection{Comparison with AGB Model Predictions} 
\label{subsec:AGBmodels}

The chemical yields of an AGB star depends on the initial mass ($M_{\rm ini}$) of its main-sequence progenitor and metallicity ($Z$).  For an AGB star with $M_{\rm ini}\sim$1--8\,$M_{\odot}$, its surface chemical composition can be changed due to the recurrent mixing events that bring the synthesized material to the surface \citep[e.g.][]{2005ARA&A..43..435H,Karakas_2014}.  Depending on $Z$, AGB stars with $M_{\rm ini}\gtrsim$3--4\,$M_{\odot}$ experience the second dredge-up and hot bottom burning (HBB), resulting in a significant increase in the surface nitrogen content at the expense of carbon and oxygen \citep[e.g.][]{Karakas_2010}.  The relative abundances of a PN are thus indicative of its $M_{\rm ini}$.  A number of theoretical models have been published to investigate the stellar yields from AGB nucleosynthesis at various $M_{\rm ini}$ and $Z$ cases \citep[e.g.][]{Karakas_2010, Karakas_etal2010, Karakas_etal2014, KL_2016, Ventura_2013, Ventura_2014, Fishlock_2014, Pignatari_2016}.  

In Figure\,\ref{fig12}, we compare the He/H, O/H, N/O and Ne/H ratios of the two PNe, as measured from our GTC deep spectroscopy, with those predicted by the AGB nucleosynthesis models at different $M_{\rm ini}$ and $Z$ cases (sources of the AGB models used in comparison are mentioned in the caption of Figure\,\ref{fig12}).  Depending on $M_{\rm ini}$, the mixing processes during the evolution of a low- to intermediate-mass star increase its surface nitrogen by a certain amount \citep{2005ARA&A..43..435H}.  However, the N/O ratios of both PNe are lower than the lowest levels of the model predictions of \citet{Karakas_2010} and \citet{KL_2016}, whose model yields are insensitive to stellar mass $\leq$3\,$M_{\odot}$ (Figure\,\ref{fig12}-top).  Given that these AGB models are unlikely to over-predict the N/O ratios \citep[][see discussion in Section\,4.3 therein]{2018ApJ...853...50F}, the low-mass progenitors of PN\,G048.5+04.2 and PN\,G068.7+14.8 were probably indeed born in the ISM with very low N/O ratios. 

By contrast, the N/O ratio predicted by the AGB models of \citet{Ventura_2013,Ventura_2014} extends to a lower level of 0.02 (i.e.\ $\log{\rm (N/O)}$ = $-$1.68), and is very sensitive to $M_{\rm ini}$ (Figure\,\ref{fig12}-bottom).  The differences in model predictions are mainly due to the different initial compositions adopted in the two sets of AGB models, although prescriptions for convection are also different -- the third dredge-up and HBB were considered to occur at $\sim$3\,$M_{\odot}$ in Ventura's models, while at $\gtrsim$4--5\,$M_{\odot}$ in the AGB models of Karakas et al.  The initial masses of both PN\,G048.5+04.2 and PN\,G068.7+14.8 are generally constrained by Ventura's models to to be $\lesssim$2.5\,$M_{\odot}$, as discerned from the $\log{\rm (N/O)}$ v.s.\ $\log{\rm He/H}$ relation in Figure\,\ref{fig12} (bottom-left).  Using the $\log{\rm (N/O)}$ v.s.\ $\log{\rm (O/H)}$ relation, the initial mass of PN\,G068.7+14.8 can be constrained to be $<$1.5\,$M_{\odot}$ (consistent with our photoionization model result for this PN in Table\,\ref{tab:Central stars}) if Ventura's AGB models at $Z$=0.004 is adopted.

\subsection{Comparison with Galactic Abundance Gradients} 
\label{subsec:gradient} 

There are two hypotheses of the formation of galaxy disks, one is the inside-out model through accretion of infalling gas \citep{1976MNRAS.176...31L}, however some models suggest that galaxy disks can form outside-in \citep{2004ApJ...606...32R}.  Galactic elemental abundance gradients can help us to constrain the mechanism of disk formation and to understand the evolution of the Galactic disk.  Thanks to the widespread distribution of Galactic PNe and their strong emission lines, we can use PNe to derive the abundance gradients of Galactic disk to very high accuracy.  The chemical compositions of PNe reflect the abundances of the ISM where the PNe progenitors formed; thus we can investigate the evolution of Galactic abundance gradients with time using the PNe that evolved from progenitors of different lifetimes. 

In Figure\,\ref{fig13}, we locate PN\,G048.5+04.2 and PN\,G068.7+14.8 on the Galactic radial abundance gradients of oxygen exhibited by the Galactic PNe \citep[samples adopted from][]{Bucciarelli_2023} with progenitors younger than 1\,Gyr (YPPNe) and those with progenitors older than 7.5\,Gyr (OPPNe), as defined by \citet{SH_2018}.  The elemental abundances derived in Section\,\ref{subsec:element} were used in this comparison.  It is always difficult to determine accurate distances to Galactic PNe (mainly due to extinction in the disk), in particular, to the compact PNe.  Recently, \citet{Bucciarelli_2023} obtained revised statistical distances of Galactic PNe based on an astrometrically defined sample of central stars from the \emph{Gaia} DR3 measurements as calibrators.  These new distances of the two PNe ($D$ = 7.905$^{+0.800}_{-0.726}$\,kpc for PN\,G048.5+04.2 and 14.251$^{+1.457}_{-1.322}$\,kpc for PN\,G068.7+14.8) were adopted in Figure\,\ref{fig13}. 

Our two PNe visually agree with the oxygen gradients of both YPPNe and OPPNe samples, within measurement uncertainties.  This might be related to the fact that the main-sequence ages of the two PNe (1.35\,Gyr for PN\,G048.5+04.2 and 2.80\,Gyr for PN\,G068.7+14.8; Table\,\ref{tab:Central stars}) are between those of YPPNe and OPPNe.

\section{Summary and Conclusions} 
\label{sec:conclusion}

We report deep long-slit spectroscopy of two Galactic compact PNe, PN\,G048.5+04.2 and PN\,G068.7+14.8, using the OSIRIS spectrograph on the 10.4\,m GTC.  The spectroscopy covers $\sim$3630--10,370\,{\AA} in wavelength, enabling detection of the nebular emission lines that are critical for plasma diagnostics and subsequent ionic abundance determinations.  Also detected in the GTC spectrum of PN\,G068.7+14.8 are broad features of C\,{\sc iii} and C\,{\sc iv} probably due to stellar emission, indicating this PN has a [WC]-type central star.  We carefully measured nebular emission lines to obtain their fluxes, and then carried out detailed spectral analysis to obtain the physical and chemical properties of the two PNe; empirical ICFs from the literature were adopted to derive elemental abundances.  In both PNe, $T_{\rm e}$(CEL)'s are higher than $T_{\rm e}$(H\,{\sc i} PJ), a common phenomenon found in numerous Galactic PNe and H\,{\sc ii} regions; a classical explanation of this discrepancy in temperature determination, in particular for H\,{\sc ii} regions, is temperature fluctuation. 

Spectral analyses show that PN\,G048.5+04.2 is a high-excitation PN with moderate density ($N_{\rm e}\sim$2000--3000\,cm$^{-3}$), while PN\,G068.7+14.8 is of relatively low excitation but probably with higher density ($N_{\rm e}\gtrsim$10$^{4}$\,cm$^{-3}$).  Compared to other Galactic PNe, both PN\,G048.5+04.2 and PN\,G068.7+14.8 have moderate abundance levels, but their N/O ratios are close to Type\,II.  Both PNe are sub-solar in oxygen.  PN\,G068.7+14.8 is obviously carbon-rich (compared to the Sun) in nebular abundance, which is consistent with its [WC]-type central star. 

The archival \emph{Spitzer}/IRS mid-IR spectra of the two PNe were also analyzed, and abundances of the ionic species (of O, Ne and S) that were undetected in the GTC optical-NIR spectra were derived.  Elemental abundances were then determined using the \emph{Spitzer}+GTC combined data, and were used as initial inputs for the follow-up photoionization modeling of the two PNe.  In both PNe, we found inconsistency in the Ne$^{2+}$ (and also S$^{2+}$) ionic abundances derived using the mid-IR and optical CELs, at odds with recent findings in H\,{\sc ii} regions that IR CELs and optical CELs yield consistent abundances. 

Photoionization models of the two PNe were constructed using the {\sc cloudy} code, and their central-star parameters ($T_{\rm eff}$ and $L_{\ast}$) were constrained by comparing the model-output line fluxes with those measured from the GTC optical-NIR spectra.  Grids of stellar parameters were built for model running.  By adjusting the input model parameters and then comparing the model-predicted and the observed line-flux ratios (which reflect the physical and chemical properties of a PN), we obtained for the two PNe the best {\sc cloudy} models that are in excellent agreement ($\lesssim$2.0\%) with observations.  The $T_{\rm eff}$ and $L_{\ast}$ yielded by our best photoionization models were adopted as the parameters of the PN central stars.  By comparing the central-star locations in the H-R diagram with the H-burning post-AGB evolutionary model tracks, we obtained the core masses of the two PNe.  Using the initial-final mass relations of white dwarfs, we estimated the initial masses of the progenitors for the two PNe and consequently, their main-sequence ages. 

Our population analyses, based on the assumption of single-star evolution, indicate that both PNe evolved from low-mass stars ($M_{\rm ini}<$2\,$M_{\odot}$) probably with main-sequence ages $<$3\,Gyr.  PN\,G048.5+04.2's progenitor is slightly more massive than that of PN\,G068.7+14.8.  Although both PNe are compact in angular size, their central stars are in different evolutionary stages in the H-R diagram:  the central star of PN\,G068.7+14.8 is still in the process of heating up, while that of PN\,G048.5+04.2 evolves faster and is already on the white dwarf cooling track. 

We compared the abundance ratios (in particular, N/O) of the PNe with those predicted by the AGB nucleosynthesis models, and found that both PNe likely evolved from low-mass progenitors, consistent with our estimation through photoionization modeling.  We also compared the oxygen abundances of the two PNe with the radial abundance gradients (with linear fits) exhibited by the Galactic PNe with young and old progenitors (defined as $t_{\rm Progenitor}<$1\,Gyr and $t_{\rm Progenitor}>$7.5\,Gyr, respectively), using the most recent Galactocentric distances reported in the literature.  Both PNe agree well with the oxygen gradients of the Galactic samples, although our sample is too small to draw any definite conclusion.  A new work based on the analysis of $\sim$20 Galactic compact PNe is being underway (Fang et al.\ in preparation), which will incorporate a comparison study with the abundance gradients exhibited by Galactic PNe and H\,{\sc ii} regions.

\begin{acknowledgments}
We are very grateful to the anonymous referee, whose insightful comments and suggestions greatly improved this article.  We also thank Dr.\ Yuguang Chen (at the Chinese University of Hong Kong) for in-depth discussion during paper revision.  This work was supported by the National Key R\&D Program of China (Grant No. 2023YFA1607902), and China Manned Space Program with grant No.\ CMS-CSST-2025-A14.  X.H.L.\ acknowledges support from the Natural Science Foundation of Xinjiang Uygur Autonomous Region (No. 2024D01E37) and the National Science Foundation of China (grant No. \#12473025).  X.F.\ acknowledges support from the ``Tianchi Talents'' Program (2023) of the Xinjiang Autonomous Region, China.  J.F.L.\ acknowledges support from the New Cornerstone Science Foundation through the New Cornerstone Investigator Program and the XPLORER PRIZE. 
\end{acknowledgments}

\section*{Data Availability}

The \emph{HST} WFC3 F502N images of PN\,G048.5+04.2 and PN\,G068.7+14.8 presented in Figure\,\ref{fig1} were obtained from the Mikulski Archive for Space Telescopes (MAST) at the Space Telescope Science Institute (STScI), and can be accessed via \dataset[doi:10.17909/dy8k-8f57]{https://hla.stsci.edu/hlaview.html}.




\begin{table*}
\begin{center}
\tablenum{1}
\caption{Fluxes and Intensities of the Emission Lines Detected in the GTC Optical Spectra} 
\label{tab:lines}
\begin{tabular}{lllcccc}
\hline\hline
Ion & $\lambda_{\rm lab}$ & Transition & \multicolumn{2}{c}{\underline{~~~~~PN\,G048.5$+$04.2~~~~~}} & \multicolumn{2}{c}{\underline{~~~~~PN\,G068.7$+$14.8~~~~~}} \\
  & ({\AA}) & (Lower -- Upper) & $F$($\lambda$) & $I$($\lambda$) & $F$($\lambda$) & $I$($\lambda$) \\
\hline
$[$O\,{\sc ii}$]$ & 3727.4\tablenotemark{\rm{\scriptsize a}} & \textrm{$2p^3\ ^4S^{\rm{o}}-2p^3\ ^2D^{\rm{o}}$} & 5.07 & 9.47$\pm 2.47$ & 45.1 & 57.1$\pm 1.5$ \\
H\,{\sc i} & 3770.6 & \textrm{$2p\ ^2P^{\rm{o}}-11d\ ^2D$} & 1.86 & 3.42$\pm2.30$ & 2.62 & 3.30$\pm0.37$ \\
H\,{\sc i} & 3797.9 & \textrm{$2p\ ^2P^{\rm{o}}-10d\ ^2D$} & 2.39 & 4.34$\pm2.56$ & 3.53 & 4.42$\pm0.34$ \\
He\,{\sc i} & 3819.6 & \textrm{$2p\ ^3P^{\rm{o}}-6d\ ^3D$} & $\cdots$ & $\cdots$ & 1.24 & 1.55$\pm0.20$ \\
H\,{\sc i} & 3835.4 & \textrm{$2p\ ^2P^{\rm{o}}-9d\ ^2D$} & 3.64 & 6.51$\pm0.74$ & 5.73 & 7.13$\pm0.29$ \\
$[$Ne\,{\sc iii}$]$& 3868.8 & \textrm{$2p^4\ ^3P_2-2p^4\ ^1D_2$} & 54.9 & 96.6$\pm1.3$ & 15.8 & 19.6$\pm0.3$ \\
H\,{\sc i} & 3889.1\tablenotemark{\rm{\scriptsize b}} & \textrm{$2p\ ^2P^{\rm{o}}-8d\ ^2D$} & 9.77 & 17.04$\pm0.94$ & 16.40 & 20.23$\pm0.36$ \\
$[$Ne\,{\sc iii}$]$& 3967.5\tablenotemark{\rm{\scriptsize c}} & \textrm{$2p^4\ ^3P_1-2p^4\ ^1D_2$} & 27.4 & 29.7$\pm0.8$ & 20.7 & 6.7$\pm0.9$ \\
N\,{\sc iii} & 4003.7 & \textrm{$4d\ ^2D-5f\ ^2F^{\rm{o}}$} & 0.98 & 1.62$\pm0.53$ & $\cdots$ & $\cdots$ \\
He\,{\sc i} & 4026.2 & \textrm{$2p\ ^3P^{\rm{o}}-5d\ ^3D$} & 1.06 & 1.73$\pm 0.47$ & 2.40 & 2.88$\pm0.35$ \\
$[$S\,{\sc ii}$]$ & 4068.6\tablenotemark{\rm{\scriptsize d}} & \textrm{$3p^3\ ^4S_{3/2}^{\rm{o}}-3p^3\ ^2P_{3/2}^{\rm{o}}$} & 1.27 & 2.02$\pm0.44$ & 1.42 & 1.69$\pm0.31$ \\
H\,{\sc i} & 4101.7 & \textrm{$2p\ ^2P^{\rm{o}}-6d\ ^2D$} & 19.03 & 27.54$\pm0.52$ & 23.48 & 27.71$\pm0.43$ \\
He\,{\sc i} & 4143.8 & \textrm{$2p\ ^1P^{\rm{o}}-6d\ ^1D$} & $\cdots$ & $\cdots$ & 0.28 & 0.33$\pm0.17$ \\
O\,{\sc ii} & 4189.8 & \textrm{$3p\ ^2F^{\rm{o}}-3d\ ^2G$} & $\cdots$ & $\cdots$ & 0.23 & 0.27$\pm0.14$ \\
He\,{\sc ii} & 4199.8 & \textrm{$4f\ ^2F^{\rm{o}}-11g\ ^2G$} & 0.80 & 1.08$\pm0.40$ & $\cdots$ & $\cdots$ \\
C\,{\sc ii} & 4267.3 & \textrm{$3d\ ^2D-4f\ ^2F^{\rm{o}}$} & 0.11 & 0.16$\pm :$ & 1.24 & 1.41$\pm0.15$ \\
H\,{\sc i} & 4340.5\tablenotemark{\rm{\scriptsize e}} & \textrm{$2p\ ^2P^{\rm{o}}-5d\ ^2D$} & 32.91 & 44.62$\pm0.35$ & 40.11 & 44.99$\pm0.42$ \\
$[$O\,{\sc iii}$]$ & 4363.2 & \textrm{$2p^2\ ^1D_2-2p^2\ ^1S_0$} & 14.57 & 19.48$\pm0.31$ & 4.49 & 5.01$\pm0.23$ \\
He\,{\sc i} & 4387.9 & \textrm{$2p\ ^1P_1^{\rm{o}}-5d\ ^1D_2$} & 0.08 & 0.11$\pm :$ & 0.48 & 0.53$\pm0.13$ \\
He\,{\sc i} & 4437.6 & \textrm{$2p\ ^1P_1^{\rm{o}}-5s\ ^1S_0$} & $\cdots$ & $\cdots$ & 0.38 & 0.42$\pm0.08$ \\
He\,{\sc i} & 4471.5 & \textrm{$2p\ ^3P^{\rm{o}}-4d\ ^3D$} & 1.47 & 1.84$\pm0.22$ & 4.58 & 4.98$\pm0.15$ \\
He\,{\sc ii} & 4541.6 & \textrm{$4f\ ^2F^{\rm{o}}-9g\ ^2G$} & 1.98 & 2.37$\pm0.17$ & $\cdots$ & $\cdots$ \\
N\,{\sc iii} & 4640.6\tablenotemark{\rm{\scriptsize f}} & \textrm{$3p\ ^2P_{3/2}^{\rm{o}}-3d\ ^2D_{5/2}$} & 3.85 & 4.35$\pm0.52$ & $\cdots$ & $\cdots$ \\
C\,{\sc iii} & 4649.6\tablenotemark{\rm{\scriptsize g}} & \textrm{$3s\ ^3S–3p\ ^3P^o$} & $\cdots$ & $\cdots$ & 11.04 & 11.54$\pm$0.22 \\
He\,{\sc ii} & 4685.7 & \textrm{$3d\ ^2D-4f\ ^2F^{\rm{o}}$} & 65.1 & 71.7$\pm0.2$ & 2.18 & 2.26$\pm0.16$ \\
$[$Ar\,{\sc iv}$]$ & 4711.4\tablenotemark{\rm{\scriptsize h}} & \textrm{$3p^3\ ^4S_{3/2}^{\rm{o}}-3p^3\ ^4D_{5/2}^{\rm{o}}$} & 9.66 & 9.71$\pm0.16$ & $\cdots$ & $\cdots$ \\
He\,{\sc i} & 4713.2\tablenotemark{\rm{\scriptsize i}} & \textrm{$2p\ ^3P^{\rm{o}}-4d\ ^3S$} & $\cdots$ & $\cdots$ & 0.91 & 0.94$\pm0.07$ \\
$[$Ne\,{\sc iv}$]$ & 4724.8\tablenotemark{\rm{\scriptsize j}} & \textrm{$2p^3\ ^2D_{3/2}^{\rm{o}}-2p^3\ ^2P_{3/2}^{\rm{o}}$} & 0.75 & 0.81$\pm0.11$ & $\cdots$ & $\cdots$ \\
$[$Ar\,{\sc iv}$]$ & 4740.2 & \textrm{$3p^3\ ^4S_{3/2}^{\rm{o}}-3p^3\ ^4D_{3/2}^{\rm{o}}$} & 7.77 & 8.30$\pm0.21$ & $\cdots$ & $\cdots$ \\
H\,{\sc i} & 4860.6\tablenotemark{\rm{\scriptsize e}} & \textrm{$2p\ ^2P^{\rm{o}}-4d\ ^2D$} & 100 & 100 & 100 & 100 \\
He\,{\sc i} & 4921.9 & \textrm{$2p\ ^1P_1^{\rm{o}}-4d\ ^1D_2$} & 0.77 & 0.75$\pm0.40$ & 1.47 & 1.45$\pm0.14$ \\
$[$O\,{\sc iii}$]$ & 4958.9 & \textrm{$2p^2\ ^1P_1-2p^2\ ^1D_2$} & 430  & 409$\pm01$ & 205 & 201$\pm1$ \\
$[$O\,{\sc iii}$]$ & 5006.8 & \textrm{$2p^2\ ^1P_2-2p^2\ ^1D_2$} & 1360 & 1263$\pm3$ & 625 & 608$\pm1$ \\
He\,{\sc ii} & 5411.5 & \textrm{$4f\ ^2F^{\rm{o}}-7g\ ^2G$} & 7.50 & 5.87$\pm0.33$ & 0.22 & 0.20$\pm0.06$ \\
$[$Cl\,{\sc iii}$]$ & 5517.7 & \textrm{$3p^3\ ^4S_{3/2}^{\rm{o}}-3p^3\ ^2D_{5/2}^{\rm{o}}$} & 0.54 & 0.41$\pm0.07$ & 0.20 & 0.18$\pm0.03$ \\
$[$Cl\,{\sc iii}$]$ & 5537.9 & \textrm{$3p^3\ ^4S_{3/2}^{\rm{o}}-3p^3\ ^2D_{3/2}^{\rm{o}}$} & 0.37 & 0.28$\pm0.06$ & 0.32 & 0.29$\pm0.04$ \\
C\,{\sc iii} & 5695.9 & \textrm{$3p\ ^1P_1^{\rm{o}}-3d\ ^1D_2^{\rm{o}}$} & $\cdots$ & $\cdots$ & 0.91 & 0.80$\pm$0.13 \\
$[$N\,{\sc ii}$]$ & 5754.6 & \textrm{$2p^2\ ^1D_2-2p^2\ ^1S_0$} & $\cdots$ & $\cdots$ & 0.55 & 0.48$\pm0.08$ \\
C\,{\sc iv} & 5805.0 & \textrm{$3s\ ^2S-3p\ ^2P^o$} & $\cdots$ & $\cdots$ & 3.68 & 3.19$\pm$0.16 \\
He\,{\sc i} & 5875.6 & \textrm{$2p\ ^3P^{\rm{o}}-3d\ ^3D$} & 8.13 & 5.47$\pm0.25$ & 17.79 & 15.33$\pm0.06$ \\
He\,{\sc ii} & 6233.8 & \textrm{$5g\ ^2G-17h\ ^2H^{\rm{o}}$} & 0.45 & 0.27$\pm0.11$ & $\cdots$ & $\cdots$ \\
$[$O\,{\sc i}$]$& 6300.3 & \textrm{$2p^4\ ^3P_2-2p^4\ ^1D_2$} & $\cdots$ & $\cdots$ & 0.92 & 0.76$\pm0.03$ \\
$[$S\,{\sc iii}$]$ & 6312.1\tablenotemark{\rm{\scriptsize k}} & \textrm{$3p^2\ ^1D_2-3p^2\ ^1S_0$} & 1.77& 0.86$\pm0.09$ & 0.58 & 0.48$\pm0.03$ \\
$[$O\,{\sc i}$]$& 6363.8 & \textrm{$2p^4\ ^3P_2-2p^4\ ^1D_2$} & $\cdots$ & $\cdots$ & 0.33 & 0.27$\pm0.02$ \\
\hline
\end{tabular}
\caption{(Continued)}
\end{center}
\end{table*}

\addtocounter{table}{-1}
\begin{table*}
\begin{center}
\tablenum{1}
\caption{(Continued)}
\label{tab:lines}
\begin{tabular}{lllcccc}
\hline\hline
Ion & $\lambda_{\rm lab}$ & Transition & \multicolumn{2}{c}{\underline{~~~~~PN\,G048.5$+$04.2~~~~~}} & \multicolumn{2}{c}{\underline{~~~~~PN\,G068.7$+$14.8~~~~~}} \\
  & ({\AA}) & (Lower -- Upper) & $F$($\lambda$) & $I$($\lambda$) & $F$($\lambda$) & $I$($\lambda$) \\
\hline
He\,{\sc ii} & 6406.4 & \textrm{$5g\ ^2G-15h\ ^2H^{\rm{o}}$} & 0.59 & 0.34$\pm0.04$ & $\cdots$ & $\cdots$ \\
$[$Ar\,{\sc v}$]$ & 6435.1 & \textrm{$3p^2\ ^3P_1-3p^2\ ^1D_2$} & 0.91 & 0.53$\pm0.06$ & $\cdots$ & $\cdots$ \\
C\,{\sc ii} & 6462.0 & \textrm{$4f\ ^2F^{\rm{o}}-6g\ ^2G$}  & $\cdots$ & $\cdots$ & 0.21 & 0.17$\pm0.04$  \\
He\,{\sc ii} & 6527.1 & \textrm{$5g\ ^2G-14h\ ^2H^{\rm{o}}$} & 1.04 & 0.59$\pm0.09$ & $\cdots$ & $\cdots$ \\
$[$N\,{\sc ii}$]$ & 6548.0 & \textrm{$2p^2\ ^3P_1-2p^2\ ^1D_2$} & 1.86 & 1.05$\pm0.08$ & 8.50 & 6.84$\pm0.09$ \\
H\,{\sc i} & 6562.8\tablenotemark{\rm{\scriptsize e}} & \textrm{$2p\ ^2P^{\rm{o}}-3d\ ^2D$} & 561 & 301$\pm1$ & 342 & 275$\pm1$ \\
$[$N\,{\sc ii}$]$ & 6583.5 & \textrm{$2p^2\ ^3P_2-2p^2\ ^1D_2$} & 3.95 & 2.20$\pm0.10$ & 25.60 & 20.54$\pm0.08$ \\
He\,{\sc i} & 6678.2\tablenotemark{\rm{\scriptsize l}} & \textrm{$2p\ ^1P_1^{\rm{o}}-3d\ ^1D_2$} & 4.39& 1.38$\pm0.08$ & 4.96 & 3.94$\pm0.07$ \\
$[$S\,{\sc ii}$]$ & 6716.4 & \textrm{$3p^3\ ^4S_{3/2}^{\rm{o}}-3p^3\ ^2D_{5/2}^{\rm{o}}$} & 0.59 & 0.32$\pm0.05$ & 0.49 & 0.39$\pm0.06$ \\
$[$S\,{\sc ii}$]$ & 6730.8 & \textrm{$3p^3\ ^4S_{3/2}^{\rm{o}}-3p^3\ ^2D_{3/2}^{\rm{o}}$} & 0.80 & 0.43$\pm0.06$ & 1.10 & 0.87$\pm0.07$ \\
He\,{\sc ii} & 6890.9 & \textrm{$5g\ ^2G-12h\ ^2H^{\rm{o}}$} & 0.74 & 0.38$\pm0.12$ & $\cdots$ & $\cdots$ \\
$[$Ar\,{\sc v}$]$ & 7005.7 & \textrm{$3p^2\ ^3P_2-3p^2\ ^1D_2$} & 2.32 & 1.16$\pm0.07$ & $\cdots$ & $\cdots$ \\
He\,{\sc i} & 7065.7 & \textrm{$2p\ ^3P^{\rm{o}}-3s\ ^3S$} & 4.10 & 2.02$\pm0.06$ & 11.00 & 8.42$\pm0.12$ \\
$[$Ar\,{\sc iii}$]$ & 7135.8 & \textrm{$3p^4\ ^3P_2-3p^4\ ^1D_2$} & 18.42 & 8.89$\pm0.04$ & 5.56 & 4.23$\pm0.10$ \\
He\,{\sc ii} & 7177.5 & \textrm{$5g\ ^2G-11h\ ^2H^{\rm{o}}$} & 2.64 & 1.26$\pm0.05$ & $\cdots$ & $\cdots$ \\
C\,{\sc ii} & 7234.3 & \textrm{$3p\ ^2P^{\rm{o}}-3d\ ^2D$}  & $\cdots$ & $\cdots$ & 2.47 & 1.86$\pm0.11$  \\
$[$Ar\,{\sc iv}$]$ & 7237.4\tablenotemark{\rm{\scriptsize m}} & \textrm{$3p^3\ ^2D_{5/2}^{\rm{o}}-3p^3\ ^2P_{3/2}^{\rm{o}}$} & 0.81 & 0.38$\pm0.05$ & $\cdots$ & $\cdots$ \\
$[$Ar\,{\sc iv}$]$ & 7262.8 & \textrm{$3p^3\ ^2D_{3/2}^{\rm{o}}-3p^3\ ^2P_{1/2}^{\rm{o}}$} & 0.84 & 0.39$\pm0.06$ & $\cdots$ & $\cdots$ \\
He\,{\sc i} & 7281.4 & \textrm{$2p\ ^1P_1^{\rm{o}}-3s\ ^1S_0$} & 0.54 & 0.25$\pm0.05$ & 1.30 & 0.97$\pm0.06$ \\
$[$O\,{\sc ii}$]$ & 7325.0\tablenotemark{\rm{\scriptsize n}} & \textrm{$2p^3\ ^2D^{\rm{o}}-2p^3\ ^2P^{\rm{o}}$} & 1.94 & 0.89$\pm0.11$ & 14.08 & 10.51$\pm0.19$ \\
$[$Cl\,{\sc iv}$]$ & 7530.8 & \textrm{$3p^2\ ^3P_1-3p^2\ ^1D_2$} & 1.72 & 0.75$\pm0.06$ & $\cdots$ & $\cdots$ \\
He\,{\sc ii} & 7592.8 & \textrm{$5g\ ^2G-10h\ ^2H^{\rm{o}}$} & 4.14 & 1.78$\pm0.45$ & $\cdots$ & $\cdots$ \\
$[$S\,{\sc i}$]$ & 7725.0 & \textrm{$3p^4\ ^1D_2-3p^4\ ^1S_0$}  & $\cdots$ & $\cdots$ &0.60 & 0.43$\pm0.22$ \\
$[$Ar\,{\sc iii}$]$ & 7751.1 & \textrm{$3p^4\ ^3P_1-3p^4\ ^1D_2$} & 5.41 & 2.23$\pm0.42$ & 1.39 & 1.00$\pm0.15$ \\
$[$Cl\,{\sc iv}$]$ & 8045.6 & \textrm{$3p^2\ ^3P_2-3p^2\ ^1D_2$} & 4.72 & 1.81$\pm0.09$ & $\cdots$ & $\cdots$ \\
He\,{\sc ii} & 8236.8 & \textrm{$5g\ ^2G-9h\ ^2H^{\rm{o}}$} & 5.42 & 1.99$\pm0.12$ & $\cdots$ & $\cdots$ \\
H\,{\sc i} & 8345.5 & \textrm{$3d\ ^2D-23f\ ^2F^{\rm{o}}$} & $\cdots$ & $\cdots$ & 0.20 & 0.14$\pm0.04$ \\
H\,{\sc i} & 8359.0 & \textrm{$3d\ ^2D-22f\ ^2F^{\rm{o}}$} & $\cdots$ & $\cdots$ & 0.30 & 0.20$\pm0.04$ \\
H\,{\sc i} & 8374.5 & \textrm{$3d\ ^2D-21f\ ^2F^{\rm{o}}$} & $\cdots$ & $\cdots$ & 0.23 & 0.16$\pm0.04$ \\
H\,{\sc i} & 8392.4 & \textrm{$3d\ ^2D-20f\ ^2F^{\rm{o}}$} & $\cdots$ & $\cdots$ & 0.31 & 0.21$\pm0.05$ \\
H\,{\sc i} & 8413.3 & \textrm{$3d\ ^2D-19f\ ^2F^{\rm{o}}$} & 0.62 & 0.22$\pm0.12$ & 0.41 & 0.28$\pm0.05$ \\
H\,{\sc i} & 8438.0 & \textrm{$3d\ ^2D-18f\ ^2F^{\rm{o}}$} & 0.57 & 0.20$\pm0.09$ & 0.66 & 0.44$\pm0.07$ \\
H\,{\sc i} & 8467.3 & \textrm{$3d\ ^2D-17f\ ^2F^{\rm{o}}$} & 1.30 & 0.45$\pm0.05$ & 0.61 & 0.41$\pm0.04$ \\
H\,{\sc i} & 8502.5 & \textrm{$3d\ ^2D-16f\ ^2F^{\rm{o}}$} & 1.26 & 0.44$\pm0.03$ & 0.91 & 0.61$\pm0.05$ \\
H\,{\sc i} & 8545.4 & \textrm{$3d\ ^2D-15f\ ^2F^{\rm{o}}$} & 1.32 & 0.45$\pm0.06$ & 0.87 & 0.58$\pm0.07$ \\
H\,{\sc i} & 8598.4 & \textrm{$3d\ ^2D-14f\ ^2F^{\rm{o}}$} & 2.54 & 0.86$\pm0.09$ & 1.09 & 0.73$\pm0.07$ \\
H\,{\sc i} & 8665.0 & \textrm{$3d\ ^2D-13f\ ^2F^{\rm{o}}$} & 2.94 & 0.99$\pm0.09$ & 1.38 & 0.91$\pm0.06$ \\
H\,{\sc i} & 8750.5 & \textrm{$3d\ ^2D-12f\ ^2F^{\rm{o}}$} & 4.24 & 1.40$\pm0.18$ & 1.64 & 1.08$\pm0.07$ \\
H\,{\sc i} & 8862.8 & \textrm{$3d\ ^2D-11f\ ^2F^{\rm{o}}$} & 5.50 & 1.78$\pm0.35$ & 2.14 & 1.40$\pm0.07$ \\
H\,{\sc i} & 9014.9 & \textrm{$3d\ ^2D-10f\ ^2F^{\rm{o}}$} & 6.83 & 2.16$\pm0.26$ & 1.92 & 1.25$\pm0.05$ \\
$[$S\,{\sc iii}$]$ & 9068.6 & \textrm{$3p^2\ ^3P_1-3p^2\ ^1D_2$} & 28.4 & 8.9$\pm0.2$ & 9.57 & 6.19$\pm0.10$ \\
H\,{\sc i} & 9229.0 & \textrm{$3d\ ^2D-9f\ ^2F^{\rm{o}}$} & 10.13 & 3.12$\pm0.20$ & 3.96 & 2.54$\pm0.08$ \\
He\,{\sc ii} & 9344.9 & \textrm{$5g\ ^2G-8h\ ^2H^{\rm{o}}$} & 11.36 & 3.44$\pm0.33$ & 0.64 & 0.41$\pm0.08$ \\
$[$S\,{\sc iii}$]$ & 9530.6\tablenotemark{\rm{\scriptsize o}} & \textrm{$3p^2\ ^3P_2-3p^2\ ^1D_2$} & 51.34 & 15.19$\pm0.42$ & 5.80 & 3.66$\pm0.08$ \\
\hline
\end{tabular}
\addtocounter{table}{-1}
\caption{(Continued)}
\end{center}
\end{table*}

\addtocounter{table}{-1}
\begin{table*}
\begin{center}
\tablenum{1}
\caption{(Continued)}
\label{tab:lines}
\begin{tabular}{lllcccc}
\hline\hline
Ion & $\lambda_{\rm lab}$ & Transition & \multicolumn{2}{c}{\underline{~~~~~PN\,G048.5$+$04.2~~~~~}} & \multicolumn{2}{c}{\underline{~~~~~PN\,G068.7$+$14.8~~~~~}} \\
  & ({\AA}) & (Lower -- Upper) & $F$($\lambda$) & $I$($\lambda$) & $F$($\lambda$) & $I$($\lambda$) \\
\hline
H\,{\sc i} & 10049.3\tablenotemark{\rm{\scriptsize o}} & \textrm{$3d\ ^2D-7f\ ^2F^{\rm{o}}$} & 5.32 & 1.48$\pm0.11$ & 1.98 & 1.22$\pm0.07$ \\
\\
$c$(H$\beta$) & &                         & \multicolumn{2}{c}{0.845} & \multicolumn{2}{c}{0.319} \\
\multicolumn{2}{l}{$\log$\,$F$(H$\beta$)\tablenotemark{\rm{\scriptsize p}}}  &                         & \multicolumn{2}{c}{$-$13.76} & \multicolumn{2}{c}{$-$12.27} \\
\hline
\end{tabular}
\begin{description}
NOTE. -- Fluxes and intensities are normalized such that H$\beta=100$.  ``:'' indicates that the uncertainty in line intensity is large (>100\%);  ``$\cdots$'' means the line was undetected.  Intensities are the extinction-corrected fluxes.  Some line-blending cases have been corrected for. \\
\vspace{2mm}
\tablenotemark{\rm{\scriptsize a}} A blend of the [O\,{\sc ii}] $\lambda3726$ (\textrm{$2p^3\ ^4S_{3/2}^{\rm{o}}-2p^3\ ^2D_{3/2}^{\rm{o}}$}) and $\lambda3729$ (\textrm{$2p^3\ ^4S_{3/2}^{\rm{o}}-2p^3\ ^2D_{5/2}^{\rm{o}}$}) doublet. \\
\tablenotemark{\rm{\scriptsize b}} Blended with the He\,{\sc i} $\lambda3888$ (\textrm{$2s\ ^3S–3p\ ^3P^o$}) line. \\
\tablenotemark{\rm{\scriptsize c}} Corrected for the flux from the blended H\,{\sc i} $\lambda3970$ (\textrm{$2p\ ^2P^o–7d\ ^2D$}) line and He\,{\sc i} $\lambda3965$ (\textrm{$2s\ ^1S–4p\ ^1P^o$}) line. Fluxes from possible blended C\,{\sc ii} lines were not corrected. \\
\tablenotemark{\rm{\scriptsize d}} Blended with [S\,{\sc ii}] $\lambda4076$; probably also blended with the weak O\,{\sc ii} M10 \textrm{$3p\ ^4D^o–3d\ ^4F$} and C\,{\sc iii} M16 \textrm{$4f ^3F^o–5g ^3G$} lines. \\
\tablenotemark{\rm{\scriptsize e}} Corrected for the flux from the blended He\,{\sc ii} line. \\
\tablenotemark{\rm{\scriptsize f}} Blended with the N\,{\sc iii} $\lambda\lambda$4634, 4642 lines; probably also blended with O\,{\sc ii} M1 $\lambda\lambda$4639, 4642. \\
\tablenotemark{\rm{\scriptsize g}} Blended with the O\,{\sc ii} M1 \textrm{$3s ^4P–3p ^4D^o$} lines. \\
\tablenotemark{\rm{\scriptsize h}} Corrected for the flux from the blended He\,{\sc i} $\lambda4713$ (\textrm{$2p ^3P^o–4s ^3S$}) line and [Ne\,{\sc iv}] $\lambda4714$ (\textrm{$2p^3\ ^2D_{5/2}^{\rm{o}}-2p^3\ ^2P_{3/2}^{\rm{o}}$}) and $\lambda4716$ (\textrm{$2p^3\ ^2D_{5/2}^{\rm{o}}-2p^3\ ^2P_{1/2}^{\rm{o}}$}) doublet. \\
\tablenotemark{\rm{\scriptsize i}} No other [Ar\,{\sc iv}] line was detected in PN\,G068.7+14.8, so we considered it as He\,{\sc i} $\lambda$4713. \\
\tablenotemark{\rm{\scriptsize j}} A blend of the [Ne\,{\sc iv}] $\lambda4724$ (\textrm{$2p^3\ ^2D_{3/2}^{\rm{o}}-2p^3\ ^2P_{3/2}^{\rm{o}}$}) and $\lambda4726$ (\textrm{$2p^3\ ^2D_{3/2}^{\rm{o}}-2p^3\ ^2P_{1/2}^{\rm{o}}$}) doublet.\\
\tablenotemark{\rm{\scriptsize k}} Corrected for the flux from the blended He\,{\sc ii} $\lambda6311$ (\textrm{$5g ^2G–16h ^2H^o$}) line. \\
\tablenotemark{\rm{\scriptsize l}} Corrected for the flux from the blended He\,{\sc ii} $\lambda6683$ (\textrm{$5g ^2G–13h ^2H^o$}) line. \\
\tablenotemark{\rm{\scriptsize m}} Blended with the C\,{\sc ii} $\lambda7234$ (\textrm{$3p\ ^2S^o–3d\ ^2D$}) line. \\
\tablenotemark{\rm{\scriptsize n}} A blend of the [O\,{\sc ii}] $\lambda7320$ (\textrm{$2p^3\ ^2D_{5/2}^{\rm{o}}-2p^3\ ^2P_{3/2}^{\rm{o}}$}) and $\lambda7330$ (\textrm{$2p^3\ ^2D_{3/2}^{\rm{o}}-2p^3\ ^2P_{3/2}^{\rm{o}}$}) doublet. \\
\tablenotemark{\rm{\scriptsize o}} Flux underestimated due to the second-order contamination beyond 9200 $\AA$. \\
\tablenotemark{\rm{\scriptsize p}} In units of erg cm$^{-2}$ s$^{-1}$, as measured from the extracted 1D GTC spectrum.
\end{description}
\end{center}
\end{table*}

\bibliography{references}{}
\bibliographystyle{aasjournal}
\end{document}